\documentclass[structabstract]{aa}  

\usepackage{graphicx,aalongtable}
\usepackage{epstopdf}

\usepackage{txfonts}

\usepackage{natbib}
\bibpunct[, ]{(}{)}{;}{a}{}{,}
\newcommand{\griz}{$g^\prime r^\prime i^\prime z^\prime$~}
\newcommand{\JHK}{$JHK_S$~}
\newcommand{\gK}{$g^\prime r^\prime i^\prime z^\prime JHK_S$~}

\begin{document}

  \title{Optical and near-infrared follow-up observations of four \textit{Fermi}/LAT GRBs : Redshifts, afterglows, energetics and host galaxies \thanks{Based on observations made with the ESO Telescopes at the La Silla Paranal Observatories under programme ID 083.D-0903 and 283.D-5059, the MPG/ESO 2.2~m Telescope at La Silla Observatory and the Schmidt telescope of the Th\"uringer Landessternwarte Tautenburg.} }

 \author{S. McBreen
          \inst{1,2}
          \and
          T. Kr\"{u}hler 
          \inst{2,3}
          \and
          A. Rau
          \inst{2}
          \and 
          J. Greiner
          \inst{2}
          \and
          D. A. Kann \inst{4}
          \and 
          S. Savaglio\inst{2}
          \and
          P. Afonso \inst{2}
         \and
           C. Clemens \inst{2}
		  \and
          R. Filgas \inst{2}
          \and
          S. Klose \inst{4}
          \and
                A. K\"{u}pc\"{u} Yolda\c{s} \inst{5}
          \and
          F. Olivares E.  \inst{2}
          \and
          A. Rossi \inst{4}
          \and
          G. P. Szokoly \inst{6}
          \and
          A. Updike \inst{7}
          \and
          A. Yolda\c{s} \inst{2}
          }

  \institute{School of Physics, University College Dublin, Dublin 4, Ireland. \\
              \email{Sheila.McBreen@ucd.ie}              
             \and
      Max-Planck-Institut f\"{u}r extraterrestrische Physik, 85748 Garching, Germany.
            \and
            Universe Cluster, Technische Universit\"{a}t M\"{u}nchen, Boltzmannstrasse 2, D-85748, Garching,
Germany.
            \and
          Th\"uringer Landessternwarte Tautenburg, Sternwarte 5,  D-07778 Tautenburg, Germany.
           \and
           European Southern Observatory, 85748 Garching, Germany.
                      \and
           Institute of Physics, E\"{o}tv\"{o}s University, P\'{a}zm\'{a}ny P. s. 1/A, 1117 Budapest, Hungary.
           \and
                      Department of Physics and Astronomy, Clemson University, Clemson, SC 29634, USA.
             }
                       
\authorrunning{S. McBreen et al.} 
\titlerunning{Optical and NIR follow-up of \textit{Fermi}/LAT GRBs}

\date{} 

 
  \abstract
   {} 
   { \textit{Fermi} can measure the spectral properties of gamma-ray bursts over a very large energy range and is opening a new window on the prompt emission of these energetic events.  Localizations by the instruments on  \textit{Fermi}  in combination with  follow-up by \textit{Swift}  provide accurate positions for observations at longer wavelengths leading to the determination of redshifts,  the true energy budget, host galaxy properties and facilitate comparison with pre-Fermi bursts.
   }
{Multi-wavelength follow-up observations were performed on the afterglows of four bursts with high energy emission   detected by {\textit{Fermi}/LAT} :  GRB\,090323, GRB\,090328, GRB\,090510 and GRB\,090902B. They were obtained in the optical/near-infrared bands with GROND 
mounted at the MPG/ESO 2.2\ m  telescope
 and additionally of GRB\,090323 in the optical with the 2\ m telescope in Tautenburg, Germany. Three of the events are classified as long bursts while GRB\,090510 is a well localized short GRB with GeV emission. In addition, host galaxies were detected for three of the four bursts. Spectroscopic follow-up was initiated with the VLT   for GRB\,090328 and GRB\,090510. 
      }
{
The afterglow observations in 7 bands are presented for all bursts   and  their host galaxies are investigated.             Knowledge of the distance and the local dust extinction enables comparison of the afterglows of LAT-detected GRBs with the general sample. The spectroscopic redshifts of GRB\,090328 and GRB\,090510  were determined to be $z=0.7354\pm0.0003$        and $z=0.903\pm0.001$ and dust corrected star-formation rates of $4.8$ M$_\odot$ yr$^{-1}$ and $0.60$ M$_\odot$ yr$^{-1}$ were derived for their host galaxies, respectively. 
}
   {  
   The afterglows of long bursts exhibit power-law decay indices ($\alpha$) from less than 1  to $\sim$2.3 and spectral  indices ($\beta_{\rm opt}$) values   from 0.65 to $\sim$1.2 which are fairly standard for  GRB afterglows. 
   Constraints are placed on the jet half opening angles of $\lesssim  2.1^\circ$ to  $\gtrsim 6.4^\circ$, which  allows limits to be placed on the beaming corrected energies. These range from  $\lesssim $ $5 \times 10^{50}$ erg to the one of the highest values ever recorded, $\gtrsim 2.2 \times 10^{52}$ erg  for GRB~090902B,  and are not consistent with a standard candle.
The extremely energetic long \textit{Fermi} bursts have optical afterglows which lie in the top half  of the brightness distribution of all optical afterglows detected in the \textit{Swift} era or even in the top 5 \% if incompleteness is considered.  
The properties of the host galaxies  of these LAT detected bursts in terms of extinction, star formation rates
 and  masses do not appear to differ from previous samples. 
 }


  \maketitle


\section{Introduction}

The follow-up of gamma-ray bursts (GRBs) detected by the \textit{Swift} satellite  \citep{2004ApJ...611.1005G} has led to the determination of the distance scale for a large sample of bursts. The Burst Alert Telescope \citep[BAT,][]{2005SSRv..120..143B} is sensitive in the energy range 15$-$150~keV and has good localization capabilities with typical uncertainties in the arcminute range.
Rapid follow-up by \textit{Swift's} narrow field instruments in the X-rays \citep[XRT,][]{2005SSRv..120..165B} and optical/UV \citep[UVOT,][]{2005SSRv..120...95R} have lead to  the arcsecond localizations 
required for ground-based observers
 and in turn to spectroscopic redshift measurements of a large sample of GRBs \citep[e.g.,][]{2009arXiv0907.3449F} and investigation of their host galaxies \citep[e.g.,][]{2009ApJ...691..182S,2009AJ....138.1690P}. To date distances to $\sim$ 200 GRB sources have been established with redshifts ranging from $z = 0.0085$ \citep[GRB 980425:][]{1998IAUC.6896....3W,1998Natur.395..670G} to $z\sim8.2$  \citep[GRB090423:][]{2009arXiv0906.1577T,2009arXiv0906.1578S}.  The BAT has a narrow spectral range and is not able to determine the spectral parameters of the prompt emission of GRBs over a broad energy range. Since the launch of the \textit{Fermi} Gamma-Ray Space Telescope \citep{1994NIMPA.342..302A,1996SPIE.2806...31M} there is now the possibility to investigate  
 the spectrum over seven decades in energy.  These events can be localized  by instruments from \textit{Fermi}  and followed up by  \textit{Swift} and ground-based obervatories, enabling investigation of their afterglows and host properties and facilitating comparison to a general sample. 
In some fortuitous cases instruments on the \textit{Fermi} and \textit{Swift}  satellites may trigger on the same event with high-energy emission enabling both rapid follow-up and broadband prompt emission coverage (e.g. GRB~090510, see Section 2.3).
The high-energy spectral properties of GRBs were investigated previously by instruments on-board the \textit{Compton Gamma-ray Observatory}, however the redshifts of these events are unknown. 
The spectra of these GRBs were described by an extrapolation of the low energy spectra to energies $>$100 MeV \citep[e.g.,][]{1994A&A...285..161H},  and in some cases an additional component  at high-energies 
\citep{2003Natur.424..749G,2008ApJ...677.1168K} and long-lived GeV emission \citep{1995Natur.374...94H}. 
Recently a   photometric redshift of $z=1.8^{+0.4}_{-0.3}$ was reported by \citet{2008A&A...491L..29R} for GRB~080514B which was detected at energies up to 300~MeV by instruments on the AGILE satellite \citep{2008A&A...491L..25G}.

\textit{Fermi} hosts two instruments that detect GRBs: 
 the Gamma-ray Burst Monitor \citep[GBM:][]{2009arXiv0908.0450M} sensitive to photons from $\sim$8 keV to $\sim$40 MeV, and the Large Area Telescope with a spectral coverage from $\sim$ 30 MeV to $\sim$100 GeV \citep[LAT:][]{2009ApJ...697.1071A}. 
 Together they provide valuable information  on the spectral properties of the prompt emission
 \citep[e.g.][]{2009arXiv0908.1832F,2009arXiv0910.4192F,090902B_PAPER,2009Sci...323.1688A}, the presence or absence of intrinsic spectral cut-offs or those due to the optical depth of the universe to high-energy $\gamma$-rays  due to pair production on infrared diffuse  Extragalactic Background Light  \citep[e.g.,][]{2002ApJ...566..738D,2003A&A...407..791M,2004A&A...413..807K,2005Natur.438...45K,2006ApJ...648..774S,2008A&A...487..837F,2009arXiv0905.1115F,090902B_PAPER} 
 and test for quantum gravity effects \citep[e.g.,][]{1998Natur.393..763A,2008ApJ...673..972S,2005LRR.....8....5M,2009PhRvD..80h4017A,2009Sci...323.1688A,2009arXiv0908.1832F}.
 
Up to the end of January 2010,  fourteen GRBs
have been detected by both instruments and localized by the LAT  : 
GRB 080825C \citep[][]{LAT_080825C,2009arXiv0910.4192F}, 
GRB 080916C  \citep[][]{2009Sci...323.1688A},
GRB 081024B \citep[][]{2008GCN..8407....1O}, 
GRB 081214 \citep[][]{2008GCN..8723....1W},
GRB 090217 \citep[][]{LAT_090217},
GRB 090323 \citep[][]{2009GCN..9021....1O},
GRB 090328 \citep[][]{2009GCN..9044....1M},
GRB 090510 \citep[][]{2009GCN..9334....1O,2009arXiv0908.1832F}, 
GRB 090626 \citep[][]{090626_LAT},
GRB 090902B \citep[][]{LAT_090902B,090902B_PAPER},
GRB 090926A \citep[][]{GBM_090926,LAT_090926},
GRB 091003 \citep[][]{McEnery09_GCN9985},
GRB 091031 \citep[][]{Palma09_GCN10163},
and
GRB 100116A \citep[][]{McEnery10_GCN10333}.
Twelve are long duration GRBs and two have reported durations compatible with the short burst class (GRBs~081024B and 090510).  
The redshifts of five of these bursts have been determined 
and range from $z=0.736$ to $z=4.35\pm0.15$  \citep{2009A&A...498...89G,2009GCN..9026....1U,2009GCN..9028....1C,2009GCN..9053....1C,2009_arne,Z_090902B,Xshooter_090926}. 

Among the most impressive of these events are GRB 080916C \citep{2009Sci...323.1688A}  and GRB 090510 \citep{2009arXiv0908.1832F}.  GRB~080916C is a long bright GRB for which the prompt emission spectrum could be fit over six decades in energy by the empirical Band function \citep[{see}][{for detailed results}]{2009Sci...323.1688A}. The burst was found to have a high photometric redshift of $z=4.35\pm0.15$ via ground-based follow up observations \citep{2009A&A...498...89G}. 
The distance information enabled the determination of the rest frame properties, energetics, and the placing of lower limits on the bulk Lorentz factor of the outflow ($\gtrsim 900-1100$) 
\citep{2009Sci...323.1688A,2009A&A...498...89G}.
GRB 090510 is a short burst  with a T$_{\rm 90}$ of 2.1 s   and from which a 31~GeV photon was detected \citep{2009arXiv0908.1832F}.   The distance determination \citep{2009_arne} and high energy emission allow a limit to be placed on photon dispersion  \citep{2009arXiv0908.1832F}. Moreover, very recently  a 33 GeV photon was detected from GRB~090902B \citep{090902B_PAPER} and this burst is at redshift $z = 1.822$ \citep{Z_090902B}. Clearly, the distance determination via optical spectroscopy is crucial to the interpretation of these events.

We report on the optical and near-infrared (NIR) observations of the afterglow of four bursts with high energy emission, GRB\,090323, GRB\,090328, GRB\,090510 and GRB\,090902B and compare them to the sample of GRB afterglows to date. Furthermore we also report on the spectroscopic redshift determination for GRB~090328 and GRB~090510. Throughout the paper, we adopt concordance $\Lambda$CDM cosmology ($\Omega_M=0.27$, $\Omega_{\Lambda}=0.73$, $H_0=71$~km/s/Mpc), and the convention that the flux density of the GRB afterglow can be described as $F_\nu (t)\propto \nu^{-\beta}t^{-\alpha}$


\section{Observations}


\subsection{GRB 090323}

At 00:02:42.63 UT on 23 March 2009, \textit{Fermi} GBM triggered and located the long burst GRB 090323 
\citep{2009GCN..9021....1O}. An excess of 5 $\sigma$ was detected by the LAT and the burst was localized to  R.A.(J2000), decl.(J2000)=190.69, +17.08 degrees with a 68\% containment radius of 0.09$^{\circ}$.  
Furthermore, it was reported that emission continued up to a few kiloseconds post trigger \citep{2009GCN..9021....1O}.
A triangulation by the Inter Planetary Network localized the burst to an arc which intersected the GBM and LAT positions  \citep{2008IPN090323}.  \textit{Swift} carried out a Target of Opportunity observation $\sim19$~hours post burst and 
a fading afterglow was found \citep{2009GCN..9024....1K,2009GCN..9031....1P}. 
\citet{2009GCN..9026....1U} observed the position of the X-ray afterglow with the Gamma-ray Burst Optical/Near-infrared Detector (GROND) in 7 bands $\sim$ 27 hours after the burst and detected a source for which they reported a preliminary photometric redshift of $4.0\pm0.3$ (see below). \citet{2009GCN..9028....1C} reported a spectroscopic redshift of  $z=3.57$ based on observations of the optical afterglow using the  Gemini Multi-Object Spectrograph (GMOS) mounted on the Gemini South Telescope. The GRB was also detected in the radio band
\citep{2009GCN..9043....1H,2009GCN..9047....1V}.

Using the reported spectral parameters \citep{2010arXiv1002.4194B}
and a redshift of $z = 3.57$ \citep{2009GCN..9028....1C} $E_{\gamma, \rm{iso}}$ is $4.1\times10^{54}$~erg in the 1~keV to 10~MeV range, or $5.1\times 10^{54}$~erg in the 1~keV to 10~GeV range with a restframe peak energy of $2.7\pm0.2$~MeV. The $E_{\gamma, \rm{iso}}$ is one of the highest ever measured even rivals GRB~080916C for the most energetic GRB.

\begin{figure}[t]
\centering
\includegraphics[width=0.99\columnwidth]{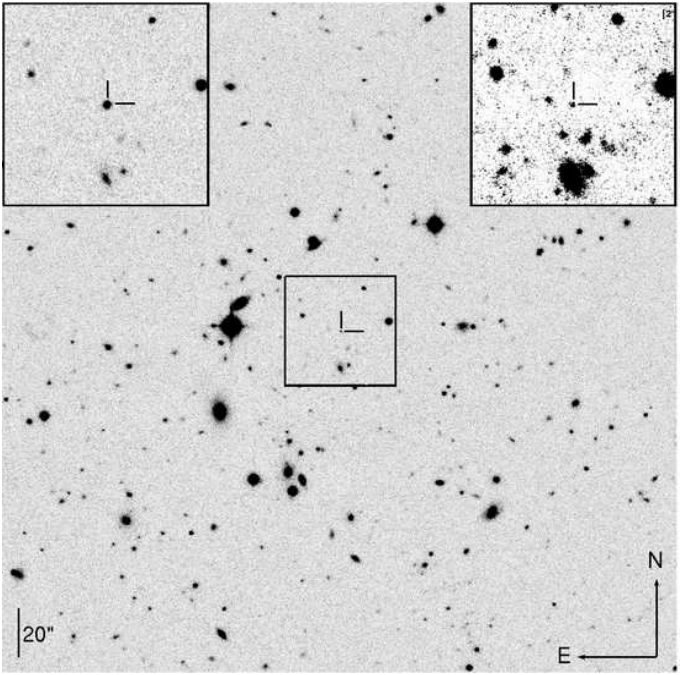}
\caption{GROND finding chart for the afterglow of GRB~090323. The main panel shows the field of GRB~090323 taken 3.5 days post burst. The field shown has 4\arcmin $\times$ 4\arcmin , where North is up and East is to the left. The afterglow position is indicated with two lines in each image. The inset in the top left shows a zoom into the afterglow region at the earliest epoch, where the afterglow was still bright. All images are $i^\prime$ band, except the host image   ($r^\prime$,  top right inset), which is also enhanced in contrast for demonstration purposes.}
\label{090323fc}
\end{figure}

The field of GRB~090323 was observed simultaneously in 7 bands (\gK) by GROND, mounted at the 2.2~m MPG/ESO telescope at La Silla Observatory, Chile \citep{2008PASP..120..405G} and using the Schmidt-camera at the Alfred-Jensch telescope of the Th\"uringer Landessternwarte Tautenburg (TLS), Germany. GROND and TLS observations started on 24th March 2009, 27 hours and 45 hours after the GBM trigger, respectively \citep{2009GCN..9026....1U, 2009GCN..9033....1K}.  Observations continued in the upcoming nights \citep{2009GCN..9041....1K, 2009GCN..9063....1K} and the last GROND epoch was obtained $\sim 100$~days after $T_{0}$. 
The best optical position of the afterglow measured in the $r^\prime$ band against the SDSS catalogue is R.A.(J2000)=12:42:50.28, decl.(J2000)=$+$17:03:11.9 with systematic and statistical uncertainties of 
$0\farcs1$
in each coordinate. A finding chart of the afterglow is shown in Fig.~\ref{090323fc}.

\subsection{GRB 090328}
At 09:36:46 UT on 28 March 2009, \textit{Fermi} GBM triggered and located a long burst GRB 090328.  The event was significantly detected up to a few GeV at the 5 $\sigma$ level by the \textit{Fermi}/LAT and was localized to R.A.(J2000), decl.(J2000)=90.87, $-$41.95 with a 68\% containment radius of 0.11 deg \citep{2009GCN..9044....1M}. 
The emission in the LAT was reported to last until about 900 seconds post trigger by \citet{2009GCN..9077....1C}.
A \textit{Swift} Target of Opportunity observation was initiated and XRT observations commenced  $\sim$16 hours post burst \citep{2009GCN..9045....1K} and  X-ray and bright UV/optical afterglow candidates were subsequently reported by \citet{2009GCN..9046....1K} and \citet{2009GCN..9048....1O}.
A spectrum of the optical afterglow was taken with the GMOS instrument mounted on the Gemini South Telescope \citep{2009GCN..9053....1C} 
and  the redshift was reported to be $z=0.736$. Further photometric observations of the afterglow were reported in the optical \citep{2009GCN..9058....1A} and optical/NIR with GROND \citep[][see below]{2009GCN..9054....1U} and additionally in the radio band \citep{2009GCN..9060....1F}.

The time averaged spectrum ($T_0+3.1$ to $T_0+29.7$~s) of the prompt emission was reported by GBM to be best fit by a Band function with indices $\alpha=-0.93 \pm0.02$ and $\beta=-2.2 \pm 0.1$, a peak energy of $E_{\rm peak}=653 \pm 45$~keV with an event fluence of ($9.5 \pm 1.0$) $\times 10^{-5}$~{erg/cm}$^2$ in the 8~keV to 40~MeV band \citep{2009GCN..9057....1R}. \citet{2010arXiv1002.4194B} report that
the spectrum ($T_0+0$ to $T_0+66.6$~s) is best fit by a power-law with exponential cut off. The index of the power-law is $-1.07\pm{0.02}$
and the peak energy is 744$^{+50}_{-47}$~keV.  At a redshift of 0.736 
, these spectral parameters correspond to $E_{\gamma, \rm{iso}}$ in the 1~keV to 10~MeV range of $1.0\times10^{53}$~erg and $1.0\times 10^{53}$~erg in the 1~keV to 10 GeV band with a restframe peak energy of $E_{\rm peak}^{\rm rest} = 1.3^{+0.09}_{-0.08}$~MeV.

{Optical/NIR follow-up observations  by GROND started 1.6 days post trigger. The afterglow was detected in all seven bands 
 consistent with the reported redshift  \citep{2009GCN..9054....1U}. In addition, GROND imaged the field of GRB~090328 at 2.5, 3.5, 4.5, 6.5 and 11.5 days after the trigger. The best optical position of the afterglow measured in the $r^\prime$ band against the USNO-B1 catalog is R.A.(J2000)=06:02:39.69, decl.(J2000)=$-$41:42:54.9 with systematic uncertainties of $0\farcs3$
  in each coordinate. 
The position of the host galaxy is R.A.(J2000)=06:02:39.69, decl.(J2000)=$-$41:52:55.1 and therefore the distance of the
afterglow from the  host is $0\farcs14$. 
  A finding chart of the afterglow is shown in Fig.~\ref{090328fc}. 
  
Spectroscopy of the afterglow was performed on 2009 March 30.01 UT ($\sim1.6$\,days post burst) using the    FOcal    Reducer   and    low    dispersion   Spectrograph    1 \citep[FORS1;][]{1998Msngr..94....1A}   at   the   8\,m  ESO-VLT   UT2 telescope (Programme ID:  083.D-0903). Observations were obtained using the  $600B$   and  $600R$   grisms,  providing  a spectral   resolution  of roughly 5\,\AA\ full width at half maximum. Two  $1800$\,s integrations were taken in each of the two grisms.

\begin{figure}[t]
\centering
\includegraphics[width=0.99\columnwidth]{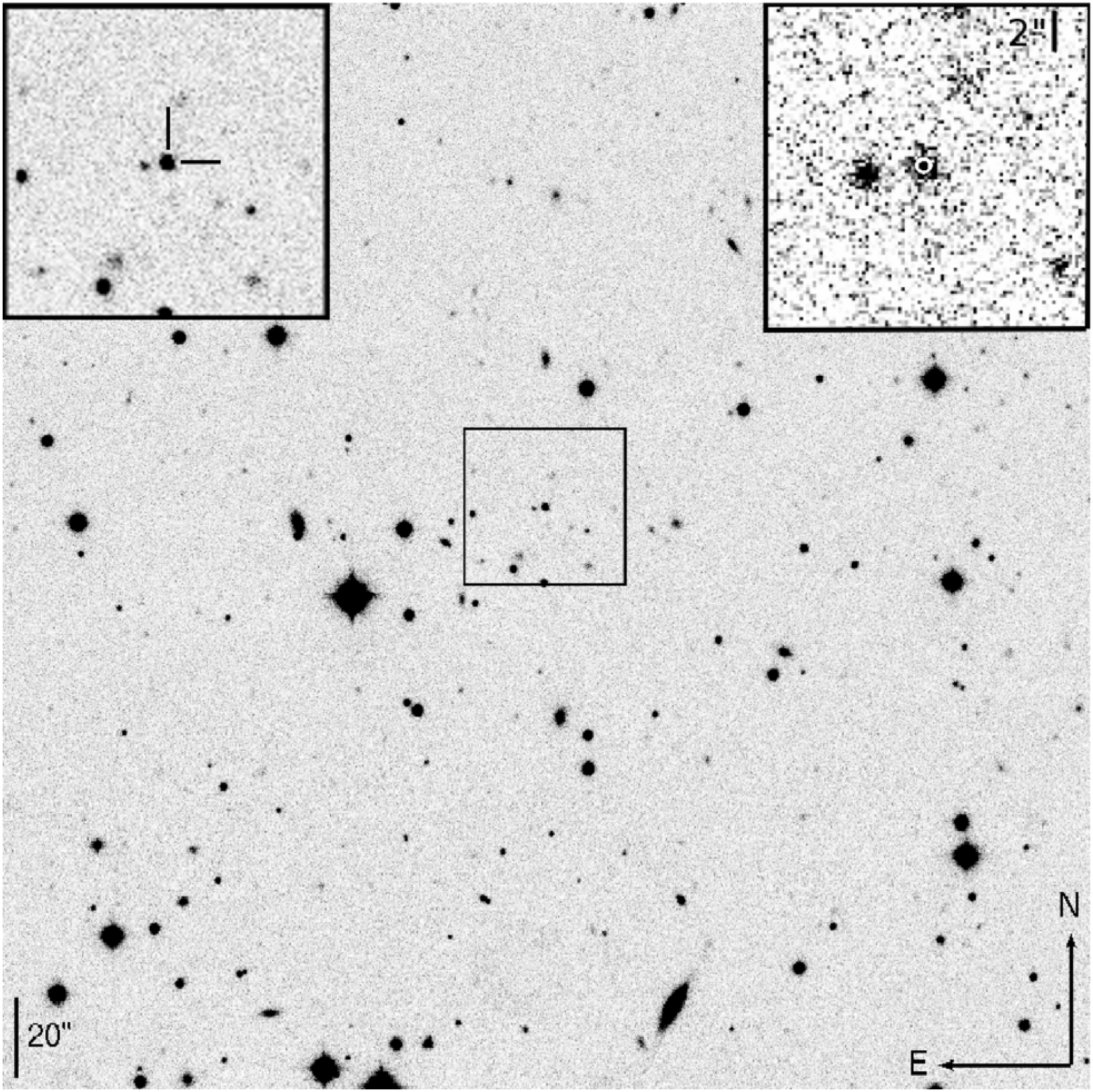}
\caption{GROND finding chart for the afterglow of GRB~090328. The main panel shows the field of GRB~090328 taken 1.5 days post burst. The field shown has 4\arcmin   $\times$ 4\arcmin , where North is up and East is to the left. The inset in the top left shows a zoom into the afterglow region at the same epoch, where the afterglow is indicated with two lines.
 The inset on the top right contains a 15\arcsec$\times$15\arcsec region and shows a host image taken 10 days post burst. The afterglow position is shown by a small white circle 
  of radius $0\farcs3$. All images are in the $i^\prime$ band.}
\label{090328fc}
\end{figure}


\subsection{GRB 090510}

At 00:23:00 UT on 10 May 2009  the \textit{Swift} BAT triggered and located GRB 090510 
 \citep{2009GCN..9331....1H}.   The burst also triggered \textit{Fermi}/GBM  
 \citep{2009GCN..9336....1G} and other instruments  \citep{2009GCN..9344....1P,2009GCN..9355....1O,2009GCN..9343....1L,2009arXiv0908.1908G}. 
The duration ($T_{\rm 90} \sim 2.1$~sec)   and the negligible spectral lags below $\sim$1~MeV 
of the event make it  
consistent with the class of short bursts \citep{2009arXiv0908.1832F}. 
Significant emission was detected by the LAT \citep{2009GCN..9334....1O,2009GCN..9350....1O}.
The time integrated spectrum using \textit{Fermi} GBM and LAT data is best fit by two spectral components, a Band function  with the Epeak = $3.9\pm 0.3$~MeV and a power-law at high energies.  The total isotropic energy is ($1.08\pm0.06)\times 10^{53}$~erg (10~keV -- 30~GeV)
\citep{2009arXiv0908.1832F}.
The event was also detected at 5$\sigma$ significance at energies above 100 MeV by AGILE  GRID  \citep{2009GCN..9343....1L,2009arXiv0908.1908G}. 

\textit{Swift} slewed after 91 seconds and observations with the XRT and UVOT  revealed an uncatalogued source \citep{2009GCN..9332....1M}.  The XRT light curve was reported to be fading \citep{2009GCN..9341....1G} and the UVOT afterglow candidate was confirmed by \citet{2009GCN..9338....1O} using the Nordic Optical Telescope. \citet{2009GCN..9351....1K} report that the afterglow of GRB 090510 is detected in almost all UVOT filters which implies that the redshift is less than about 1.5 \citep{2009GCN9342......1K}. See also \citet{2009arXiv0910.1629D}.

{GROND imaged the field of GRB~090510 starting 6.18~h after the \textit{Swift} and \textit{Fermi} triggers. At that point the position of the GRB was becoming visible above the pointing constraints of the telescope \citep{2009GCN..9352....1O}. Therefore GROND images taken in the first two hours suffer from high airmass and moderate image quality. Observations continued for 4~h until local twilight and were resumed in the following night for 1.5~h. 
In the first epoch, the afterglow is detected in the \griz optical channels, while the \JHK band yield only upper limits and the second epoch only yield upper limits on the afterglow flux.
The position of the optical transient associated with GRB~090510 is measured against USNO-B1 to R.A.(J2000)=22:14:12.54, decl.(J2000)=$-$26:34:59.1 with errors of $0\farcs5$  in each coordinate. In both epochs 
a nearby, extended object with coordinates of R.A.(J2000)=22:14:12.56, decl.(J2000)=$-$26:34:58.7 at a distance of $1\farcs2$ 
 with respect to the afterglow is clearly detected. Given the low spatial separation both objects are blended in the first epoch. A finding chart of the field of GRB 090510 is shown in Fig.~\ref{090510fc}.}  

Spectroscopy of  the host galaxy  of GRB~090510 was performed  ($\sim2.3$\,days   post   burst)   using   FORS2 \citep{1998Msngr..94....1A}   at  the   8\,m  ESO-VLT   UT1  telescope (Programme ID:  083.D-0903(A)). Observations were  obtained using  the 300$I$ grism,  providing  a  spectral  resolution of  $\approx7$\,\AA\  FWHM.
Three $1800$\,s integrations were taken. The preliminary analysis and redshift $z=0.903$ was reported by \citet{2009_arne}.  We note that the redshift  of this galaxy is consistent with the afterglow colours detected by UVOT \citep{2009GCN..9351....1K}. Additionally, spectroscopy of the source located  5\arcsec south of the GRB host galaxy was performed $\sim1.3$\,days post burst using FORS2 (300$V$  grism), which yields a redshift consistent with that of the GRB host.
Two $1800$\,s  integrations were taken of this object. 

\begin{figure}[t]
\centering
\includegraphics[width=0.99\columnwidth]{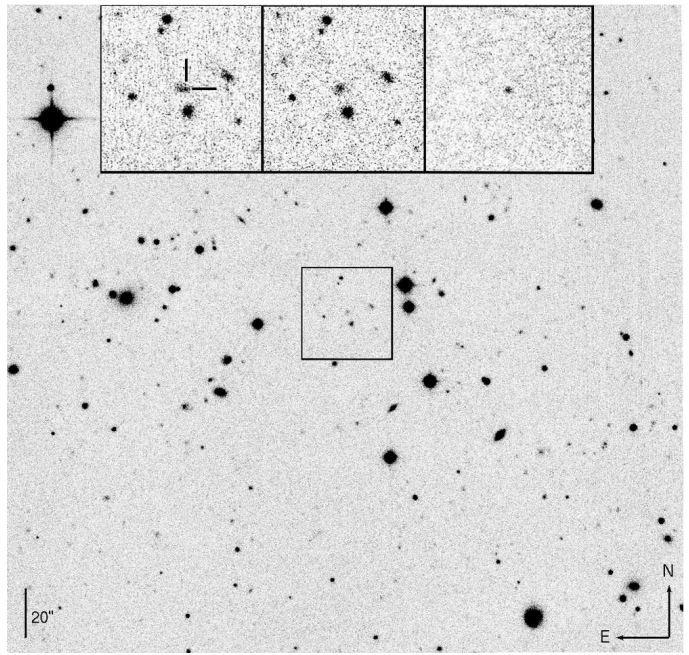}
\caption{GROND finding chart for the afterglow of GRB~090510. The main panel shows the field of GRB~090510 taken 1.4 days post burst. The field shown has 4\arcmin  $\times$ 4 \arcmin , where North is up and East is to the left. The inset in the top left shows a zoom into the afterglow region at 8.5 hours after the burst, where the afterglow is indicated with two lines. The middle inset contains the same region with the host 1.4 days post burst. 
The top right inset is a difference image between the first and second epoch. All images are $i^\prime$ band.}
\label{090510fc}
\end{figure}


\subsection{GRB 090902B}
At 11:05:08.31 UT on 2 September 2009  \textit{Fermi} GBM triggered on a long, bright, hard burst GRB 090902B (trigger 273582310 / 090902462) \citep{Betta_090902B}.  
 The burst was localized by $\textit{Fermi}$/LAT to  R.A.(J2000), decl.(J2000) = 265.00,  27.33 with a statistical uncertainty of only 0.06 degrees \citep{LAT_090902B} and Target of Opportunity observations were initiated with the near-field instruments on \textit{Swift} $\sim$12.5 hours post-trigger. A candidate X-ray afterglow within the LAT error circle was reported by \citet{XRT_090902B} and subsequently confirmed to be fading \citep{XRT2_090902B}.  Optical detections  were reported by a number of observers \citep{UVOT_090902B,UVOT2_090902B,Perley_090902B,Guidorzi_090902B,Pandey_090902B}, in the NIR \citep{Olivares_090902B} and in the radio  \citep{Alexander_090902B,Chandra_090902B}. The afterglow redshift of $z = 1.822$ was obtained from an afterglow absorption spectrum by \citet{Z_090902B} using the  GMOS spectrograph. Based on the spectral parameters \citep{LAT2_090902B}, the isotropic energy released is $E_{\gamma, \rm{iso}} = 2.26 \times 10^{54}$ in the 1~keV to 10~MeV range and $3.54\times 10^{54}$~erg in the 1~keV to 10 GeV band. 
An excess of emission in addition to the Band function at both low energies $\lesssim$ 50~keV and above $\gtrsim$100~MeV was reported by \citet{090902B_PAPER}.

GROND imaged the field of GRB~090902B starting $\sim 13$~h after the \textit{Fermi} trigger when the position of the GRB was becoming visible above the pointing constraints of the telescope. At the time of the first epoch observations, only the LAT localization with 3.5 arcmin uncertainty was available. The afterglow location was outside the narrower (5.4\arcmin   $\times$ 5.4\arcmin ) optical field of view (FoV)  but within the larger NIR FoV (10\arcmin   $\times$  10\arcmin) \citep{Olivares_090902B}. 
Observations continued in the following nights by GROND and  an additional VLT/FORS2  $R$  band observation was executed  at $\sim$ 23 days post trigger (ESO DDT Proposal number : 283.D-5059).
A finding chart of the field of GRB ~090902B is shown in Fig.~\ref{090902Bfc}.

 \begin{figure}[t]
\centering
\includegraphics[width=0.99\columnwidth]{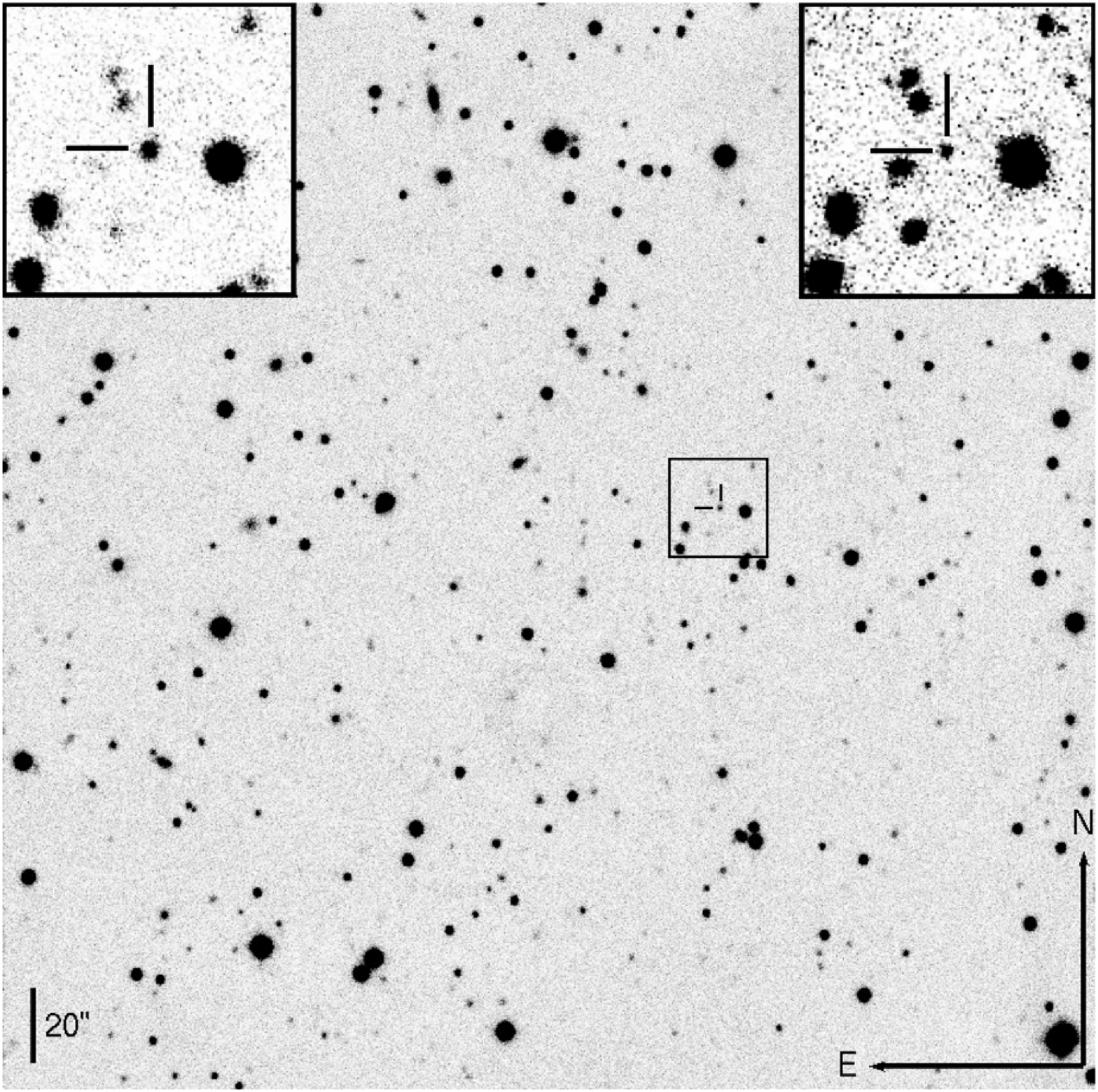}
\caption{GROND/VLT finding chart for the afterglow of GRB~090902B. The main panel shows the field of GRB~090902B in the $i^\prime$ band taken 1.5 days post burst. The shown field has 4\arcmin  x 4\arcmin, where North is up and East is to the left. The afterglow position is indicated with two lines in each image. The inset in the top left shows a zoom into the afterglow region at the same epoch, where the afterglow was still bright. The top right inset shows a VLT $R$-band image taken 23 days after the trigger. Both insets were enhanced in contrast for demonstration purposes.}
\label{090902Bfc}
\end{figure}


\section{Analysis}

{GROND optical/NIR data were reduced and analyzed in standard manner using pyraf/IRAF tasks \citep{1993ASPC...52..173T}. A detailed description of the reduction and relative photometry can be found in \citet{2008ApJ...685..376K}. In the afterglow-dominated regime we used PSF fitting techniques to derive the brightness of the optical transient. These were complemented by standard aperture photometry, which yields consistent results. The magnitudes of the GRB host galaxies were measured using aperture photometry with an appropriate aperture correction. In addition, SExtractor \citep{1996A&AS..117..393B} was used to perform host photometry, again with consistent results to the reported aperture photometry. 

Absolute photometry in $g^\prime r^\prime i^\prime z^\prime$-bands was obtained against the \griz magnitudes of the Sloan Digital Sky Survey (SDSS DR7, \citealp{2009ApJS..182..543A}) for all bursts. SDSS field stars in the vicinity of GRB~090323 \citep{2009GCN..9026....1U} were used for GRBs 090323 and 090328. Both bursts were observed with GROND consecutively in 5 different nights, out of which 4 where photometric. The obtained calibration is consistent between all epochs. The field of GRB~090510 was calibrated against the primary Sloan Standards SA {114-750 and 114-650} and their SDSS field stars. Absolute photometry of the afterglow of GRB~090902B was obtained against a nearby SDSS field observed under photometric conditions. \JHK absolute photometry was measured against the magnitudes of selected stars in the field of GRBs~090323, 090328, 090510 and 090902B from the 2MASS catalogue \citep{skr06}. This procedure results in a typical absolute accuracy of 0.04~mag in \griz, 0.06~mag in $JH$ and 0.08 in $K_S$, which was added quadratically to the statistical error in the analysis of the broadband spectral energy distribution. 
TLS data of GRB~090323 was reduced in a standard manner with MIDAS and IRAF and calibrated against six SDSS stars in the field-of-view, where we derived $R_C$ magnitudes using the transformations of Lupton from 2005{\footnotemark[1]
\footnotetext[1]{\tiny{http://www.sdss.org/dr7/algorithms/sdssUBVRITransform.html{\#}Lupton2005}}.  
The standard stars were measured with SExtractor, and afterglow magnitudes were derived using aperture photometry.
All magnitudes are corrected for the expected Galactic reddening
 in the direction of the bursts \citep{sch98}, which is $E_{B-V}=0.03$~mag for GRB~090323, $E_{B-V}=0.06$~mag for GRB~090328, $E_{B-V}=0.02$~mag for GRB~090510, and $E_{B-V}=0.04$~mag for GRB~090902B, respectively}.
 All magnitudes are quoted in the AB system throughout this paper.

The  FORS1+2  spectroscopy  data   were  reduced  with  standard  IRAF
routines,    and   spectra   were    extracted   using    an   optimal
(variance-weighted)  method.  The  observations were  obtained  with a
long  slit of  1\farcs0 width  and corrected  for slit  losses  due to
finite slit width.   Spectro-photometric calibrations were carried out
using  observations   of  the  standard  stars   EG~274  and  LTT~7379
\citep{1992PASP..104..533H}    for  the observations of GRB~090328    and    GRB~090510,
respectively.


\section{Results}

\subsection{The afterglow of GRB 090323}

\subsubsection{Light curve}

The GROND and TLS multicolor light curve of the optical/NIR afterglow starting at $T_0+95$~ks is presented  in Fig.~\ref{090323lc} and is well described with a single power-law plus a superimposed bump component and a constant host contribution dominating at later times ($\chi^2 = 161/138$ d.o.f). All available optical/NIR data were fitted simultaneously, allowing for a color change in the bump and host dominated regime. The initial power-law decay slope is $\alpha=1.90\pm0.01$, with no sign of an additional break at later times. The host magnitudes in the $g^\prime z^\prime JHK_S$ bands are not constrained by the data and have been fixed. Similarly, the flare is  unconstrained in the $H$ and $K_S$ bands.
The host galaxy of GRB~090323 was detected in the 
$r^\prime$ and $ i^\prime$ bands but the stellar mass is not constrained by the optical identification (magnitudes $r^\prime = 24.87 \pm 0.15$ and $i^\prime = 24.25 \pm 0.18$) obtained by GROND  because observations probe the rest-frame UV, where the mass-to-light ratio can vary by a factor of more than 100.

\begin{figure}[t]
\centering
\includegraphics[angle=270,width=\columnwidth]{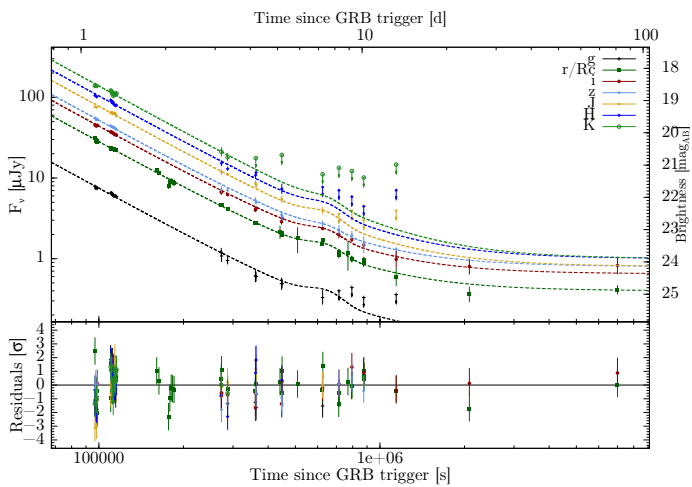}
\caption{GROND optical/NIR and TLS optical light curve of the afterglow of GRB~090323. The data are fitted with a power-law, superimposed bump and constant host component. Late upper limits are not shown to enhance clarity.
}
\label{090323lc}
\end{figure}

\subsubsection{Spectral energy distribution}

The multi-band spectral energy distribution (SED) of the afterglow of GRB~090323 is shown in Fig.~\ref{seds}. It is well 
 fit with a power-law of index  $\beta_{\rm opt} = 0.65\pm0.13$ and a small amount of Small Magellanic Cloud (SMC)-like dust extinction \citep{bou85} with $A_{V}^{\rm{host}}=0.14^{+0.04}_{-0.03}$ ($\chi^2 = 3.3 / 4$~d.o.f.). A pure power-law does not provide an acceptable fit ($\chi^2 = 26 / 5$~d.o.f.), and Large Magellanic Cloud (LMC) \citep{fit86} and Milky Way (MW) \citep{sea79}  dust models are strongly ruled out because of the absence of the 2200~\AA~feature, which would be in the $z^{\prime}$ band at this redshift. The flux difference between the extrapolation of the SED defined from $i^{\prime}$ to the $K_S$ and the $g^\prime$ and $r^\prime$ bands is nicely consistent with the spectroscopic redshift of $z=3.57$. 
 The  final photo-$z$ obtained with the calibrated GROND data is $3.44^{+0.18}_{-0.16}$.

\begin{figure}[t]
\centering
\includegraphics[angle=270,width=\columnwidth]{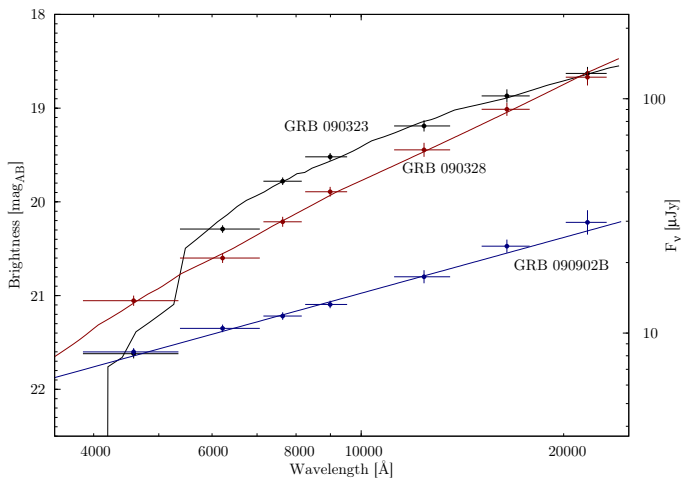}
\caption{GROND broadband spectral energy distribution of the afterglows of  GRB~090323, GRB~090328 and GRB~090902B. 
Note that the $g^\prime$ band points for GRB~090323 and GRB~090902B are overlapping.
 }
\label{seds}
\end{figure}

In the standard fireball model (see \citealp[e.g.][for reviews]{2004RvMP...76.1143P, 2006RPPh...69.2259M}) the spectral index, $\beta$ of the afterglow is connected to the temporal index, $\alpha$, via the closure relations. These depend on the type of circumburst medium, location of the characteristic frequencies in the synchrotron spectrum, and the evolutionary stage of the afterglow \citep[e.g.][]{1998ApJ...496L...1R,1998ApJ...497L..17S,2000ApJ...536..195C,2004IJMPA..19.2385Z,2005MNRAS.362..921P,2006MNRAS.366.1357P,2006ApJ...642..354Z,2009ApJ...698...43R}. However, afterglow light curves often display complex temporal behaviour with shallow decays, plateaus, rebrightenings, flares or smooth breaks such that limited sampling strongly affects the inferred light-curve slope \citep[e.g.][]{2004ApJ...606..381L, 2006ApJ...642..389N, 2009MNRAS.397.1177E, 2009ApJ...693.1912G}. Hence, here and later, we derive the spectral index from the optical to NIR SED, and use the temporal index from the light curve fitting and the closure relations to obtain constraints on the jet properties. In particular, for a constant spectral index  and value of p, the expected light curve slopes significantly differ ($\Delta \alpha$ is between 0.5 and 1.3 depending on the model) between the pre and post jet break evolution, which facilitates a reliable discrimination between these two regimes.

For GRB~090323, the obtained spectral index $\beta_{\rm opt}=0.65\pm0.13$ is compatible with the observational frequency $\nu_{\rm opt}$ being between the typical synchrotron frequency $\nu_{\rm m}$ and the cooling frequency $\nu_{\rm c}$  for an ISM (density $n =$ constant) or wind ($n \propto r^{-2}$ ) environment.

The power-law decay index of the light curve, $\alpha$, is expected to be  $\sim$1 in the pre-jet break, ISM case and is not compatible with the observations. A value of $\alpha$ $\sim$ 2.3 is obtained in a post-jet break evolution with significant lateral spreading of the ejecta,
\citep[both ISM and wind][]{2000ApJ...536..195C}), and $ \alpha$ $\sim$ 1.7 (ISM) or $\alpha \sim 2.0$ (wind) if lateral spreading of the ejecta is not considered \citep{2005MNRAS.362..921P}. 
The latter case (post-jet break without spreading) is compatible within $1\sigma$ with the observed $\alpha$ while the case with spreading is compatible at the 2$\sigma$ level. In the above cases the electron spectral index $p$ can be derived from $\beta=\frac{p-1}{2}$ and a value of $p=2.3\pm0.3$ is obtained.

The  value of  $\alpha$ for GRB 090323 indicates that the jet break occurred before the start of the observations. However it is not possible to distinguish between the ISM and wind environments. 
The post-jet evolution implies
 that the half opening angle   of the jet is smaller than $\theta_{\rm jet}^{\rm ISM}  \lesssim 2.1^\circ$ in an ISM-type circumburst environment \citep{1999ApJ...519L..17S,2001ApJ...562L..55F}. The beaming corrected energy emitted in $\gamma$-rays $E_\gamma$ (1~keV to 10~GeV) of GRB 090323 is thus $E_\gamma \lesssim 3.3 \times 10^{51}$~erg.
In a wind environment \citep{2000ApJ...536..195C} the opening angle of the jet would be 
$\theta_{\rm jet}^{\rm wind}  \lesssim 1.1^\circ$   and the beaming corrected energy would be 
$E_\gamma \lesssim 1.0 \times 10^{51}$~erg.  (1~keV to 10~GeV). 
The following values are assumed throughout the paper $A^*, n_1, \eta_0.2 = 1$.

\subsection{The afterglow and host galaxy of GRB 090328}

\subsubsection{Light curve}

The GROND multicolor light curve of the afterglow of GRB~090328  starting at T$_0$+137~ks is presented in Fig.~\ref{090328lc}  and is remarkably similar to that of  GRB~090323 (Fig.~\ref{090323lc}). It is well described with a single power-law of index $\alpha= 2.27\pm0.04$, which is typical of a post-jet break slope \citep{2006ApJ...637..889Z}, plus a superimposed bump component and a constant host contribution dominating at later times ($\chi^2$=35/39 degrees of freedom). 
There is minor evidence for a slightly redder spectra in the bump component peaking at $\sim$350~ks post burst, but given the uncertainties in the measurement however, the spectral index is consistent with the afterglow slope at the $2\sigma$ level.

\begin{figure}[t]
\centering
\includegraphics[angle=270,width=\columnwidth]{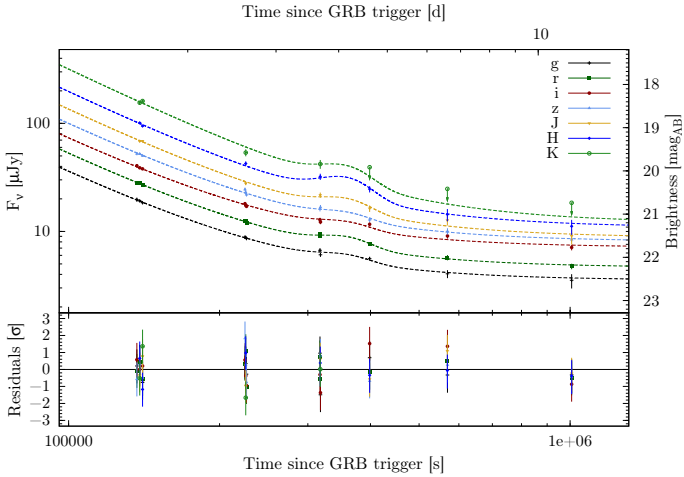}
\caption{GROND optical/NIR light curve of the afterglow of GRB~090328. The data are fitted with a power-law, superimposed bump and constant host component. All parameters are left free in the fit except the $K_S$ band host magnitude, which is not constrained by the data.}
\label{090328lc}
\end{figure}

\subsubsection{Spectral energy distribution}
\label{sec090328sed}

{The broadband SED of the afterglow was constructed from data taken in the first epoch where the last epoch was used as a reference frame. Subtracting the host fluxes was performed using two methods: firstly the fluxes from the light curve fitting were used and directly subtracted from the first epoch magnitudes, where measurement errors were propagated accordingly. For the $K_S$ band, where the host is undetected, we used the mean flux between the $H$ band detection and $K_S$ band upper limits with errors over the complete range. The effect of the host magnitude error on the afterglow SED, however, is negligible as its uncertainty is dominated by the absolute photometric accuracy. Secondly, the reference frame was subtracted from the 1st epoch image in each band, and magnitudes were derived using the difference image directly. Both methods resulted in consistent results. As the second method tends to underestimate the true photon noise in the image, results from the direct flux difference with propagated errors are used in the following.}

{
The afterglow SED was fitted using a power-law modified by dust reddening as shown in Fig.~\ref{seds}. A pure power-law provides a reasonably good fit to the data with a $\chi^2$ of 4.5/5 d.o.f. The resulting power-law index of $\beta_{\rm opt}=1.46^{+0.07}_{-0.08}$, however would be surprisingly red, both theoretically \citep{1998ApJ...497L..17S} and observationally \citep{2006ApJ...641..993K,2007arXiv0712.2186K} if it were intrinsic to the afterglow. In fact, a small amount of extinction of  $A_{V}^{\rm{host}}=0.22^{+0.06}_{-0.18}$\,mag with a SMC type dust attenuation law provides a better fit ($\chi^2$ = 2.6/4 d.o.f) and a more typical spectral power-law index of $\beta_{\rm opt}^{\rm SMC}=1.19^{+0.21}_{-0.19}$. Though statistically not necessarily required, it seems thus very likely that there is mild reddening by dust in the circumburst environment. Due to the redshift of $z=0.7354$, the optical data obtained hardly probe the rest frame UV regime and do not probe the region of the 2200~\AA ~bump. Therefore, LMC and MW dust models cannot be distinguished and return comparable results, with spectral indices of $\beta_{\rm opt}^{\rm LMC}=1.07^{+0.29}_{-0.12}$, $\beta_{\rm opt}^{\rm MW}=1.16^{+0.24}_{-0.21}$ and $A_{V}^{\rm{host}}$ between 0 and 0.4\,mag.}

A spectral index of $\beta_{\rm opt} \sim 1.2$ is only consistent with temporal index $\alpha = 2.27\pm0.04$ in a post jet break evolution and the ISM or wind, slow cooling case in the spectral regime above the cooling frequency $\nu_{\rm opt}>\nu_c$ \citep{1998ApJ...497L..17S,2000ApJ...536..195C,2004IJMPA..19.2385Z}. In this case the electron energy index $p$ is $p=2\beta= 2.4^{+0.4}_{-0.4}$ as obtained from the spectral index. 
 Hence, the observations indicate, that the jet break must have occurred before the first GROND observations. This yields an upper limit of $\theta_{\rm jet}^{\rm ISM} \lesssim 5.5^\circ$ or $\theta_{\rm jet}^{\rm wind} \lesssim 4.2^\circ$ \citep{1999ApJ...519L..17S,2001ApJ...562L..55F}, which constrains $E_\gamma$ (1~keV to 10~GeV) to $E_\gamma \lesssim 4.6 \times 10^{50}$~erg for an ISM or $E_\gamma \lesssim 2.7 \times 10^{50}$~erg for a wind type environment.

\subsubsection{The host galaxy of GRB~090328}

The optical spectrum of the afterglow of GRB~090328 is presented in Fig.~\ref{spec_28} and
shows a strong [OII]$\lambda3727$ emission line, with a flux $f_{\rm [OII]} = (25.9\pm0.8)\times 10^{-17}$~erg~cm$^{-2}$ s$^{-1}$. 
We possibly detect a weak, if there, [NeIII]$\lambda3868$ emission with a flux nearly consistent with zero within the large uncertainties $f_{\rm [NeIII]} = (0.96\pm0.83)\times 10^{-17}$~erg~cm$^{-2}$ s$^{-1}$. 

The [OII] luminosity $L_{\rm [OII]} = (6.47\pm0.21)\times10^{41}$ {erg s}$^{-1}$ would give a star formation rate SFR $=3.6$ M$_\odot$ yr$^{-1}$, not corrected for dust extinction (the SFR conversion derived for GRB hosts is from Savaglio et al.\ 2009). A mild dust extinction correction with 
$A_{V}^{\rm{host}} = 0.22$
 gives SFR $=4.8$ M$_\odot$ yr$^{-1}$. This is about a factor of two higher than the mean $< {\rm SFR}> =2.5$  M$_\odot$ yr$^{-1}$ derived for a sample 43 GRB hosts in the redshift interval $0<z<3.4$ \citep{2009ApJ...691..182S}, and about one order of magnitude higher than in the LMC.
The spectrum as been corrected for slit losses, however we note that the star formation rate quoted above is a fraction of the total and that the true SFR  may be slightly higher. 

The spectrum has a number of absorption lines at the same redshift of the [OII] line, associated with FeII, MnII, MgII, MgI and CaII. We derived column densities using the curve-of-growth analysis \citep{1978ppim.book.....S}, and cross-checked the results with the apparent optical-depth method \citep{1991ApJ...379..245S}. Identification, rest-frame equivalent width EW$_r$ and measured column densities are in Table~\ref{tabs}. The effective Doppler parameter given by the best fit is, independently for FeII and MnII, $b\simeq 78$ km s$^{-1}$. We used this value to derive the column density for MgI and CaII.
We estimate a very uncertain iron-to-manganese relative abundance [Fe/Mn] $=0.08^{+0.53}_{-0.37}$.
 No intervening absorption systems have been identified.
The redshift obtained from the optical spectrum is $z = 0.7354\pm0.0003$.

\begin{figure}
\centering
\includegraphics[angle=270,width=\columnwidth]{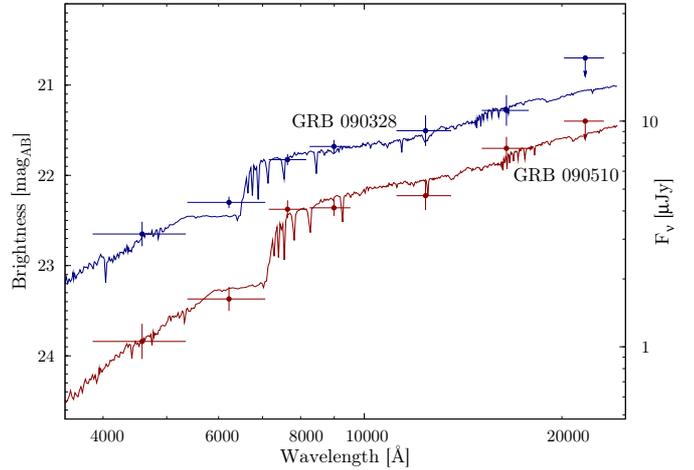}
\caption{Broadband spectral energy distributions of the host galaxies of GRB~090328 and GRB~090510 and host galaxy templates from $hyperZ$ \citep{2000A&A...363..476B}.}
\label{090328host}
\end{figure}

\begin{table}
 \caption[t1]{Absorption lines in the afterglow spectrum of GRB~090328.}
\begin{center} 
\begin{tabular}{lcccc} 
\hline\hline
	&	&  	&		&   \\
Line  & $\lambda_{\rm obs}$ & z & EW$_r$ (\AA) & $\log N_{\rm}$  (cm$^{-2}$)\\
\hline
	&	&  	&		&   \\
 FeII $\lambda2344$ & 4067.0 & 0.7351 & $2.71\pm0.15$ & $16.07^{+0.52}_{-0.36}$ \\
 FeII $\lambda2374$ & 4120.4 & 0.7356 & $2.20\pm0.15$ & . . . \\
 FeII $\lambda2381$ & 4133.9 & 0.7360 & $2.82\pm0.15$ & . . . \\
 FeII $\lambda2586$ & 4488.0 & 0.7355 & $2.69\pm0.18$ & . . . \\
 FeII $\lambda2600$ & 4511.2 & 0.7351 & $3.26\pm0.18$ & . . . \\
 MnII $\lambda2576$ & 4470.4 & 0.7354 & $1.16\pm0.16$ & $13.93\pm0.10$ \\
 MnII $\lambda2594$ & 4500.9 & 0.7351 & $0.91\pm0.14$ & . . . \\
 MnII $\lambda2606$ & 4523.1 & 0.7356 & $0.79\pm0.15$ & . . . \\
 MgII $\lambda2796$ & 4851.7 & 0.7352 & $3.99\pm0.18$ & $>15.7$ \\
 MgII $\lambda2803$ & 4864.3 & 0.7353 & $3.87\pm0.18$ & . . . \\
 MgI  $\lambda2852$ & 4949.6 & 0.7355 & $1.60\pm0.19$ & $13.38\pm0.12$ \\
 CaII $\lambda3934$ & 5315.9 & 0.7353 & $>1.2$ & . . . \\
 CaII $\lambda3969$ & 5362.9 & 0.7353 & $1.68\pm0.19$ & $13.76\pm0.11$ \\
 	&	&  	&		&   \\
 \hline
\end{tabular}
\end{center}
\label{tabs}
\end{table}

  \begin{figure*}[t]
   \centering
{\rotatebox{90}{\includegraphics[width=0.4\textwidth]{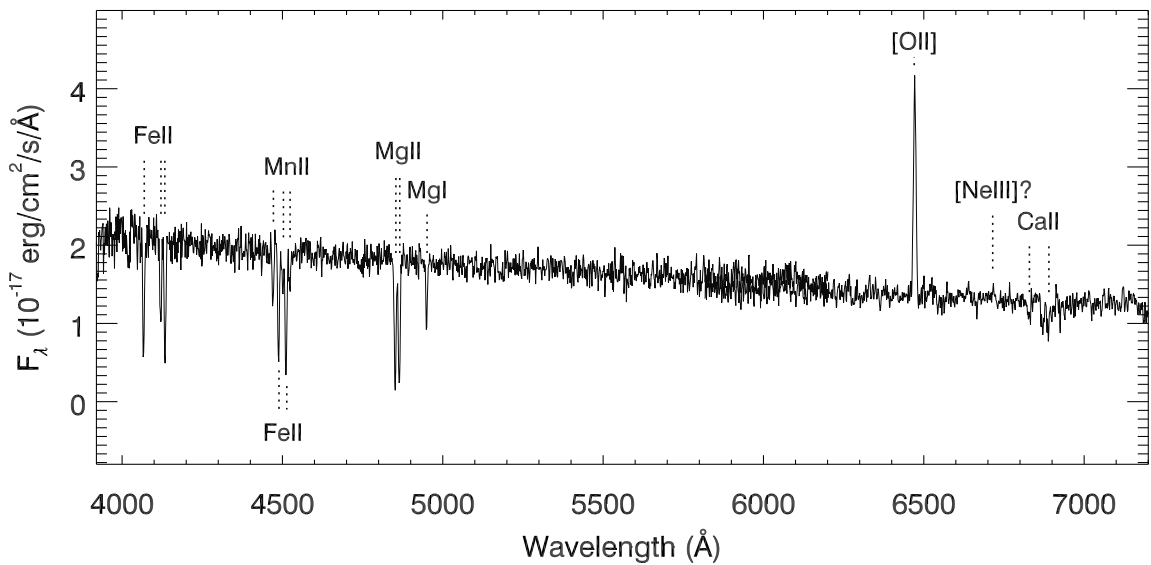}}}
\caption{VLT/FORS1 spectrum of the afterglow of  GRB~090328 obtained $\sim1.6$\,days post-burst using the  $600B$   and  $600R$   grisms.  Prominent  absorption and emission lines are indicated and listed in Table\,\ref{tabs}. }
     \label{spec_28}
   \end{figure*}

\label{090328hostsec}

The SED of the host galaxy of GRB~090328 is presented in Fig.~\ref{090328host} and was fit using $hyperZ$ \citep{2000A&A...363..476B} and models from \citet{2003MNRAS.344.1000B}. The fitted galaxy spectral templates include elliptical, different types of early and late spirals, irregular and starburst galaxies at various ages and different star formation rates and solar metallicity. In addition, the spectral templates were modified by an extinction term, resembling the dust attenuation law of the SMC. The redshift of the galaxy was fixed at $z=0.7354$, which is consistent with the apparent 4000 \AA~Balmer break being within the $r^\prime$ band. 

The host galaxy of GRB~090328 is best fit with a starburst galaxy template, but all other templates provide acceptable fits. Similar to the afterglow SED, there is evidence for dust reddening from the host observations of the order of $A_V^{\rm host}= 0.5_{-0.3}^{+0.5}$ in the starburst galaxy
(although $A_V^{\rm host}$ is not well constrained ), with a best-fit age of the dominant stellar population of $\sim$50~Myr. By folding the restframe best-fit template spectrum with a typical $B$ and $K$ filter response, we derive absolute AB magnitudes of the host galaxy of GRB~090328 of $M_B =-20.67$ and $M_K = -20.17$, well within the samples of previous long GRB host galaxies \citep[e.g.,][]{2004A&A...425..913C, 2009ApJ...691..182S}.

The stellar mass of the host can be approximated from the best fitting template  to $M_\ast \simeq 10^{9.4}$ M$_\odot$. 
When using the empirical relation derived by \citet{2009ApJ...691..182S} between $K$-band absolute magnitude and stellar mass:

\begin{equation}\label{emk_m}
 \log M_\ast =  -0.467 \times M_K  -  0.179 ,
\end{equation}

\noindent 
we derive $M_\ast \simeq 10^{9.24}$ M$_\odot$ with a typical uncertainty of a factor 2, consistent with the previous estimate and similar to the stellar mass of the LMC. Using the dust-corrected star formation rate SFR $=4.8$ M$_\odot$ yr$^{-1}$ (see note above on slit loss) and the above stellar mass, the specific star formation rate per unit mass rate is 2.8 Gyr$^{-1}$ again well within the sample of previous long burst host galaxies \citep{2009ApJ...691..182S}.

Using the galaxy distribution function in the corresponding redshift interval from $z=0.6-0.8$ from \citet{2006ApJ...647..853W} to compare the host's absolute magnitudes to the population of field galaxies, the luminosity of the GRB host is estimated to $\sim 0.46 - 0.50 L_{*}$. We obtain a dust-corrected star formation rate per unit luminosity of SFR / $L_B$ $\sim$ 10 M$_\odot$ yr$^{-1} L_*^{-1}$  for this host galaxy.
 Hence, the specific star formation rate in the host galaxy of GRB~090328 is consistent with the values for the long bursts presented by \citet{2009ApJ...690..231B}.

\begin{figure*}
\centering
{\rotatebox{90}{ \includegraphics[width=0.5\textwidth]{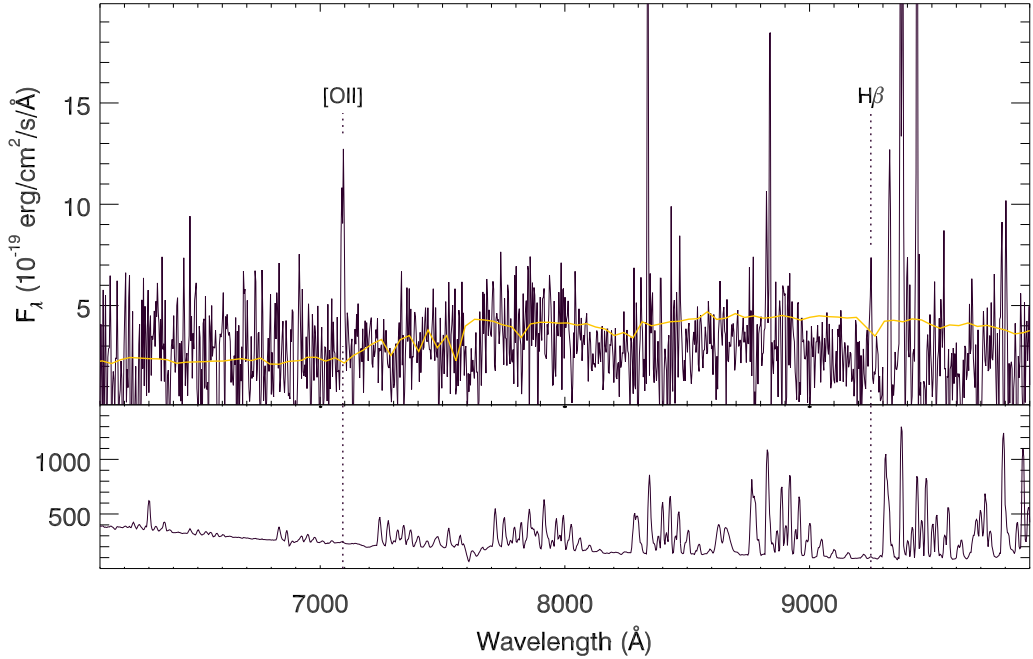}}}
\caption{VLT/FORS2 300I spectrum of  the host galaxy of GRB~090510 obtained $\sim2.3$\,days   post-burst (top panel) and corresponding noise spectrum (lower panel). The Sd galaxy template \citep[yellow line; ][]{2007ApJ...663...81P} in the top panel indicates the location of the 4000\,\AA\ break. The [OII] and H$\beta$ emission lines are marked and their fluxes listed in Table\,\ref{Lines_090510}.}
\label{spec_0510}
\end{figure*}

\subsection{The afterglow and host galaxy of  GRB~090510}

\subsubsection{Afterglow and host association}

The properties of the afterglow of GRB~090510 are not well constrained
due its faintness   at the time of the GROND observation and the blending with the nearby galaxy. Fitting the afterglow SED with a reddened power-law results in an unconstrained power-law index, and a 3$\sigma$ upper limit for the intrinsic extinction of $A_{V}^{\rm{host}}\leq 1.1$~mag, assuming an SMC-type extinction law. MW- or LMC-type extinction cannot be distinguished and result in comparable values for $\beta_{\rm opt}$ and $A_V$.

The putative host galaxy at a distance of  1\farcs2\ ,  corresponding to a projected distance of 9.4~kpc from the afterglow to the brightest emission in the nearby galaxy,  has  $r^\prime$- and $z^\prime$-band magnitude of 23.4 and 22.3~mag, respectively and extends to the afterglow position in the GROND images of  0\farcs9\ seeing (Fig. \ref{090510fc}).
According to object counts from the Hubble Ultra Deep Field \citep{2006AJ....132.1729B} and the GOODS \citep{2004ApJ...600L..93G} one would statistically expect less than 0.008 unrelated objects of similar brightness at this distance to the afterglow, corresponding to a chance probability of $\lesssim 0.8$\%. For comparison, the chance coincidence probability for the galaxies $\sim$5\arcsec\ south 
and $\sim$8\arcsec\ west of the afterglow  are  $\sim$7\%, and $\sim$30\% respectively.

In the merger of two compact objects, the progenitors may be kicked  outside their host galaxy at scales of 1-100~kpc, with a distribution peaking at several kpc for different galaxies and merger conditions \citep{1999ApJ...526..152F}. Most of the convincing host galaxy associations however are found for host positions which coincide or overlap with the afterglow \citep[e.g.,][]{2005Natur.437..845F, 2006ApJ...648L...9L, 2008arXiv0804.1959K,2008MNRAS.385L..10T}. Short burst host samples show a typical offset of 0\arcsec$-$3\arcsec\ with clustering at small values \citep{2007ApJ...664.1000B, 2009A&A...498..711D} indicating that a large kick might not be a dominant process. 
Recently, based on HST observations of ten short burst hosts, \citet{2010ApJ...708....9F} report that the median offset of short GRBs from their host galaxies is 5~kpc and based on a larger sample of short bursts they report that 25\% of bursts have projected offsets of less than  10~kpc. 
The host galaxy of GRB~090510 certainly falls within this distribution.
 \citet{2007ApJ...664.1000B} argue  that even fainter, undetected galaxies at low redshifts $z\lesssim 0.5$ are very unlikely to host the dominant fraction of short GRBs.  Also in the case of GRB~090510, direct observations of the optical afterglow place a strong constraint of $z<1.5$ of the redshift \citep{2009GCN..9351....1K}. Given that the observed properties of the closeby galaxy fit very well into previous samples of short burst host galaxies in terms of brightness and redshift \citep[e.g.,][]{2007ApJ...664.1000B, 2009ApJ...691..182S}, and the probability of chance association is very low, this galaxy can be confidently associated with the burst  and is hence very likely the host of GRB~090510.

We further report that the redshift of the galaxy  $\sim$5\arcsec\ south   
obtained using   VLT/FORS
 is the same as that of the host galaxy based on the detection of [OII] and H$\beta$ emission lines. Using a distance scale of 7.8 kpc/\arcsec\ this translates to a  projected distance of $\sim 40$~kpc and suggests that the host and this galaxy are part of a group or cluster association. Consequently in the unlikely event that this second galaxy  were indeed the host galaxy, the distance to the source would therefore remain unchanged. The source to the east
  $\sim$ 8\arcsec\ away appears stellar and all the remaining  nearby bright galaxies have chance coincidence probabilities of order $\sim$100\%.

\subsubsection{The host  galaxy of GRB~090510}

The spectrum is presented in Fig.~\ref{spec_0510} and has a relatively weak [OII]$\lambda3727$ emission line, with a flux $f_{\rm [OII]} = (1.30\pm0.15)\times 10^{-17}$ {erg cm}$^{-2}$ s$^{-1}$ 
(Table~\ref{Lines_090510}). 
 The weaker H$\beta$ flux is $f_{\rm H\beta} = (0.50\pm0.15)\times 10^{-17}$ {erg cm}$^{-2}$ s$^{-1}$. 
 The redshift obtained from the optical spectrum is $0.903\pm0.001$.

The [OII] and H$\beta$ luminosities are $L_{\rm [OII]} = (5.4\pm0.6) \times10^{40}$ {erg s}$^{-1}$ and $L_{\rm H\beta} = (2.1\pm0.6) \times10^{40}$ erg s$^{-1}$, respectively. The star formation rate, not corrected for dust extinction, derived from the [OII] luminosity is SFR $=0.30$ M$_\odot$ yr$^{-1}$, consistent with the SFR $=0.26$ M$_\odot$ yr$^{-1}$ derived from the H$\beta$ luminosity.  
The SFR conversions are those proposed by \citet{2009ApJ...691..182S} for GRB hosts. The derived value is one order of magnitude lower than the mean SFR measured in long GRB hosts \citep{2009ApJ...691..182S} and consistent with relatively faint UV emission derived from the SED of the host (Fig.~\ref{090328host}) and at the lower end of the distribution of SFRs of short GRBs  \citep{2009ApJ...690..231B}.  
Correcting for dust extinction yields a  star formation rate of $\sim0.60$ M$_\odot$ yr$^{-1}$.  We note as in the case of GRB~090328, the spectrum as been corrected for slit losses, however we note that the star formation rate quoted above is a fraction of the total and that the true SFR  may be slightly higher.

The SED fitting for the host of GRB~090510 has been performed in a similar manner to that of
 GRB~090328  (Section \ref{090328hostsec}). The redshift of the galaxy was fixed to the spectroscopic value of $z=0.903$. Again all galaxy templates provide an acceptable fit, where the best fit is obtained with an elliptical galaxy. 
 We note that the SFR is in the range of local star-forming elliptical galaxies \citep{2009MNRAS.398.1651H}.
 From the best fit we derive AB absolute magnitudes $M_B =-19.91$ and $M_K = -20.51$, and slight evidence for a dust extinction reddening similar to the SMC with a visual extinction of $A_{V}^{\rm{host}}= 0.7^{+0.2}_{-0.4}$.
The stellar mass of the host can be estimated  
directly from the galaxy templates to $M_\ast \simeq 10^{9.6-10.2}\rm{M}_\odot$. 
The dust-corrected star formation rate per unit mass is  0.04 to 0.15 Gyr $^{-1}$, significantly below 
that of the host galaxy of GRB~090328.

Using the galaxy distribution function in the corresponding redshift interval from $z=0.8 -1$ from  \citet{2006ApJ...647..853W}, the luminosity of the GRB host is estimated to $\sim 0.26-0.30L_*$, and
we obtain a  dust-corrected star formation rate per unit luminosity of SFR / $L_B$ $\sim$ 2 M$_\odot$ yr$^{-1} L_*^{-1}$  for this host galaxy, consistent with the albeit wide range of SFR / $L_*$ values  for the short bursts presented by \citet{2009ApJ...690..231B}.

\begin{table}
 \caption{Emission lines detected in the spectrum of GRB~090510 corrected for Galactic extinction.}
 \label{Lines_090510}
\begin{center} 
\begin{tabular}{lcccc} 
\hline\hline
	&	&  	&		&   \\
Line  & $\lambda_{\rm obs}$ & $z$ & Flux \\
 & & & ($10^{-17}$ erg cm$^{-2}$ s$^{-1}$) \\
 \hline\hline
  	&	&  	&		&   \\
\ [OII]$\lambda3727$ & 7092.6 & 0.9030 & $1.30\pm0.15$ \\
\ H$\beta$ & 9249.7 & 0.9028 & $0.50\pm0.15$ \\
\ [OIII]$\lambda5007$ & n.d. & . . . & $<0.3$ \\ 
 	&	&  	&		&   \\
\hline
	&	&  	&		&   \\
\end{tabular}
\end{center}
\end{table}

\subsection{The afterglow of GRB 090902B}

\subsubsection{Light curve}

The early ($t<T_0+1.10$~Ms) GROND multi-color light curve of the optical/NIR afterglow of GRB~090902B is well fit by a single power-law of index $\alpha_1=0.94 \pm 0.02$. A late epoch of imaging obtained with the VLT 23~days after the burst indicates a break in the light curve at the $2\sigma$ level. 
No further follow-up observations could be performed due to the proximity of the GRB field  to the Sun. Consequently the light curve coverage at late times is sparse and the evolution not fully conclusive. Fixing the late slope to a conventional post jet break slope of $\alpha_2 = 2.2$ with a relatively sharp break as shown in Fig.~\ref{090902Blc} results in a break time of $t_{\break}\sim 1.5$~Ms ($\chi^2 = 43 / 35$ d.o.f).

\begin{figure}
\centering
\includegraphics[angle=270,width=\columnwidth]{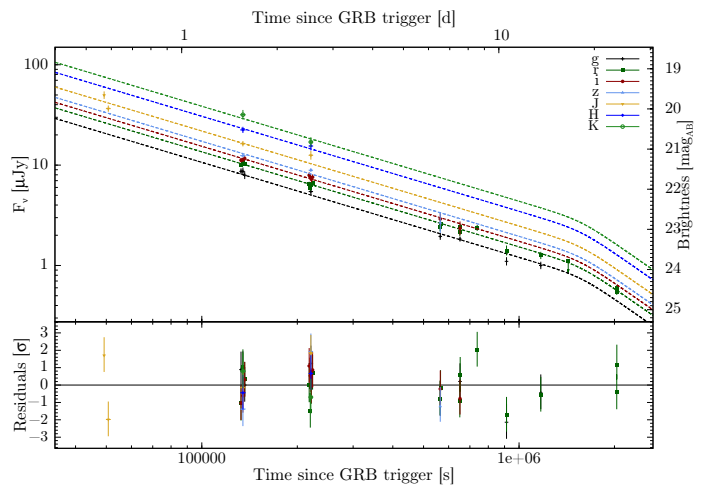}
\caption{GROND and VLT optical/NIR light curve of GRB~090902B. The data are fitted with a broken power-law. Due to the proximity of the source to the sun, the light curve coverage is sparse at later times, and the evolution is ambigous. There is however indication of a break at around $150$~Ms post burst. Shown is the fit with a fixed late slope and relatively sharp break.}
\label{090902Blc}
\end{figure}

\subsubsection{Spectral energy distribution}

The optical/NIR SED of the afterglow of GRB~090902B is well described with a simple power-law template (Fig.~\ref{seds}). The spectral index is $\beta_{\rm opt} = 0.79^{+0.05}_{-0.19}$, with 3$\sigma$ upper limits of the intrinsic visual extinction of $A_{V}^{\rm{host}} \lesssim 0.36$ with an SMC/LMC type reddening and $A_{V}^{\rm{host}} \lesssim 0.24$ assuming a MW-like dust attenuation law. 
 This is in agreement with the light curve slope of $\alpha\sim 0.94$ in the fireball model, which is typical for a pre-jet break evolution with an ISM surrounding and with an observational frequency between $\nu_{\rm m}$ and $\nu_{\rm c}$.  Following \citet{1999ApJ...519L..17S} and \citet{2001ApJ...562L..55F}, the half opening angle for a break time of $t_{\rm \break}\sim 150$~Ms would be $7.2^\circ$, or setting $t_{\rm break}\gtrsim 1.1$~Ms as a conservative lower limit $\theta_{\rm jet} \gtrsim 6.4^\circ$. Together with the energetics of the prompt emission, this implies a beaming corrected energy of $E_{\gamma} \gtrsim 2.2 \times 10^{52}$~ergs. We note that a wind environment would significantly relax the lower bounds on the opening angle and thus $E_{\gamma}$,  however  it is not consistent with the expectations for $\alpha$ and $\beta_{\rm opt}$ in the standard models. The power-law index $p$ of the electron population is then $p = 2.58^{+0.10}_{-0.38}$ as obtained from the spectral index which is in reasonable agreement with the values estimated from the early LAT emission under the assumption that it is generated by an external forward shock \citep{2009arXiv0910.5726K}.


\begin{figure*}[t]
\centering
{\rotatebox{0}{ \includegraphics[width=0.7\textwidth]{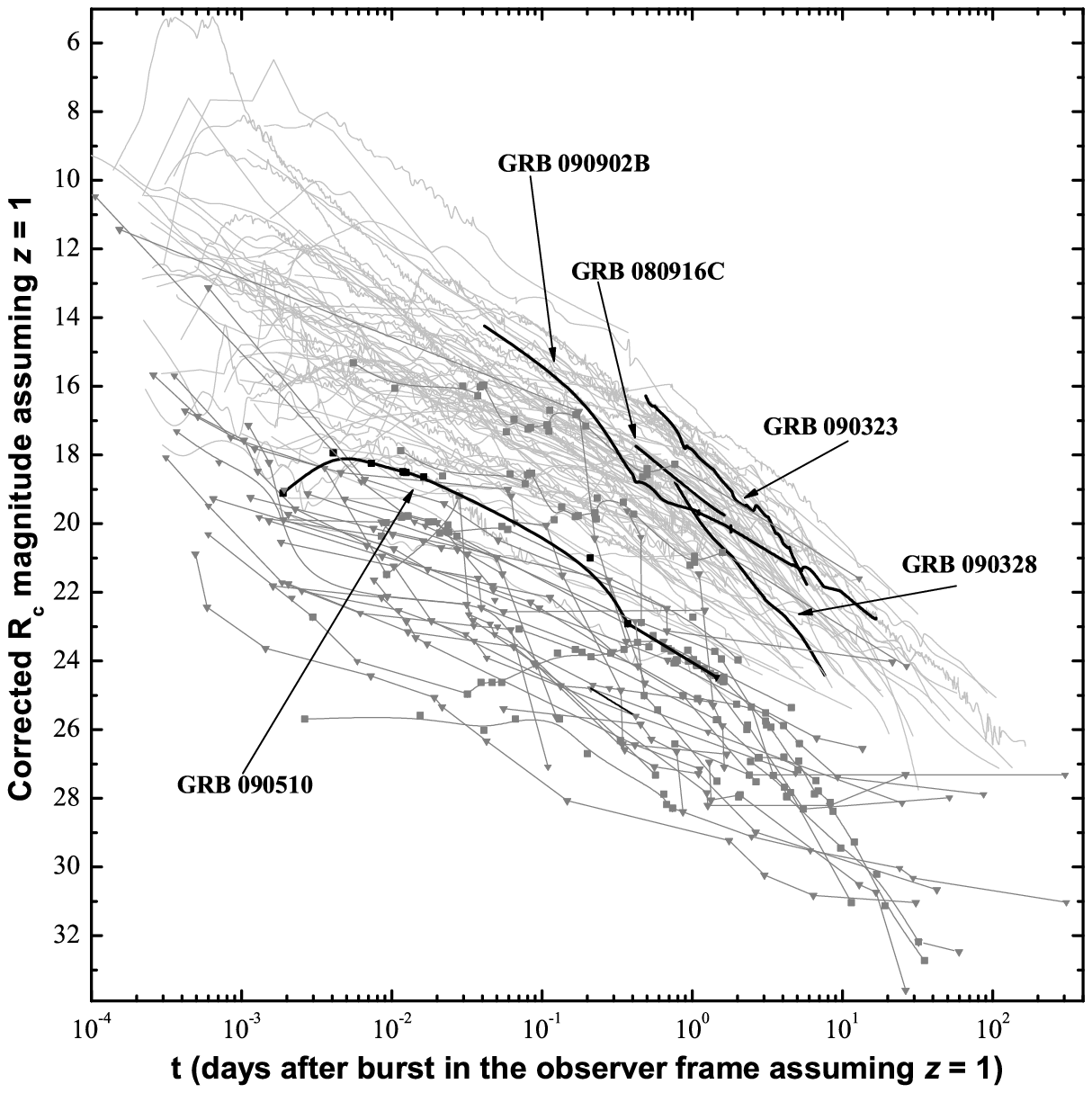}}}
\caption{The afterglows of the \emph{Fermi}/LAT-detected GRBs in comparison with a sample of over 120 well 
observed afterglows detected until May 2009 \citep{2006ApJ...641..993K,2007arXiv0712.2186K,2008arXiv0804.1959K}. GROND and TLS data only 
are presented for the the long \textit{Fermi} bursts while the early data points from GRB 090510 are from 
UVOT   \citep{2009GCN9342......1K} and include a point from \citet{2009GCN..9338....1O}.  All afterglows have 
been corrected for Galactic extinction and for host-galaxy contribution, where applicable. All afterglows have been additionally corrected for host-galaxy extinction, 
and have been shifted to a redshift of $z=1$. Long GRBs are shown as light grey lines, whereas 
short GRBs are thicker grey lines with symbols. Squares are detections, and downward pointing 
triangles upper limits.
}
\label{BigFig}
\end{figure*}

\section{Discussion}

We have presented the optical and NIR follow-up observations of four bursts with high energy emission detected by \textit{Fermi}/LAT and spectroscopic investigations  of two of those events. Since the end of  the commissioning phase \textit{Fermi}/GBM has detected several hundred GRBs, many of which were in the field of view of the LAT instrument,  however only fourteen of those have been also detected in by LAT (up to the end of January 2010).  Follow-up of these bursts is of special interest due to the broadband coverage  of the prompt emission  (keV to GeV)  which allows the spectral properties and energetics of these events to be constrained.  It is noteworthy that the redshifts of  bursts detected in the GeV regime by the LAT  range from the relatively low-redshift GRB~090328 to the high-reshift of GRB~080916C and that  the long bursts  have high fluences and very high isotropic gamma-ray energies (although GRB~090328 has a smaller isotropic energy release due to its  lower redshift).
The properties of these events presented in this paper and another \textit{Fermi}/LAT event, GRB~080916C, including their afterglows and energetics  are summarized in Table\,\ref{summary_table}. 

Three of the bursts presented in this  paper belong to the class of long gamma-ray bursts while GRB~090510 is a short burst. 
The existence of two classes of gamma-ray bursts has been established for some time \citep[e.g.,][]{1981Ap&SS..80..119M,1984Natur.308..434N,1992AIPC..265....3H,1993ApJ...413L.101K}. It is generally accepted that long GRBs have their origins in massive star progenitors because of their   association with core-collapse supernovae (SNe) \citep[e.g.][]{1998Natur.395..670G,2003Natur.423..847H,
2004ApJ...609L...5M,2004ApJ...609..952Z}  
and occurrence in star-forming galaxies \citep{2002AJ....123.1111B} and in highly star-forming regions therein \citep{2006Natur.441..463F}.   
For a recent review of gamma-ray bursts see \citet{2009grbb.book.....V}.
  
The origin of short GRBs is still open, with mergers of compact objects being the leading concept \citep[e.g.,][]{1989Natur.340..126E,2005Natur.437..851G,2005Natur.437..859H,2005Natur.437..845F,2006MNRAS.368L...1L}. For a recent review of short bursts and their progenitors see \citet{2007PhR...442..166N} and \citet{2007NJPh....9...17L}. In contrast to long GRBs, the afterglows of the short bursts are generally fainter  \citep{2008arXiv0804.1959K,2009ApJ...701..824N}, making it difficult to obtain an absorption line spectrum of the afterglow itself. Almost exclusively, the distances to short bursts are obtained by associations with and observations of the (putative) host galaxies \citep[e.g.,][]{2009ApJ...690..231B,2009ApJ...698.1620G}.  See however \citet{2007MNRAS.378.1439L}. One possible exception is the case of GRB 090426 at $z=2.609$ \citep{2009arXiv0907.1661L,2009arXiv0911.0046A,Zhang080913} for which absorption line spectroscopy of the afterglow was successfully obtained, but may be considered on balance by the authors to be associated with the population of long gamma-ray bursts. 
The host galaxies of the short bursts were found initially to be mostly elliptical galaxies, e.g.,
GRB 050509B \citep{2005Natur.437..851G,2006ApJ...638..354B} and GRB 050724 
 \citep{2005Natur.438..988B,2006A&A...450...87G} and that an old progenitor population was required.
However, as the sample has since increased, all types of galaxies, from elliptical to star-forming, have been found to be associated with short bursts and the majority of short bursts are now found to reside in star forming galaxies \citep[e.g.,][]{2007ApJ...664.1000B,2009ApJ...690..231B,2006ApJ...642..989P}. 
A comparison of the short and long burst hosts by  \citet{2009ApJ...690..231B}  revealed that the former have
higher luminosities, and that it is unlikely that both samples are drawn from the same underlying
galaxy distribution ($p=10^{-3}$) and that the short burst hosts properties match well to field galaxies in the range $z \sim 0.1 - 1$. 
 
\begin{table*}
\begin{minipage}{\textwidth}
\begin{center}
\caption{Summary of the burst properties including the redshifts, isotropic gamma-ray energy ($E_{\gamma, \rm{iso}}$), the 
power-law decay index of the afterglow light curve ($\alpha$), the afterglow spectral index ($\beta_{\rm opt}$), the template employed to constrain the host galaxy extinction ($A_{V}^{\rm{host}}$), limits on the half opening angle 
 ($\theta_{\rm jet}^{\rm ISM}$) and on the beaming corrected gamma-ray energies  ($E_{\rm{\gamma}}$ ) and finally the star formation rates obtained for two of the host galaxies. 
} 
\label{summary_table}
\renewcommand{\thefootnote}{\alph{footnote}}
\renewcommand{\footnoterule}{}  
\begin{tabular}{lccccccccc} 
\hline\hline
 	&	&  	&		&	&		& & 	  &     &  \\
Burst  &  Redshift  & $E_{\gamma, \rm{iso}}$\footnotemark[1]   & $\alpha$ &  $\beta_{\rm opt}$  &  Template & $A_{V}^{\rm{host}}$ &  $\theta_{\rm jet}^{\rm ISM}$  &$E_{\rm{\gamma}}$ \footnotemark[1]   & SFR\footnotemark[8]   \\
 	&	& \textit{erg}	&		&	&		& & 	 {\it $^\circ$}  &  {(\it erg)}    & {\it (M$_\odot$ yr$^{-1}$)}  \\
\hline\hline
 	&	&  	&		&	&		& & 	  &     &  \\
GRB\,090323\footnotemark[2] & 3.57\footnotemark[3] & $5.1\times 10^{54}$ & $1.90\pm0.01$ & $0.65\pm0.13$ & SMC & $0.14^{+0.04}_{-0.03}$ & $\lesssim 2.1^\circ$ & $\lesssim 3.3 \times 10^{51}$ & $\cdots$ \\
GRB\,090328 & 0.7354 & $1.0\times 10^{53}$ & $2.27\pm{0.04}$ & $1.19^{+0.24}_{-0.21}$ & SMC\footnotemark[4] & $0.22^{+0.06}_{-0.18}$ & $\lesssim 5.5^\circ$ & $\lesssim 4.6 \times 10^{50}$	&  4.8  \\
GRB\,090902B & 1.822\footnotemark[5] & $3.5\times 10^{54}$ & $0.94 \pm 0.02$ & $0.79^{+0.05}_{-0.19}$ & SMC/LMC & $\lesssim 0.34$  & $\gtrsim 6.4^\circ$ &  $\gtrsim 2.2 \times 10^{52}$ 	& $\cdots$    \\
GRB\,080916C\footnotemark[2],\footnotemark[6],\footnotemark[7]  & 4.35\footnotemark[7] & $8.8\times 10^{54}$ & $ 1.40 \pm 0.05$ & $0.38^{+0.20}_{-0.19}$ & $\cdots$ &  $\cdots$  & $\gtrsim 6^\circ$ &  $\gtrsim 4.8 \times 10^{52}$ 	& $\cdots$    \\
GRB\,090510 & 0.903 & $1.1\times 10^{53}$ & $\cdots$ & $\cdots$ & $\cdots$ & $\leq$1.1 & $\cdots$ & $\cdots$ & 0.30 \\
\hline
\footnotetext[1]{$E_{\gamma, \rm{iso}}$ values are  quoted in the energy range 1~keV -- 10~GeV, with the exception of 
GRB~080916C (10~keV--10~GeV: Abdo et al. 2009a) and GRB~090510  (10~keV --30~GeV : Abdo et al. 2009c). }
\footnotetext[2]{The opening angle and $E_{\rm{\gamma}}$  are lower in a wind medium for GRB~090323 and 080916C.}
\footnotetext[3]{ \citet{2009GCN..9028....1C}.}
\footnotetext[4]{LMC and MW templates fit the data equally well and it is not possible to distinguish between models. }
\footnotetext[5]{\citet{Z_090902B}.}
\footnotetext[6]{A value of $A_{V}^{\rm{host}}$ $\lesssim 0.23$ is obtained using the MW template.}
\footnotetext[7]{ Values of $\alpha$, $\beta_{\rm opt}$ and the jet break lower limit are taken from \citet{2009A&A...498...89G}. } 
\footnotetext[8]{The true SFR may be larger than measured because of slit losses.} 
\end{tabular}
\end{center}
\end{minipage}
\end{table*}

\subsection{Afterglow properties and energetics}

The bursts were followed up in the optical and NIR starting at times ranging from 6 hours to 1.6 days post trigger. In all cases a bright afterglow  was detected and followed for many days, with the exception of GRB~090510 which was only detected on the first night post-trigger.  The afterglow light curves of the three long bursts are well sampled in the seven GROND bands. The two March events are well fit by power-laws with indices consistent with post-jet break models constraining the opening angle to be before the start of GROND observations. In the case of GRB~090323 the opening angle ($\theta_{\rm jet}^{\rm ISM}$) is less than $2.1^\circ$ while for GRB~090328 the observations began later and the opening angle 
($\theta_{\rm jet}^{\rm ISM}$) is less than $5.5^\circ$. The slope of the afterglow decay of GRB~090902B is flatter and consistent with a pre-jet break value, constraining the opening angle ($\theta_{\rm jet}^{\rm ISM}$) of the jet to be $\gtrsim 6.4^\circ$. 

Knowledge of the opening angle allows us to correct the isotropic energy and calculate limits on $E_{\gamma}$, the beaming corrected energy release in $\gamma$-rays \citep{1999ApJ...519L..17S,2001ApJ...562L..55F}. The $E_{\gamma}$ of GRBs~090323 and 090328 is in a range compatible with the majority of GRBs observed to date.  
In the case of GRB~090902B, 
the late  break places at least a lower limit on the opening angle and therefore a  very large beaming corrected energy output of $\gtrsim 2.2 \times 10^{52}$~{erg}  is obtained. To date this is one of the largest values of $E_{\gamma}$ measured and indicates that the large $E_{\gamma, \rm{iso}}$ values of \textit{Fermi}/LAT GRBs are not always due to a narrow opening angle where we happen to be located in the cone. The value of $E_{\gamma}$ obtained for the three long bursts ranges from less than $5 \times 10^{50}$  {erg}  to greater than $2 \times 10^{52}$  {erg}, spanning over a factor of at least  30. 
Values in excess of $10^{52}$ {erg} have been reported for another \textit{Fermi}/LAT detected burst \citep{2009A&A...498...89G} and a number of \textit{Swift} bursts \citep{2009arXiv0905.0690C} and indicate that the distribution of  $E_{\gamma}$ is broad and not compatible with a standard candle \citep{2001ApJ...562L..55F,2003ApJ...594..674B}. Furthermore these very luminous GRBs with high values of the redshift corrected $E_{\rm peak}$ are not compatible with the $E_{\rm peak, i}$ -- $E_{\rm \gamma}$ relationship as shown in Fig~\ref{Amaghi} which was obtained for GRBs with in general lower values of the peak energy \citep{2004ApJ...616..331G,2007A&A...472..395C}. While being consistent with the Amati ($E_{\rm peak, i}$ -- $E_{\rm iso}$) relation at the $2\sigma$ level (see also \citealp{2009A&A...508..173A}), albeit with a large scatter (see also \citealp[e.g.][]{2007ApJ...671..656B}), all four LAT bursts are inconsistent with the Ghirlanda relation \citep{2004ApJ...616..331G} at $\gtrsim 2\sigma$ using a standard approach. Two GRBs (080916C and 090902B) have too large beaming corrected $E_{\gamma}$ for their $E_{\rm peak, i}$, while $E_{\rm peak, i}$ for GRBs 090323 and 090328 is too high to fit the Ghirlanda relation. In the earlier case, subtracting the energy in the additional power-law component of the prompt emission \citep{090902B_PAPER}, a wind environment, or a two-component model could possibly relax the constraints on $E_{\gamma}$ directly (extra-component) or indirectly by reducing the lower limit on the opening angle (Wind-case), or by disconnecting the beaming of the prompt emission from the late afterglow (two-component model). Although somewhat contrived, multi-component jets \citep[e.g.][]{2003Natur.426..154B, 2008Natur.455..183R, 2009A&A...508..593K} would allow the $\gamma$-ray emission to be tighter collimated than the late afterglow, and hence reduce the measure on the opening angle obtained from the afterglow light-curve to an upper limit. Similar arguments can not be invoked for GRBs~090323 and 090328, where the latter one in particular is a significant outlier from the $E_{\rm peak, i}$ -- $E_{\rm \gamma}$ relationship (Fig.~\ref{Amaghi}).

 \begin{figure}[t]
\centering
{\rotatebox{0}{ \includegraphics[width=1\columnwidth]{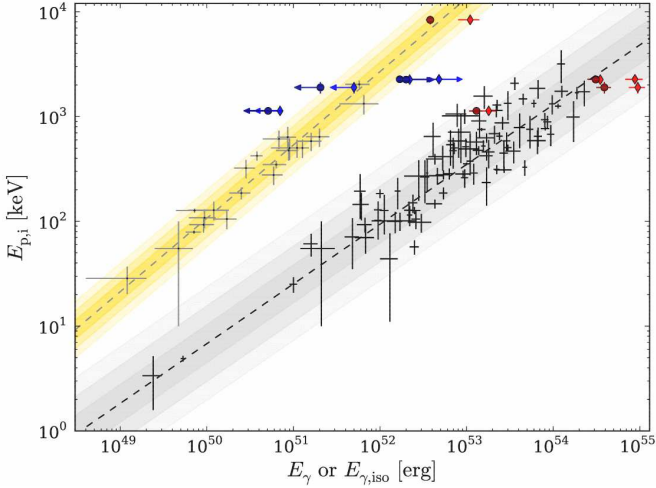}}}
\caption{
Amati and Ghirlanda relation challenged by \textit{Fermi}/LAT bursts. The black crosses, the black dashed line, and the grey shaded areas represent the data and best-fit $E_{\rm peak, i}$ -- $E_{\rm iso}$ relation and its $1\sigma$,  $2\sigma$ and $3\sigma$ scatter as compiled in \citealp{2008MNRAS.391..577A}. The grey crosses, the grey dashed line, and the golden shaded areas represent the data and best-fit $E_{\rm peak, i}$ -- $E_{\gamma}$ relation and its $1\sigma$,  $2\sigma$ and $3\sigma$ scatter as compiled in \citealp{2007A&A...466..127G}. Red points are the 5 \textit{Fermi}/LAT bursts with respect to the Amati relation. The red circles are calculated in the bolometric energy range (1~keV to 10~MeV) and the red diamonds are calculated over the energy ranges quoted in Table~\ref{summary_table}.  The four long LAT bursts are consistent with it at the $\lesssim2\sigma$ level. 
Blue points with arrows are the limits on the four long \textit{Fermi}/LAT bursts with respect to the Ghirlanda relation. The blue circles symbols indicate the data in the bolometric energy range (1~keV to 10~MeV) and the diamonds are again the values from Table~\ref{summary_table}.  Over the bolometric energy range all four are inconsistent with the relation at $\gtrsim 2 \sigma$ level and over the wider energy range, three are inconsistent with it at the $\gtrsim 3 \sigma$ level and GRB~090323 is consistent.
}
\label{Amaghi}
\end{figure}

\subsection{Host properties}

Host galaxies were detected for two of the long bursts, GRBs~090323 and 090328, and the short burst GRB~090510.  
The host galaxy of GRB~090323 was detected in the $r^\prime$ and $ i^\prime$ bands but further constraints could not be obtained  due to the faintness of the source and the high redshift. The host galaxy of GRB~090328  is  best fit with a burst galaxy template (although all other templates provide acceptable fits) and the absolute  magnitudes are comparable with  previous long GRB host galaxies \citep[e.g.,][]{2004A&A...425..913C, 2009ApJ...691..182S}.  A dust-corrected star-formation rate of $4.8$ M$_\odot$ yr$^{-1}$, as well as the star formation rate per unit mass  (2.8 Gyr$^{-1}$) and per unit luminosity  ( $\sim$ 10 M$_\odot$ yr$^{-1} L_*^{-1}$) derived from the afterglow spectrum are  consistent with the long burst sample \citep{2009ApJ...691..182S,2009ApJ...690..231B}. The host galaxy of GRB~090510 was detected by GROND and a spectrum was obtained by the VLT. A star formation rate of  $0.30$ M$_\odot$ Gyr$^{-1}$ (not corrected for dust) was derived and the host galaxy is best fit by an elliptical template, however all templates provide acceptable fits. The dust-corrected star formation rate per unit luminosity of $\sim$ 2 M$_\odot$ yr$^{-1} L_*^{-1}$   for this host galaxy is consistent with the values for the short bursts presented by \citet{2009ApJ...690..231B}. The properties of the host galaxies of two bursts with high-energy emission detected by \textit{Fermi}/LAT, the long burst GRB~090328 and the short burst GRB~090510, are not  exceptional with respect to pre-\textit{Fermi} samples.  

\subsection{Optical afterglows of LAT-detected GRBs in comparison with the previous sample}

\begin{figure}[t]
\centering
{\rotatebox{0}{ \includegraphics[width=1\columnwidth]{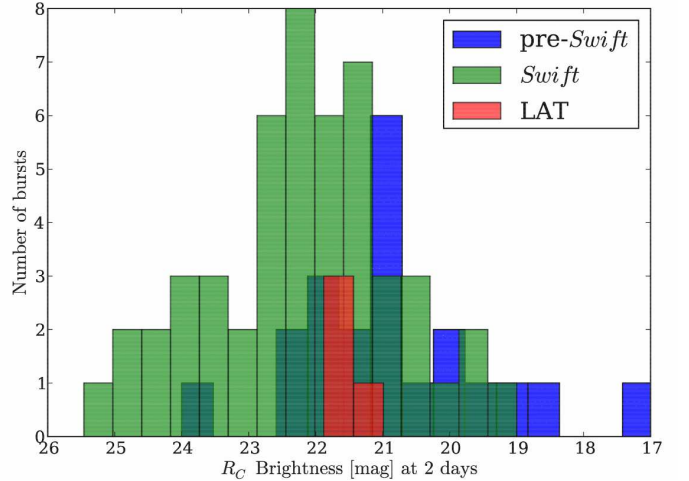}}}
\caption{
The afterglow brightness at day 2 of four long \emph{Fermi}/LAT-detected GRBs in comparison with a sample of over 70 well 
observed afterglows of long GRBs detected until May 2009 (Kann et al. 2006, 2009a). The comparison sample comprises 50 \textit{Swift}-era and 21 pre-\textit{Swift} afterglows.}
\label{Brightness_dist}
\end{figure}

With knowledge of the distance from the redshift determination and the local dust extinction from the SED fitting, we are able to compare the afterglows of our sample of LAT-detected GRBs \citep[where we also include GRB 080916C, ][]{2009A&A...498...89G} with a large sample of well observed GRB afterglows presented in \cite{2006ApJ...641..993K} (pre-\textit{Swift} long GRBs), \cite{2007arXiv0712.2186K} (\textit{Swift}-era long GRBs) and \cite{2008arXiv0804.1959K}(\textit{Swift}-era short GRBs). Using the method detailed in \cite{2006ApJ...641..993K}, the afterglows are corrected for host-galaxy extinction and all shifted to a common redshift of $z=1$, so they can be compared directly both in terms of temporal evolution as well as luminosity. The LAT-detected-GRB afterglows are shown in comparison to the total sample in Fig. \ref{BigFig}.

In comparison to the sample of long GRB afterglows, the three afterglows presented in this work as well as GRB~080916C are seen to be diverse. The optical afterglow of GRB~090323, especially at early times, is one of the most luminous afterglows ever detected. The steep decay from discovery on gives it a more typical luminosity at later times. The afterglow of GRB~080916C lies close to the mean of the sample distribution. At discovery, the afterglow of GRB 090328 has a similar luminosity, but again, a steep decay from the onset implies that it becomes much less luminous at later times. The afterglow of GRB~090902B has a flatter slope and the very late break causes it to be one of the most luminous afterglows known at late times.

Observationally, however, the four optical afterglows of the long \textit{Fermi}/LAT bursts are exceptionally bright. The histogram of observed magnitudes from the sample in Fig.~\ref{BigFig} is shown in Fig.~\ref{Brightness_dist} at a common time of 2~days after the trigger in the observers frame. At this time, all four bursts are brighter than the mean brightness ($R_C = 22.2$) of the \textit{Swift}-era comparison sample. Furthermore, the sample is not complete, comprising 50 out of a total of 370 bursts with X-ray afterglows as a rough indicator for observability between the launch of \textit{Swift} and May 2009. Accounting for the observational bias, and assuming in the extreme case that the comparison sample is complete at the bright end of the distribution, then all four LAT afterglows would lie in the brightest 5\%, with GRB~090328 being the brightest (top 2\%, i.e. 8th 
brightest of \textit{Swift}-era or 16th of all afterglows).

In comparison to the afterglows of other short GRBs (grey lines), it can be seen that the afterglow of GRB 090510 is among the most luminous (in agreement with its extreme prompt energy release, \citealt{2009arXiv0908.1832F}, and the correlation between prompt energy release and afterglow luminosity, \citealt{2008arXiv0804.1959K, 2009ApJ...701..824N,2008ApJ...689.1161G}), at least at early times. After just a few hours, it goes over into a steep, probably post-jet break decay, and at one day, its luminosity is comparable to the few other short GRB afterglows detected at this time. The upper limit derived from the second GROND epoch is not restrictive. 

The host galaxy extinction found in all three long GRB afterglows lies within the range of $A_V\approx0.2$ which is typical for bright optical/NIR afterglows  \citep{2007arXiv0712.2186K}, with no extinction detected in the case of GRB 080916C \citep{2009A&A...498...89G}. 

The intrinsic spectral slopes $\beta_{\rm opt}$ are also diverse, with that of GRB 090328 being a typical value for an afterglow with a cooling break $\nu_c$ redwards of the optical/NIR regime, while the obtained optical/NIR data for GRBs 090323 and 090902B very probably probe the spectral region of $\nu_{\rm m} < \nu_{\rm opt} < \nu_{\rm c}$.

\subsection{Bumps in the afterglow light curves}

Variability superimposed on to a power-law decay as seen in GRBs~090323 and 090328 (Figs.~\ref{090323lc} and \ref{090328lc}) is frequently observed in well sampled optical afterglow light curves \citep[e.g.,][]{2002A&A...396L...5L,2003Natur.426..157G,2004ApJ...606..381L, 2008ApJ...685..361U, 2008ApJ...672..449P, 2009ApJ...697..758K}. Given the low amplitude, smooth structure, time, and roughly equal color to the power-law component, the observed variability in the light curve of GRB~090323 is most likely related to a process intrinsic to the generic forward shock or the circumburst medium, and not to late inner engine activity. The most commonly inferred mechanism for producing variability as observed in the optical afterglow of GRB~090323 is additional energy injected into the afterglow via refreshed shocks \citep{1998ApJ...496L...1R}, although an internal origin, i.e. an optical flare related to late central engine activity can not be ruled out. In GRB~090328 there is minor evidence for a slightly redder spectra in the bump component peaking at $\sim$350~ks post burst, which would support a different origin of the observed variability than forward shock emission. Given the uncertainties in the measurement however, the spectral index is consistent with the afterglow slope at the $2\sigma$ level.
 
Various models are emerging which explain delayed GeV with respect to the keV - MeV emission in \textit{Fermi} bursts, among others an external shock model \citep[e.g.][]{2009arXiv0910.2459G,2009arXiv0905.2417K} or a hadronic model  \citep{2009arXiv0908.0513R}. \citet{2009arXiv0908.0513R} demonstrate that synchrotron radiation from cosmic-ray  protons accelerated in a GRB, delayed by the proton synchrotron cooling timescale in a jet of magnetically-dominated shocked plasma moving at highly relativistic speeds, explains the delayed GeV emission as observed in many \textit{Fermi}/LAT bursts. A second generation prompt electron synchrotron component from attenuated proton synchrotron radiation makes enhanced  soft X-ray to MeV gamma-ray emission, which is, however, too weak to be detected. In this scenario the implicit assumption is made that the outflow impacts either a previously ejected shell or a pre-existing uniform density medium with an extent of $ r \sim 6 \times 10^{16} ({\Gamma}/{1000})^{2} ({\delta t}/{10s})$ cm.
For a 3 second  
delay in GRB~090323/28 this corresponds to a light travel time of 8-10 days, surprisingly similar to the timing of the bump in the optical/NIR light curve as observed with GROND. It is conceivable  that the delayed GeV emission and the late, likely achromatic bumps in the afterglow light curve of GRBs could be related. One possible scenario would be Compton scattering of this proton synchrotron emission at the shell material, thus creating an achromatic bump with a power-law of photon index -2 (or even steeper) depending on the efficiency of the scattering. We note however, that density enhancements in the circumburst medium alone can not account for significant rebrightnings or strong bumps observed in optical afterglow light curves \citep{2007MNRAS.380.1744N}. To reproduce the observed variability in optical afterglow light curves, a second process is required, possibly the variation of the microphysical shock parameters along the shock condition as suggested by \citet{2010MNRAS.402..409K}.

\section{Conclusion}

We have presented the multi-colour afterglow observations of four bursts with high energy emission including three long and one short burst.  In addition, we present spectroscopic observations and redshift determinations of two of these bursts. 
It is now possible for the first time to combine the high energy information from the prompt emission with the afterglow, host galaxy and redshifts of these sources.  Follow-up of GRBs detected by \textit{Fermi} with the seven band imager, GROND, allows simultaneous determination of the temporal, $\alpha$, and spectral, $\beta_{\rm opt}$, indices of the burst afterglows. The long bursts exhibit power-law decay indices ($\alpha$) from less than 1  to $\sim$2.3 and spectral  indices ($\beta_{\rm opt}$)   from 0.65 to $\sim$1.2 which are fairly standard for  GRB afterglows. 
 Moreover, an estimation of the jet break time is vital to determine the broadband properties and energetics of these events. The redshifts of the long bursts span the range from 0.7354 to 3.57 and the beaming corrected energies differ by a factor of at least $\sim$30 with a value of $E_{\rm{\gamma}}$ $\gtrsim 2.2 \times 10^{52}$ {erg}  for GRB~090902B and are not compatible with a standard candle.  Interestingly, the higher redshift bursts detected by the LAT are very luminous in the keV-MeV region of the prompt spectrum, have high $E_{\gamma, \rm{iso}}$ and $E_{\rm{\gamma}}$ and also exceptionally bright  afterglows. The lower redshift GRB~090328 is the exception in this respect, with a more standard $E_{\gamma, \rm{iso}}$ as compared to \textit{Swift} GRBs.
 The host galaxies of GRB~090328 and GRB~090510 fit well within the distributions of the long and short burst host galaxies respectively. 
 Future photometric and spectroscopic follow-up of these rare  \textit{Fermi} bursts with high-energy emission is crucial to further investigate the energetics  and host galaxies of a larger sample.

\begin{acknowledgements}
Based on observations made with ESO Telescopes at the Paranal Observatories under programme ID 083.D-0903 and 283.D-5059.
We thank the staff at the European Southern Observatory for the generous allocation of observing time and execution of observations.  In particular we thank Tim de Zeeuw, Andreas Kaufer,  Michael Sterzik, Cedric Ledoux, Thomas Rivinius, Chris Lidman, F. J. Selman and I. Condor. Part of the funding for GROND (both hardware as well as personnel) was generously granted from the Leibniz-Prize to Prof. G. Hasinger (DFG grant HA 1850/28-1). SMB acknowledges support of the  a European Union Marie Curie European Reintegration Grant within the 7th Program under contract number PERG04-GA-2008-239176. TK acknowledges support by the DFG cluster of excellence Origin and Structure of the Universe. DAK acknowledges support by the Th\"uringer Landessternwarte Tautenburg, and thanks S. Stecklum for observing time, as well as U. Laux and F. Ludwig for performing the observations.  We thank the anonymous referee for comments which helped to improved the paper.

\end{acknowledgements}



\begin{thebibliography}{199}
\expandafter\ifx\csname natexlab\endcsname\relax\def\natexlab#1{#1}\fi

\bibitem[{{Abazajian} {et~al.}(2009){Abazajian}, {Adelman-McCarthy},
  {Ag{\"u}eros}, {Allam}, {Allende Prieto}, {An}, {Anderson}, {Anderson},
  {Annis}, {Bahcall}, {Bailer-Jones}, {Barentine}, {Bassett}, {Becker},
  {Beers}, {Bell}, {Belokurov}, {Berlind}, {Berman}, {Bernardi}, {Bickerton},
  {Bizyaev}, {Blakeslee}, {Blanton}, {Bochanski}, {Boroski}, {Brewington},
  {Brinchmann}, {Brinkmann}, {Brunner}, {Budav{\'a}ri}, {Carey}, {Carliles},
  {Carr}, {Castander}, {Cinabro}, {Connolly}, {Csabai}, {Cunha}, {Czarapata},
  {Davenport}, {de Haas}, {Dilday}, {Doi}, {Eisenstein}, {Evans}, {Evans},
  {Fan}, {Friedman}, {Frieman}, {Fukugita}, {G{\"a}nsicke}, {Gates},
  {Gillespie}, {Gilmore}, {Gonzalez}, {Gonzalez}, {Grebel}, {Gunn},
  {Gy{\"o}ry}, {Hall}, {Harding}, {Harris}, {Harvanek}, {Hawley}, {Hayes},
  {Heckman}, {Hendry}, {Hennessy}, {Hindsley}, {Hoblitt}, {Hogan}, {Hogg},
  {Holtzman}, {Hyde}, {Ichikawa}, {Ichikawa}, {Im}, {Ivezi{\'c}}, {Jester},
  {Jiang}, {Johnson}, {Jorgensen}, {Juri{\'c}}, {Kent}, {Kessler}, {Kleinman},
  {Knapp}, {Konishi}, {Kron}, {Krzesinski}, {Kuropatkin}, {Lampeitl},
  {Lebedeva}, {Lee}, {Lee}, {Leger}, {L{\'e}pine}, {Li}, {Lima}, {Lin}, {Long},
  {Loomis}, {Loveday}, {Lupton}, {Magnier}, {Malanushenko}, {Malanushenko},
  {Mandelbaum}, {Margon}, {Marriner}, {Mart{\'{\i}}nez-Delgado}, {Matsubara},
  {McGehee}, {McKay}, {Meiksin}, {Morrison}, {Mullally}, {Munn}, {Murphy},
  {Nash}, {Nebot}, {Neilsen}, {Newberg}, {Newman}, {Nichol}, {Nicinski},
  {Nieto-Santisteban}, {Nitta}, {Okamura}, {Oravetz}, {Ostriker}, {Owen},
  {Padmanabhan}, {Pan}, {Park}, {Pauls}, {Peoples}, {Percival}, {Pier}, {Pope},
  {Pourbaix}, {Price}, {Purger}, {Quinn}, {Raddick}, {Fiorentin}, {Richards},
  {Richmond}, {Riess}, {Rix}, {Rockosi}, {Sako}, {Schlegel}, {Schneider},
  {Scholz}, {Schreiber}, {Schwope}, {Seljak}, {Sesar}, {Sheldon}, {Shimasaku},
  {Sibley}, {Simmons}, {Sivarani}, {Smith}, {Smith}, {Smol{\v c}i{\'c}},
  {Snedden}, {Stebbins}, {Steinmetz}, {Stoughton}, {Strauss}, {Subba Rao},
  {Suto}, {Szalay}, {Szapudi}, {Szkody}, {Tanaka}, {Tegmark}, {Teodoro},
  {Thakar}, {Tremonti}, {Tucker}, {Uomoto}, {Vanden Berk}, {Vandenberg},
  {Vidrih}, {Vogeley}, {Voges}, {Vogt}, {Wadadekar}, {Watters}, {Weinberg},
  {West}, {White}, {Wilhite}, {Wonders}, {Yanny}, {Yocum}, {York}, {Zehavi},
  {Zibetti}, \& {Zucker}}]{2009ApJS..182..543A}
{Abazajian}, K.~N., {Adelman-McCarthy}, J.~K., {Ag{\"u}eros}, M.~A., {et~al.}
  2009, \apjs, 182, 543

\bibitem[{{Abdo} {et~al.}(2009{\natexlab{a}}){Abdo}, {Ackermann}, {Ajello},
  {Asano}, {Atwood}, {Axelsson}, {Baldini}, {Ballet}, {Barbiellini}, {Baring},
  {Bastieri}, {Bechtol}, {Bellazzini}, {Berenji}, {Bhat}, {Bissaldi}, {Bloom},
  {Bonamente}, {Bonnell}, {Borgland}, {Bouvier}, {Bregeon}, {Brez}, {Briggs},
  {Brigida}, {Bruel}, {Burgess}, {Burnett}, {Caliandro}, {Cameron}, {Caraveo},
  {Casandjian}, {Cecchi}, {{\c C}elik}, {Chaplin}, {Charles}, {Cheung},
  {Chiang}, {Ciprini}, {Claus}, {Cohen-Tanugi}, {Cominsky}, {Connaughton},
  {Conrad}, {Cutini}, {Dermer}, {de Angelis}, {de Palma}, {Digel}, {Dingus},
  {Do Couto E Silva}, {Drell}, {Dubois}, {Dumora}, {Farnier}, {Favuzzi},
  {Fegan}, {Finke}, {Fishman}, {Focke}, {Foschini}, {Fukazawa}, {Funk},
  {Fusco}, {Gargano}, {Gasparrini}, {Gehrels}, {Germani}, {Gibby}, {Giebels},
  {Giglietto}, {Giordano}, {Glanzman}, {Godfrey}, {Granot}, {Greiner},
  {Grenier}, {Grondin}, {Grove}, {Grupe}, {Guillemot}, {Guiriec}, {Hanabata},
  {Harding}, {Hayashida}, {Hays}, {Hoversten}, {Hughes}, {J{\'o}hannesson},
  {Johnson}, {Johnson}, {Johnson}, {Kamae}, {Katagiri}, {Kataoka}, {Kawai},
  {Kerr}, {Kippen}, {Kn{\"o}dlseder}, {Kocevski}, {Kouveliotou}, {Kuehn},
  {Kuss}, {Lande}, {Latronico}, {Lemoine-Goumard}, {Longo}, {Loparco}, {Lott},
  {Lovellette}, {Lubrano}, {Madejski}, {Makeev}, {Mazziotta}, {McBreen},
  {McEnery}, {McGlynn}, {M{\'e}sz{\'a}ros}, {Meurer}, {Michelson},
  {Mitthumsiri}, {Mizuno}, {Moiseev}, {Monte}, {Monzani}, {Moretti},
  {Morselli}, {Moskalenko}, {Murgia}, {Nakamori}, {Nolan}, {Norris}, {Nuss},
  {Ohno}, {Ohsugi}, {Omodei}, {Orlando}, {Ormes}, {Ozaki}, {Paciesas},
  {Paneque}, {Panetta}, {Parent}, {Pelassa}, {Pepe}, {Pesce-Rollins},
  {Petrosian}, {Piron}, {Porter}, {Preece}, {Rain{\`o}}, {Ramirez-Ruiz},
  {Rando}, {Razzano}, {Razzaque}, {Reimer}, {Reimer}, {Reposeur}, {Ritz},
  {Rochester}, {Rodriguez}, {Roth}, {Ryde}, {Sadrozinski}, {Sanchez}, {Sander},
  {Saz Parkinson}, {Scargle}, {Schalk}, {Sgr{\`o}}, {Siskind}, {Smith},
  {Smith}, {Spandre}, {Spinelli}, {Stamatikos}, {Stecker}, {Strickman},
  {Suson}, {Tajima}, {Takahashi}, {Takahashi}, {Tanaka}, {Thayer}, {Thayer},
  {Thompson}, {Tibaldo}, {Toma}, {Torres}, {Tosti}, {Troja}, {Uchiyama},
  {Uehara}, {Usher}, {van der Horst}, {Vasileiou}, {Vilchez}, {Vitale}, {von
  Kienlin}, {Waite}, {Wang}, {Wilson-Hodge}, {Winer}, {Wood}, {Wu}, {Yamazaki},
  {Ylinen}, {Ziegler}, \& {the Fermi LAT Collaboration}}]{2009arXiv0908.1832F}
{Abdo}, A.~A., {Ackermann}, M., {Ajello}, M., {et~al.} 2009{\natexlab{a}},
  \nat, 462, 331

\bibitem[{{Abdo} {et~al.}(2009{\natexlab{b}}){Abdo}, {Ackermann}, {Asano},
  {Atwood}, {Axelsson}, {Baldini}, {Ballet}, {Band}, {Barbiellini}, {Bastieri},
  {Bechtol}, {Bellazzini}, {Berenji}, {Bhat}, {Bissaldi}, {Bloom}, {Bonamente},
  {Borgland}, {Bouvier}, {Bregeon}, {Brez}, {Briggs}, {Brigida}, {Bruel},
  {Burnett}, {Caliandro}, {Cameron}, {Caraveo}, {Casandjian}, {Cecchi},
  {Chaplin}, {Chekhtman}, {Cheung}, {Chiang}, {Ciprini}, {Claus},
  {Cohen-Tanugi}, {Cominsky}, {Connaughton}, {Conrad}, {Cutini}, {Dermer}, {de
  Angelis}, {de Palma}, {Digel}, {Silva}, {Drell}, {Dubois}, {Dumora},
  {Farnier}, {Favuzzi}, {Focke}, {Frailis}, {Fukazawa}, {Fusco}, {Gargano},
  {Gasparrini}, {Gehrels}, {Germani}, {Gibby}, {Giebels}, {Giglietto},
  {Giordano}, {Glanzman}, {Godfrey}, {Goldstein}, {Granot}, {Grenier},
  {Grondin}, {Grove}, {Guillemot}, {Guiriec}, {Hanabata}, {Harding},
  {Hayashida}, {Hays}, {Hughes}, {J{\'o}hannesson}, {Johnson}, {Johnson},
  {Kamae}, {Katagiri}, {Kataoka}, {Kawai}, {Kerr}, {Kn{\"o}dlseder},
  {Kocevski}, {Komin}, {Kouveliotou}, {Kuehn}, {Kuss}, {Latronico}, {Longo},
  {Loparco}, {Lott}, {Lovellette}, {Lubrano}, {Makeev}, {Mazziotta}, {McBreen},
  {McEnery}, {McGlynn}, {Meegan}, {Meurer}, {Michelson}, {Mitthumsiri},
  {Mizuno}, {Monte}, {Monzani}, {Moretti}, {Morselli}, {Moskalenko}, {Murgia},
  {Nakamori}, {Nolan}, {Norris}, {Nuss}, {Ohno}, {Ohsugi}, {Omodei}, {Orlando},
  {Ormes}, {Ozaki}, {Paciesas}, {Paneque}, {Panetta}, {Parent}, {Pelassa},
  {Pepe}, {Pesce-Rollins}, {Piron}, {Porter}, {Preece}, {Rain{\`o}}, {Rando},
  {Razzano}, {Razzaque}, {Reimer}, {Reposeur}, {Ritz}, {Rochester},
  {Rodriguez}, {Roth}, {Ryde}, {Sadrozinski}, {Sanchez}, {Sander}, {Saz
  Parkinson}, {Scargle}, {Sgr{\`o}}, {Siskind}, {Smith}, {Smith}, {Spandre},
  {Spinelli}, {Stamatikos}, {Strickman}, {Suson}, {Tajima}, {Takahashi},
  {Tanaka}, {Thayer}, {Thayer}, {Tibaldo}, {Torres}, {Tosti}, {Tramacere},
  {Uchiyama}, {Usher}, {van der Horst}, {Vasileiou}, {Vilchez}, {Vitale}, {von
  Kienlin}, {Waite}, {Wang}, {Wilson-Hodge}, {Winer}, {Wood}, {Ylinen}, \&
  {Ziegler}}]{2009arXiv0910.4192F}
{Abdo}, A.~A., {Ackermann}, M., {Asano}, K., {et~al.} 2009{\natexlab{b}}, \apj,
  707, 580

\bibitem[{{Abdo} {et~al.}(2009{\natexlab{c}})}]{090902B_PAPER}
{Abdo}, A.~A. {et~al.} 2009{\natexlab{c}}, ApJ, 706, L138

\bibitem[{{Abdo} {et~al.}(2009{\natexlab{d}})}]{2009Sci...323.1688A}
{Abdo}, A.~A. {et~al.} 2009{\natexlab{d}}, Science, 323, 1688

\bibitem[{{Allen} {et~al.}(2009){Allen}, {Yock}, {de Ugarte Postigo}, {Bond},
  {Hearnshaw}, {Christie}, {Kubanek}, {Castillo}, {Castro Ceron}, {Sanguino},
  {Perez-Ramirez}, {Claret}, {Garcia-Pelayo}, {Gorosabel}, {Guziy}, {Jelinek},
  {Ruiz}, \& {Castro-Tirado}}]{2009GCN..9058....1A}
{Allen}, B., {Yock}, P., {de Ugarte Postigo}, A., {et~al.} 2009, GRB
  Coordinates Network, 9058

\bibitem[{{Amati} {et~al.}(2009){Amati}, {Frontera}, \&
  {Guidorzi}}]{2009A&A...508..173A}
{Amati}, L., {Frontera}, F., \& {Guidorzi}, C. 2009, \aap, 508, 173

\bibitem[{{Amati} {et~al.}(2008){Amati}, {Guidorzi}, {Frontera}, {Della Valle},
  {Finelli}, {Landi}, \& {Montanari}}]{2008MNRAS.391..577A}
{Amati}, L., {Guidorzi}, C., {Frontera}, F., {et~al.} 2008, \mnras, 391, 577

\bibitem[{{Amelino-Camelia} {et~al.}(1998){Amelino-Camelia}, {Ellis},
  {Mavromatos}, {Nanopoulos}, \& {Sarkar}}]{1998Natur.393..763A}
{Amelino-Camelia}, G., {Ellis}, J., {Mavromatos}, N.~E., {Nanopoulos}, D.~V.,
  \& {Sarkar}, S. 1998, \nat, 393, 763

\bibitem[{{Amelino-Camelia} \& {Smolin}(2009)}]{2009PhRvD..80h4017A}
{Amelino-Camelia}, G. \& {Smolin}, L. 2009, \prd, 80, 084017

\bibitem[{{Antonelli} {et~al.}(2009){Antonelli}, {D'Avanzo}, {Perna}, {Amati},
  {Covino}, {Cutini}, {D'Elia}, {Gallozzi}, {Grazian}, {Palazzi},
  {Piranomonte}, {Rossi}, {Spiro}, {Stella}, {Testa}, {Chincarini}, {di Paola},
  {Fiore}, {Fugazza}, {Giallongo}, {Maiorano}, {Masetti}, {Pedichini},
  {Salvaterra}, {Tagliaferri}, \& {Vergani}}]{2009arXiv0911.0046A}
{Antonelli}, L.~A., {D'Avanzo}, P., {Perna}, R., {et~al.} 2009, \aap, 507, L45

\bibitem[{{Appenzeller} {et~al.}(1998){Appenzeller}, {Fricke}, {F{\"u}rtig},
  {G{\"a}ssler}, {H{\"a}fner}, {Harke}, {Hess}, {Hummel}, {J{\"u}rgens},
  {Kudritzki}, {Mantel}, {Meisl}, {Muschielok}, {Nicklas}, {Rupprecht},
  {Seifert}, {Stahl}, {Szeifert}, \& {Tarantik}}]{1998Msngr..94....1A}
{Appenzeller}, I., {Fricke}, K., {F{\"u}rtig}, W., {et~al.} 1998, The
  Messenger, 94, 1

\bibitem[{{Atwood} {et~al.}(2009){Atwood}, {Abdo}, {Ackermann}, {Althouse},
  {Anderson}, {Axelsson}, {Baldini}, {Ballet}, {Band}, {Barbiellini},
  {Bartelt}, {Bastieri}, {Baughman}, {Bechtol}, {B{\'e}d{\'e}r{\`e}de},
  {Bellardi}, {Bellazzini}, {Berenji}, {Bignami}, {Bisello}, {Bissaldi},
  {Blandford}, {Bloom}, {Bogart}, {Bonamente}, {Bonnell}, {Borgland},
  {Bouvier}, {Bregeon}, {Brez}, {Brigida}, {Bruel}, {Burnett}, {Busetto},
  {Caliandro}, {Cameron}, {Caraveo}, {Carius}, {Carlson}, {Casandjian},
  {Cavazzuti}, {Ceccanti}, {Cecchi}, {Charles}, {Chekhtman}, {Cheung},
  {Chiang}, {Chipaux}, {Cillis}, {Ciprini}, {Claus}, {Cohen-Tanugi},
  {Condamoor}, {Conrad}, {Corbet}, {Corucci}, {Costamante}, {Cutini}, {Davis},
  {Decotigny}, {DeKlotz}, {Dermer}, {de Angelis}, {Digel}, {do Couto e Silva},
  {Drell}, {Dubois}, {Dumora}, {Edmonds}, {Fabiani}, {Farnier}, {Favuzzi},
  {Flath}, {Fleury}, {Focke}, {Funk}, {Fusco}, {Gargano}, {Gasparrini},
  {Gehrels}, {Gentit}, {Germani}, {Giebels}, {Giglietto}, {Giommi}, {Giordano},
  {Glanzman}, {Godfrey}, {Grenier}, {Grondin}, {Grove}, {Guillemot}, {Guiriec},
  {Haller}, {Harding}, {Hart}, {Hays}, {Healey}, {Hirayama}, {Hjalmarsdotter},
  {Horn}, {Hughes}, {J{\'o}hannesson}, {Johansson}, {Johnson}, {Johnson},
  {Johnson}, {Johnson}, {Kamae}, {Katagiri}, {Kataoka}, {Kavelaars}, {Kawai},
  {Kelly}, {Kerr}, {Klamra}, {Kn{\"o}dlseder}, {Kocian}, {Komin}, {Kuehn},
  {Kuss}, {Landriu}, {Latronico}, {Lee}, {Lee}, {Lemoine-Goumard}, {Lionetto},
  {Longo}, {Loparco}, {Lott}, {Lovellette}, {Lubrano}, {Madejski}, {Makeev},
  {Marangelli}, {Massai}, {Mazziotta}, {McEnery}, {Menon}, {Meurer},
  {Michelson}, {Minuti}, {Mirizzi}, {Mitthumsiri}, {Mizuno}, {Moiseev},
  {Monte}, {Monzani}, {Moretti}, {Morselli}, {Moskalenko}, {Murgia},
  {Nakamori}, {Nishino}, {Nolan}, {Norris}, {Nuss}, {Ohno}, {Ohsugi}, {Omodei},
  {Orlando}, {Ormes}, {Paccagnella}, {Paneque}, {Panetta}, {Parent}, {Pearce},
  {Pepe}, {Perazzo}, {Pesce-Rollins}, {Picozza}, {Pieri}, {Pinchera}, {Piron},
  {Porter}, {Poupard}, {Rain{\`o}}, {Rando}, {Rapposelli}, {Razzano}, {Reimer},
  {Reimer}, {Reposeur}, {Reyes}, {Ritz}, {Rochester}, {Rodriguez}, {Romani},
  {Roth}, {Russell}, {Ryde}, {Sabatini}, {Sadrozinski}, {Sanchez}, {Sander},
  {Sapozhnikov}, {Parkinson}, {Scargle}, {Schalk}, {Scolieri}, {Sgr{\`o}},
  {Share}, {Shaw}, {Shimokawabe}, {Shrader}, {Sierpowska-Bartosik}, {Siskind},
  {Smith}, {Smith}, {Spandre}, {Spinelli}, {Starck}, {Stephens}, {Strickman},
  {Strong}, {Suson}, {Tajima}, {Takahashi}, {Takahashi}, {Tanaka}, {Tenze},
  {Tether}, {Thayer}, {Thayer}, {Thompson}, {Tibaldo}, {Tibolla}, {Torres},
  {Tosti}, {Tramacere}, {Turri}, {Usher}, {Vilchez}, {Vitale}, {Wang},
  {Watters}, {Winer}, {Wood}, {Ylinen}, \& {Ziegler}}]{2009ApJ...697.1071A}
{Atwood}, W.~B., {Abdo}, A.~A., {Ackermann}, M., {et~al.} 2009, \apj, 697, 1071

\bibitem[{{Atwood} {et~al.}(1994)}]{1994NIMPA.342..302A}
{Atwood}, W.~B. {et~al.} 1994, Nuclear Instruments and Methods in Physics
  Research A, 342, 302

\bibitem[{{Barthelmy} {et~al.}(2005){Barthelmy}, {Barbier}, {Cummings},
  {Fenimore}, {Gehrels}, {Hullinger}, {Krimm}, {Markwardt}, {Palmer},
  {Parsons}, {Sato}, {Suzuki}, {Takahashi}, {Tashiro}, \&
  {Tueller}}]{2005SSRv..120..143B}
{Barthelmy}, S.~D., {Barbier}, L.~M., {Cummings}, J.~R., {et~al.} 2005, Space
  Science Reviews, 120, 143

\bibitem[{{Beckwith} {et~al.}(2006){Beckwith}, {Stiavelli}, {Koekemoer},
  {Caldwell}, {Ferguson}, {Hook}, {Lucas}, {Bergeron}, {Corbin}, {Jogee},
  {Panagia}, {Robberto}, {Royle}, {Somerville}, \&
  {Sosey}}]{2006AJ....132.1729B}
{Beckwith}, S.~V.~W., {Stiavelli}, M., {Koekemoer}, A.~M., {et~al.} 2006, \aj,
  132, 1729

\bibitem[{{Berger}(2009)}]{2009ApJ...690..231B}
{Berger}, E. 2009, \apj, 690, 231

\bibitem[{{Berger} {et~al.}(2007){Berger}, {Fox}, {Price}, {Nakar}, {Gal-Yam},
  {Holz}, {Schmidt}, {Cucchiara}, {Cenko}, {Kulkarni}, {Soderberg}, {Frail},
  {Penprase}, {Rau}, {Ofek}, {Burnell}, {Cameron}, {Cowie}, {Dopita}, {Hook},
  {Peterson}, {Podsiadlowski}, {Roth}, {Rutledge}, {Sheppard}, \&
  {Songaila}}]{2007ApJ...664.1000B}
{Berger}, E., {Fox}, D.~B., {Price}, P.~A., {et~al.} 2007, \apj, 664, 1000

\bibitem[{{Berger} {et~al.}(2003){Berger}, {Kulkarni}, {Pooley}, {Frail},
  {McIntyre}, {Wark}, {Sari}, {Soderberg}, {Fox}, {Yost}, \&
  {Price}}]{2003Natur.426..154B}
{Berger}, E., {Kulkarni}, S.~R., {Pooley}, G., {et~al.} 2003, \nat, 426, 154

\bibitem[{{Berger} {et~al.}(2005){Berger}, {Price}, {Cenko}, {Gal-Yam},
  {Soderberg}, {Kasliwal}, {Leonard}, {Cameron}, {Frail}, {Kulkarni}, {Murphy},
  {Krzeminski}, {Piran}, {Lee}, {Roth}, {Moon}, {Fox}, {Harrison}, {Persson},
  {Schmidt}, {Penprase}, {Rich}, {Peterson}, \& {Cowie}}]{2005Natur.438..988B}
{Berger}, E., {Price}, P.~A., {Cenko}, S.~B., {et~al.} 2005, \nat, 438, 988

\bibitem[{{Bertin} \& {Arnouts}(1996)}]{1996A&AS..117..393B}
{Bertin}, E. \& {Arnouts}, S. 1996, \aaps, 117, 393

\bibitem[{{Bissaldi}(2009)}]{GBM_090926}
{Bissaldi}, E. 2009, GRB Coordinates Network, 9933

\bibitem[{{Bissaldi} \& {Connaughton}(2009)}]{Betta_090902B}
{Bissaldi}, E. \& {Connaughton}, V. 2009, GRB Coordinates Network, 9866

\bibitem[{{Bissaldi} {et~al.}(2010)}]{2010arXiv1002.4194B}
{Bissaldi}, E. {et~al.} 2010, ArXiv e-prints, 1002.4194

\bibitem[{{Bloom} {et~al.}(2003){Bloom}, {Frail}, \&
  {Kulkarni}}]{2003ApJ...594..674B}
{Bloom}, J.~S., {Frail}, D.~A., \& {Kulkarni}, S.~R. 2003, \apj, 594, 674

\bibitem[{{Bloom} {et~al.}(2002){Bloom}, {Kulkarni}, \&
  {Djorgovski}}]{2002AJ....123.1111B}
{Bloom}, J.~S., {Kulkarni}, S.~R., \& {Djorgovski}, S.~G. 2002, \aj, 123, 1111

\bibitem[{{Bloom} {et~al.}(2006){Bloom}, {Prochaska}, {Pooley}, {Blake},
  {Foley}, {Jha}, {Ramirez-Ruiz}, {Granot}, {Filippenko}, {Sigurdsson},
  {Barth}, {Chen}, {Cooper}, {Falco}, {Gal}, {Gerke}, {Gladders}, {Greene},
  {Hennanwi}, {Ho}, {Hurley}, {Koester}, {Li}, {Lubin}, {Newman}, {Perley},
  {Squires}, \& {Wood-Vasey}}]{2006ApJ...638..354B}
{Bloom}, J.~S., {Prochaska}, J.~X., {Pooley}, D., {et~al.} 2006, \apj, 638, 354

\bibitem[{{Bolzonella} {et~al.}(2000){Bolzonella}, {Miralles}, \&
  {Pell{\'o}}}]{2000A&A...363..476B}
{Bolzonella}, M., {Miralles}, J.-M., \& {Pell{\'o}}, R. 2000, \aap, 363, 476

\bibitem[{{Bouchet} {et~al.}(1985){Bouchet}, {Lequeux}, {Maurice}, {Prevot}, \&
  {Prevot-Burnichon}}]{bou85}
{Bouchet}, P., {Lequeux}, J., {Maurice}, E., {Prevot}, L., \&
  {Prevot-Burnichon}, M.~L. 1985, \aap, 149, 330

\bibitem[{{Bouvier} {et~al.}(2008){Bouvier}, {Band}, {Bregeon}, {Chiang},
  {Cutini}, {et~al.}}]{LAT_080825C}
{Bouvier}, A., {Band}, D., {Bregeon}, J., {et~al.} 2008, GRB Coordinates
  Network, 8183

\bibitem[{{Bruzual} \& {Charlot}(2003)}]{2003MNRAS.344.1000B}
{Bruzual}, G. \& {Charlot}, S. 2003, \mnras, 344, 1000

\bibitem[{{Burrows} {et~al.}(2005){Burrows}, {Hill}, {Nousek}, {Kennea},
  {Wells}, {Osborne}, {Abbey}, {Beardmore}, {Mukerjee}, {Short}, {Chincarini},
  {Campana}, {Citterio}, {Moretti}, {Pagani}, {Tagliaferri}, {Giommi},
  {Capalbi}, {Tamburelli}, {Angelini}, {Cusumano}, {Br{\"a}uninger}, {Burkert},
  \& {Hartner}}]{2005SSRv..120..165B}
{Burrows}, D.~N., {Hill}, J.~E., {Nousek}, J.~A., {et~al.} 2005, Space Science
  Reviews, 120, 165

\bibitem[{{Butler} {et~al.}(2007){Butler}, {Kocevski}, {Bloom}, \&
  {Curtis}}]{2007ApJ...671..656B}
{Butler}, N.~R., {Kocevski}, D., {Bloom}, J.~S., \& {Curtis}, J.~L. 2007, \apj,
  671, 656

\bibitem[{{Campana} {et~al.}(2007){Campana}, {Guidorzi}, {Tagliaferri},
  {Chincarini}, {Moretti}, {Rizzuto}, \& {Romano}}]{2007A&A...472..395C}
{Campana}, S., {Guidorzi}, C., {Tagliaferri}, G., {et~al.} 2007, \aap, 472, 395

\bibitem[{{Cenko} {et~al.}(2009{\natexlab{a}}){Cenko}, {Bloom}, {Morgan}, \&
  {Perley}}]{2009GCN..9053....1C}
{Cenko}, S.~B., {Bloom}, J.~S., {Morgan}, A.~N., \& {Perley}, D.~A.
  2009{\natexlab{a}}, GRB Coordinates Network, 9053

\bibitem[{{Cenko} {et~al.}(2009{\natexlab{b}}){Cenko}, {Frail}, {Harrison},
  {Kulkarni}, {Nakar}, {Chandra}, {Butler}, {Fox}, {Gal-Yam}, {Kasliwal},
  {Kelemen}, {Moon}, {Price}, {Rau}, {Soderberg}, {Teplitz}, {Werner}, {Bock},
  {Bloom}, {Starr}, {Filippenko}, {Chevalier}, {Gehrels}, {Nousek}, \&
  {Piran}}]{2009arXiv0905.0690C}
{Cenko}, S.~B., {Frail}, D.~A., {Harrison}, F.~A., {et~al.} 2009{\natexlab{b}},
  ArXiv e-prints,0905.0690

\bibitem[{{Chandra} \& {Frail}(2009)}]{Chandra_090902B}
{Chandra}, P. \& {Frail}, D.~A. 2009, GRB Coordinates Network, 9889

\bibitem[{{Chevalier} \& {Li}(2000)}]{2000ApJ...536..195C}
{Chevalier}, R.~A. \& {Li}, Z. 2000, \apj, 536, 195

\bibitem[{{Chornock} {et~al.}(2009){Chornock}, {Perley}, {Cenko}, \&
  {Bloom}}]{2009GCN..9028....1C}
{Chornock}, R., {Perley}, D.~A., {Cenko}, S.~B., \& {Bloom}, J.~S. 2009, GRB
  Coordinates Network, 9028

\bibitem[{{Christensen} {et~al.}(2004){Christensen}, {Hjorth}, \&
  {Gorosabel}}]{2004A&A...425..913C}
{Christensen}, L., {Hjorth}, J., \& {Gorosabel}, J. 2004, \aap, 425, 913

\bibitem[{{Cucchiara} {et~al.}(2009){Cucchiara}, {Fox}, {Tanvir}, \&
  {Berger}}]{Z_090902B}
{Cucchiara}, A., {Fox}, D.~B., {Tanvir}, N., \& {Berger}, E. 2009, GRB
  Coordinates Network, 9873

\bibitem[{{Cutini} {et~al.}(2009){Cutini}, {Vasileiou}, \&
  {Chiang}}]{2009GCN..9077....1C}
{Cutini}, S., {Vasileiou}, V., \& {Chiang}, J. 2009, GRB Coordinates Network,
  9077

\bibitem[{{D'Avanzo} {et~al.}(2009){D'Avanzo}, {Malesani}, {Covino},
  {Piranomonte}, {Grazian}, {Fugazza}, {Margutti}, {D'Elia}, {Antonelli},
  {Campana}, {Chincarini}, {Della Valle}, {Fiore}, {Goldoni}, {Mao}, {Perna},
  {Salvaterra}, {Stella}, {Stratta}, \& {Tagliaferri}}]{2009A&A...498..711D}
{D'Avanzo}, P., {Malesani}, D., {Covino}, S., {et~al.} 2009, \aap, 498, 711

\bibitem[{{de Jager} \& {Stecker}(2002)}]{2002ApJ...566..738D}
{de Jager}, O.~C. \& {Stecker}, F.~W. 2002, \apj, 566, 738

\bibitem[{{de Palma} {et~al.}(2009{\natexlab{a}}){de Palma}, {Bissaldi},
  {Tajima}, {Guiriec}, {Omodei}, {Vasileiou}, \& {Connaughton}}]{LAT2_090902B}
{de Palma}, F., {Bissaldi}, E., {Tajima}, H., {et~al.} 2009{\natexlab{a}}, GRB
  Coordinates Network, 9872

\bibitem[{{de Palma} {et~al.}(2009{\natexlab{b}}){de Palma}, {Bregeon}, \&
  {Tajima}}]{LAT_090902B}
{de Palma}, F., {Bregeon}, J., \& {Tajima}, H. 2009{\natexlab{b}}, GRB
  Coordinates Network, 9867

\bibitem[{{De Pasquale} {et~al.}(2010){De Pasquale}, {Schady}, {Kuin}, {Page},
  {Curran}, {Zane}, {Oates}, {Holland}, {Breeveld}, {Hoversten}, {Chincarini},
  {Grupe}, {Abdo}, {Ackermann}, {Ajello}, {Axelsson}, {Baldini}, {Ballet},
  {Barbiellini}, {Baring}, {Bastieri}, {Bechtol}, {Bellazzini}, {Berenji},
  {Bissaldi}, {Blandford}, {Bloom}, {Bonamente}, {Borgland}, {Bouvier},
  {Bregeon}, {Brez}, {Briggs}, {Brigida}, {Bruel}, {Burnett}, {Buson},
  {Caliandro}, {Cameron}, {Caraveo}, {Carrigan}, {Casandjian}, {Cecchi}, {{\c
  C}elik}, {Chekhtman}, {Chiang}, {Ciprini}, {Claus}, {Cohen-Tanugi},
  {Connaughton}, {Conrad}, {Dermer}, {de Angelis}, {de Palma}, {Dingus},
  {Silva}, {Drell}, {Dubois}, {Dumora}, {Farnier}, {Favuzzi}, {Fegan},
  {Fishman}, {Focke}, {Frailis}, {Fukazawa}, {Funk}, {Fusco}, {Gargano},
  {Gasparrini}, {Gehrels}, {Germani}, {Giglietto}, {Giordano}, {Glanzman},
  {Godfrey}, {Granot}, {Greiner}, {Grenier}, {Grove}, {Guillemot}, {Guiriec},
  {Harding}, {Hayashida}, {Hays}, {Horan}, {Hughes}, {Jackson},
  {J{\'o}hannesson}, {Johnson}, {Johnson}, {Kamae}, {Katagiri}, {Kataoka},
  {Kawai}, {Kerr}, {Kippen}, {Kn{\"o}dlseder}, {Kocevski}, {Kuss}, {Lande},
  {Latronico}, {Lemoine-Goumard}, {Longo}, {Loparco}, {Lott}, {Lovellette},
  {Lubrano}, {Makeev}, {Mazziotta}, {McEnery}, {McGlynn}, {Meegan},
  {M{\'e}sz{\'a}ros}, {Meurer}, {Michelson}, {Mitthumsiri}, {Mizuno}, {Monte},
  {Monzani}, {Moretti}, {Morselli}, {Moskalenko}, {Murgia}, {Nolan}, {Norris},
  {Nuss}, {Ohno}, {Ohsugi}, {Omodei}, {Orlando}, {Ormes}, {Paciesas},
  {Paneque}, {Panetta}, {Parent}, {Pelassa}, {Pepe}, {Pesce-Rollins}, {Piron},
  {Porter}, {Preece}, {Rain{\`o}}, {Rando}, {Razzano}, {Reimer}, {Reimer},
  {Reposeur}, {Ritz}, {Rochester}, {Rodriguez}, {Roth}, {Ryde}, {Sadrozinski},
  {Sander}, {Saz Parkinson}, {Scargle}, {Schalk}, {Sgr{\`o}}, {Siskind},
  {Smith}, {Spandre}, {Spinelli}, {Stamatikos}, {Starck}, {Stecker},
  {Strickman}, {Suson}, {Tajima}, {Takahashi}, {Tanaka}, {Thayer}, {Thayer},
  {Thompson}, {Tibaldo}, {Toma}, {Torres}, {Tosti}, {Tramacere}, {Uchiyama},
  {Uehara}, {Usher}, {van der Horst}, {Vasileiou}, {Vilchez}, {Vitale}, {von
  Kienlin}, {Waite}, {Wang}, {Winer}, {Wood}, {Wu}, {Yamazaki}, {Ylinen}, \&
  {Ziegler}}]{2009arXiv0910.1629D}
{De Pasquale}, M., {Schady}, P., {Kuin}, N.~P.~M., {et~al.} 2010, \apjl, 709,
  L146

\bibitem[{{Eichler} {et~al.}(1989){Eichler}, {Livio}, {Piran}, \&
  {Schramm}}]{1989Natur.340..126E}
{Eichler}, D., {Livio}, M., {Piran}, T., \& {Schramm}, D.~N. 1989, \nat, 340,
  126

\bibitem[{{Evans} {et~al.}(2009){Evans}, {Beardmore}, {Page}, {Osborne},
  {O'Brien}, {Willingale}, {Starling}, {Burrows}, {Godet}, {Vetere}, {Racusin},
  {Goad}, {Wiersema}, {Angelini}, {Capalbi}, {Chincarini}, {Gehrels}, {Kennea},
  {Margutti}, {Morris}, {Mountford}, {Pagani}, {Perri}, {Romano}, \&
  {Tanvir}}]{2009MNRAS.397.1177E}
{Evans}, P.~A., {Beardmore}, A.~P., {Page}, K.~L., {et~al.} 2009, \mnras, 397,
  1177

\bibitem[{{Finke} {et~al.}(2009){Finke}, {Razzaque}, \&
  {Dermer}}]{2009arXiv0905.1115F}
{Finke}, J.~D., {Razzaque}, S., \& {Dermer}, C.~D. 2009, ArXiv e-prints,
  0905.1115

\bibitem[{{Fitzpatrick}(1986)}]{fit86}
{Fitzpatrick}, E.~L. 1986, \aj, 92, 1068

\bibitem[{{Fong} {et~al.}(2010){Fong}, {Berger}, \&
  {Fox}}]{2010ApJ...708....9F}
{Fong}, W., {Berger}, E., \& {Fox}, D.~B. 2010, \apj, 708, 9

\bibitem[{{Fox} {et~al.}(2005){Fox}, {Frail}, {Price}, {Kulkarni}, {Berger},
  {Piran}, {Soderberg}, {Cenko}, {Cameron}, {Gal-Yam}, {Kasliwal}, {Moon},
  {Harrison}, {Nakar}, {Schmidt}, {Penprase}, {Chevalier}, {Kumar}, {Roth},
  {Watson}, {Lee}, {Shectman}, {Phillips}, {Roth}, {McCarthy}, {Rauch},
  {Cowie}, {Peterson}, {Rich}, {Kawai}, {Aoki}, {Kosugi}, {Totani}, {Park},
  {MacFadyen}, \& {Hurley}}]{2005Natur.437..845F}
{Fox}, D.~B., {Frail}, D.~A., {Price}, P.~A., {et~al.} 2005, \nat, 437, 845

\bibitem[{{Frail} {et~al.}(2009){Frail}, {Chandra}, \&
  {Cenko}}]{2009GCN..9060....1F}
{Frail}, D.~A., {Chandra}, P., \& {Cenko}, B. 2009, GRB Coordinates Network,
  9060

\bibitem[{{Frail} {et~al.}(2001){Frail}, {Kulkarni}, {Sari}, {Djorgovski},
  {Bloom}, {Galama}, {Reichart}, {Berger}, {Harrison}, {Price}, {Yost},
  {Diercks}, {Goodrich}, \& {Chaffee}}]{2001ApJ...562L..55F}
{Frail}, D.~A., {Kulkarni}, S.~R., {Sari}, R., {et~al.} 2001, \apjl, 562, L55

\bibitem[{{Franceschini} {et~al.}(2008){Franceschini}, {Rodighiero}, \&
  {Vaccari}}]{2008A&A...487..837F}
{Franceschini}, A., {Rodighiero}, G., \& {Vaccari}, M. 2008, \aap, 487, 837

\bibitem[{{Fruchter} {et~al.}(2006)}]{2006Natur.441..463F}
{Fruchter}, A.~S. {et~al.} 2006, \nat, 441, 463

\bibitem[{{Fryer} {et~al.}(1999){Fryer}, {Woosley}, \&
  {Hartmann}}]{1999ApJ...526..152F}
{Fryer}, C.~L., {Woosley}, S.~E., \& {Hartmann}, D.~H. 1999, \apj, 526, 152

\bibitem[{{Fynbo} {et~al.}(2009){Fynbo}, {Jakobsson}, {Prochaska}, {Malesani},
  {Ledoux}, {de Ugarte Postigo}, {Nardini}, {Vreeswijk}, {Wiersema}, {Hjorth},
  {Sollerman}, {Chen}, {Th{\"o}ne}, {Bj{\"o}rnsson}, {Bloom}, {Castro-Tirado},
  {Christensen}, {De Cia}, {Fruchter}, {Gorosabel}, {Graham}, {Jaunsen},
  {Jensen}, {Kann}, {Kouveliotou}, {Levan}, {Maund}, {Masetti},
  {Milvang-Jensen}, {Palazzi}, {Perley}, {Pian}, {Rol}, {Schady}, {Starling},
  {Tanvir}, {Watson}, {Xu}, {Augusteijn}, {Grundahl}, {Telting}, \&
  {Quirion}}]{2009arXiv0907.3449F}
{Fynbo}, J.~P.~U., {Jakobsson}, P., {Prochaska}, J.~X., {et~al.} 2009, \apjs,
  185, 526

\bibitem[{{Galama} {et~al.}(1998){Galama}, {Vreeswijk}, {van Paradijs},
  {Kouveliotou}, {Augusteijn}, {B{\"o}hnhardt}, {Brewer}, {Doublier},
  {Gonzalez}, {Leibundgut}, {Lidman}, {Hainaut}, {Patat}, {Heise}, {in't Zand},
  {Hurley}, {Groot}, {Strom}, {Mazzali}, {Iwamoto}, {Nomoto}, {Umeda},
  {Nakamura}, {Young}, {Suzuki}, {Shigeyama}, {Koshut}, {Kippen}, {Robinson},
  {de Wildt}, {Wijers}, {Tanvir}, {Greiner}, {Pian}, {Palazzi}, {Frontera},
  {Masetti}, {Nicastro}, {Feroci}, {Costa}, {Piro}, {Peterson}, {Tinney},
  {Boyle}, {Cannon}, {Stathakis}, {Sadler}, {Begam}, \&
  {Ianna}}]{1998Natur.395..670G}
{Galama}, T.~J., {Vreeswijk}, P.~M., {van Paradijs}, J., {et~al.} 1998, \nat,
  395, 670

\bibitem[{{Gehrels} {et~al.}(2008){Gehrels}, {Barthelmy}, {Burrows},
  {Cannizzo}, {Chincarini}, {Fenimore}, {Kouveliotou}, {O'Brien}, {Palmer},
  {Racusin}, {Roming}, {Sakamoto}, {Tueller}, {Wijers}, \&
  {Zhang}}]{2008ApJ...689.1161G}
{Gehrels}, N., {Barthelmy}, S.~D., {Burrows}, D.~N., {et~al.} 2008, \apj, 689,
  1161

\bibitem[{{Gehrels} {et~al.}(2004){Gehrels}, {Chincarini}, {Giommi}, {Mason},
  {Nousek}, {Wells}, {White}, {Barthelmy}, {Burrows}, {Cominsky}, {Hurley},
  {Marshall}, {M{\'e}sz{\'a}ros}, {Roming}, {Angelini}, {Barbier}, {Belloni},
  {Campana}, {Caraveo}, {Chester}, {Citterio}, {Cline}, {Cropper}, {Cummings},
  {Dean}, {Feigelson}, {Fenimore}, {Frail}, {Fruchter}, {Garmire}, {Gendreau},
  {Ghisellini}, {Greiner}, {Hill}, {Hunsberger}, {Krimm}, {Kulkarni}, {Kumar},
  {Lebrun}, {Lloyd-Ronning}, {Markwardt}, {Mattson}, {Mushotzky}, {Norris},
  {Osborne}, {Paczynski}, {Palmer}, {Park}, {Parsons}, {Paul}, {Rees},
  {Reynolds}, {Rhoads}, {Sasseen}, {Schaefer}, {Short}, {Smale}, {Smith},
  {Stella}, {Tagliaferri}, {Takahashi}, {Tashiro}, {Townsley}, {Tueller},
  {Turner}, {Vietri}, {Voges}, {Ward}, {Willingale}, {Zerbi}, \&
  {Zhang}}]{2004ApJ...611.1005G}
{Gehrels}, N., {Chincarini}, G., {Giommi}, P., {et~al.} 2004, \apj, 611, 1005

\bibitem[{{Gehrels} {et~al.}(2005)}]{2005Natur.437..851G}
{Gehrels}, N. {et~al.} 2005, \nat, 437, 851

\bibitem[{{Ghirlanda} {et~al.}(2004){Ghirlanda}, {Ghisellini}, \&
  {Lazzati}}]{2004ApJ...616..331G}
{Ghirlanda}, G., {Ghisellini}, G., \& {Lazzati}, D. 2004, \apj, 616, 331

\bibitem[{{Ghirlanda} {et~al.}(2007){Ghirlanda}, {Nava}, {Ghisellini}, \&
  {Firmani}}]{2007A&A...466..127G}
{Ghirlanda}, G., {Nava}, L., {Ghisellini}, G., \& {Firmani}, C. 2007, \aap,
  466, 127

\bibitem[{{Ghisellini} {et~al.}(2009){Ghisellini}, {Ghirlanda}, \&
  {Nava}}]{2009arXiv0910.2459G}
{Ghisellini}, G., {Ghirlanda}, G., \& {Nava}, L. 2009, ArXiv e-prints,
  0910.2459

\bibitem[{{Giavalisco} {et~al.}(2004){Giavalisco}, {Ferguson}, {Koekemoer},
  {Dickinson}, {Alexander}, {Bauer}, {Bergeron}, {Biagetti}, {Brandt},
  {Casertano}, {Cesarsky}, {Chatzichristou}, {Conselice}, {Cristiani}, {Da
  Costa}, {Dahlen}, {de Mello}, {Eisenhardt}, {Erben}, {Fall}, {Fassnacht},
  {Fosbury}, {Fruchter}, {Gardner}, {Grogin}, {Hook}, {Hornschemeier}, {Idzi},
  {Jogee}, {Kretchmer}, {Laidler}, {Lee}, {Livio}, {Lucas}, {Madau},
  {Mobasher}, {Moustakas}, {Nonino}, {Padovani}, {Papovich}, {Park},
  {Ravindranath}, {Renzini}, {Richardson}, {Riess}, {Rosati}, {Schirmer},
  {Schreier}, {Somerville}, {Spinrad}, {Stern}, {Stiavelli}, {Strolger},
  {Urry}, {Vandame}, {Williams}, \& {Wolf}}]{2004ApJ...600L..93G}
{Giavalisco}, M., {Ferguson}, H.~C., {Koekemoer}, A.~M., {et~al.} 2004, \apjl,
  600, L93

\bibitem[{{Giuliani} {et~al.}(2010){Giuliani}, {Fuschino}, {Vianello},
  {Marisaldi}, {Mereghetti}, {Tavani}, {Cutini}, {Barbiellini}, {Longo},
  {Moretti}, {Feroci}, {Del Monte}, {Argan}, {Bulgarelli}, {Caraveo},
  {Cattaneo}, {Chen}, {Contessi}, {D'Ammando}, {Costa}, {De Paris}, {Di Cocco},
  {Donnarumma}, {Evangelista}, {Ferrari}, {Fiorini}, {Galli}, {Gianotti},
  {Labanti}, {Lapshov}, {Lazzarotto}, {Lipari}, {Morselli}, {Pacciani},
  {Pellizzoni}, {Perotti}, {Piano}, {Picozza}, {Pilia}, {Pucella}, {Prest},
  {Rapisarda}, {Rappoldi}, {Rubini}, {Sabatini}, {Scalise}, {Striani},
  {Soffitta}, {Trifoglio}, {Trois}, {Vallazza}, {Vercellone}, {Vittorini},
  {Zambra}, {Zanello}, {Pittori}, {Verrecchia}, {Santolamazza}, {Giommi},
  {Colafrancesco}, {Antonelli}, \& {Salotti}}]{2009arXiv0908.1908G}
{Giuliani}, A., {Fuschino}, F., {Vianello}, G., {et~al.} 2010, \apjl, 708, L84

\bibitem[{{Giuliani} {et~al.}(2008){Giuliani}, {Mereghetti}, {Fornari}, {Del
  Monte}, {Feroci}, {Marisaldi}, {Esposito}, {Perotti}, {Tavani}, {Argan},
  {Barbiellini}, {Boffelli}, {Bulgarelli}, {Caraveo}, {Cattaneo}, {Chen},
  {Costa}, {D'Ammando}, {di Cocco}, {Donnarumma}, {Evangelista}, {Fiorini},
  {Fuschino}, {Galli}, {Gianotti}, {Labanti}, {Lapshov}, {Lazzarotto},
  {Lipari}, {Longo}, {Morselli}, {Pacciani}, {Pellizzoni}, {Piano}, {Picozza},
  {Prest}, {Pucella}, {Rapisarda}, {Rappoldi}, {Soffitta}, {Trifoglio},
  {Trois}, {Vallazza}, {Vercellone}, {Zanello}, {Salotti}, {Cutini}, {Pittori},
  {Preger}, {Santolamazza}, {Verrecchia}, {Gehrels}, {Page}, {Burrows},
  {Rossi}, {Hurley}, {Mitrofanov}, \& {Boynton}}]{2008A&A...491L..25G}
{Giuliani}, A., {Mereghetti}, S., {Fornari}, F., {et~al.} 2008, \aap, 491, L25

\bibitem[{{Golenetskii} {et~al.}(2009){Golenetskii}, {Aptekar}, {Mazets},
  {Pal'shin}, {Frederiks}, {et~al.}}]{2009GCN..9344....1P}
{Golenetskii}, S., {Aptekar}, R., {Mazets}, S., {et~al.} 2009, GRB Coordinates
  Network, 9344

\bibitem[{{Gonz{\'a}lez} {et~al.}(2003){Gonz{\'a}lez}, {Dingus}, {Kaneko},
  {Preece}, {Dermer}, \& {Briggs}}]{2003Natur.424..749G}
{Gonz{\'a}lez}, M.~M., {Dingus}, B.~L., {Kaneko}, Y., {et~al.} 2003, \nat, 424,
  749

\bibitem[{{Gorosabel} {et~al.}(2006){Gorosabel}, {Castro-Tirado}, {Guziy}, {de
  Ugarte Postigo}, {Reverte}, {Antonelli}, {Covino}, {Malesani},
  {Mart{\'{\i}}n-Gord{\'o}n}, {Melandri}, {Jel{\'{\i}}nek}, {Elias de La Rosa},
  {Bogdanov}, \& {Castro Cer{\'o}n}}]{2006A&A...450...87G}
{Gorosabel}, J., {Castro-Tirado}, A.~J., {Guziy}, S., {et~al.} 2006, \aap, 450,
  87

\bibitem[{{Graham} {et~al.}(2009){Graham}, {Fruchter}, {Levan}, {Melandri},
  {Kewley}, {Levesque}, {Nysewander}, {Tanvir}, {Dahlen}, {Bersier},
  {Wiersema}, {Bonfield}, \& {Martinez-Sansigre}}]{2009ApJ...698.1620G}
{Graham}, J.~F., {Fruchter}, A.~S., {Levan}, A.~J., {et~al.} 2009, \apj, 698,
  1620

\bibitem[{{Greiner} {et~al.}(2008){Greiner}, {Bornemann}, {Clemens}, {Deuter},
  {Hasinger}, {Honsberg}, {Huber}, {Huber}, {Krauss}, {Kr{\"u}hler},
  {K{\"u}pc{\"u} Yolda{\c s}}, {Mayer-Hasselwander}, {Mican}, {Primak},
  {Schrey}, {Steiner}, {Szokoly}, {Th{\"o}ne}, {Yolda{\c s}}, {Klose}, {Laux},
  \& {Winkler}}]{2008PASP..120..405G}
{Greiner}, J., {Bornemann}, W., {Clemens}, C., {et~al.} 2008, \pasp, 120, 405

\bibitem[{{Greiner} {et~al.}(2009{\natexlab{a}}){Greiner}, {Clemens},
  {Kr{\"u}hler}, {von Kienlin}, {Rau}, {Sari}, {Fox}, {Kawai}, {Afonso},
  {Ajello}, {Berger}, {Cenko}, {Cucchiara}, {Filgas}, {Klose}, {K{\"u}pc{\"u}
  Yolda{\c s}}, {Lichti}, {L{\"o}w}, {McBreen}, {Nagayama}, {Rossi}, {Sato},
  {Szokoly}, {Yolda{\c s}}, \& {Zhang}}]{2009A&A...498...89G}
{Greiner}, J., {Clemens}, C., {Kr{\"u}hler}, T., {et~al.} 2009{\natexlab{a}},
  \aap, 498, 89

\bibitem[{{Greiner} {et~al.}(2003){Greiner}, {Klose}, {Reinsch}, {Martin
  Schmid}, {Sari}, {Hartmann}, {Kouveliotou}, {Rau}, {Palazzi}, {Straubmeier},
  {Stecklum}, {Zharikov}, {Tovmassian}, {B{\"a}rnbantner}, {Ries}, {Jehin},
  {Henden}, {Kaas}, {Grav}, {Hjorth}, {Pedersen}, {Wijers}, {Kaufer}, {Park},
  {Williams}, \& {Reimer}}]{2003Natur.426..157G}
{Greiner}, J., {Klose}, S., {Reinsch}, K., {et~al.} 2003, \nat, 426, 157

\bibitem[{{Greiner} {et~al.}(2009{\natexlab{b}}){Greiner}, {Kr{\"u}hler},
  {McBreen}, {Ajello}, {Giannios}, {Schwarz}, {Savaglio}, {Yolda{\c s}},
  {Clemens}, {Stefanescu}, {Sala}, {Bertoldi}, {Szokoly}, \&
  {Klose}}]{2009ApJ...693.1912G}
{Greiner}, J., {Kr{\"u}hler}, T., {McBreen}, S., {et~al.} 2009{\natexlab{b}},
  \apj, 693, 1912

\bibitem[{{Grupe} \& {Hoversten}(2009)}]{2009GCN..9341....1G}
{Grupe}, D. \& {Hoversten}, E. 2009, GRB Coordinates Network, 9341

\bibitem[{{Guidorzi} {et~al.}(2009){Guidorzi}, {Tanvir}, {Cano}, {Steele},
  {Bersier}, {et~al.}}]{Guidorzi_090902B}
{Guidorzi}, C., {Tanvir}, N.~R., {Cano}, Z., {et~al.} 2009, GRB Coordinates
  Network, 9875

\bibitem[{{Guiriec} {et~al.}(2009){Guiriec}, {Connaughton}, \&
  {Briggs}}]{2009GCN..9336....1G}
{Guiriec}, S., {Connaughton}, V., \& {Briggs}, M. 2009, GRB Coordinates
  Network, 9336

\bibitem[{{Hamuy} {et~al.}(1992){Hamuy}, {Walker}, {Suntzeff}, {Gigoux},
  {Heathcote}, \& {Phillips}}]{1992PASP..104..533H}
{Hamuy}, M., {Walker}, A.~R., {Suntzeff}, N.~B., {et~al.} 1992, \pasp, 104, 533

\bibitem[{{Hanlon} {et~al.}(1994){Hanlon}, {Bennett}, {Collmar}, {Connors},
  {Diehl}, {van Dijk}, {Greiner}, {den Herder}, {Hermsen}, {Kippen}, {Kuiper},
  {McConnell}, {Ryan}, {Schoenfelder}, {Steinle}, {Strong}, {Varendorff},
  {Williams}, \& {Winkler}}]{1994A&A...285..161H}
{Hanlon}, L.~O., {Bennett}, K., {Collmar}, W., {et~al.} 1994, \aap, 285, 161

\bibitem[{{Harrison} {et~al.}(2009){Harrison}, {Cenko}, {Frail}, {Chandra}, \&
  {Kulkarni}}]{2009GCN..9043....1H}
{Harrison}, F., {Cenko}, B., {Frail}, D.~A., {Chandra}, P., \& {Kulkarni}, S.
  2009, GRB Coordinates Network, 9043

\bibitem[{{Hjorth} {et~al.}(2003)}]{2003Natur.423..847H}
{Hjorth}, J. {et~al.} 2003, \nat, 423, 847

\bibitem[{{Hjorth} {et~al.}(2005)}]{2005Natur.437..859H}
{Hjorth}, J. {et~al.} 2005, \nat, 437, 859

\bibitem[{{Hoversten} {et~al.}(2009){Hoversten}, {Barthelmy}, {Burrows},
  {Chester}, {Grupe}, {Kennea}, {Krimm}, {Kuin}, {Palmer}, \&
  {Ukwatta}}]{2009GCN..9331....1H}
{Hoversten}, E.~A., {Barthelmy}, S.~D., {Burrows}, D.~N., {et~al.} 2009, GRB
  Coordinates Network, 9331

\bibitem[{{Huang} \& {Gu}(2009)}]{2009MNRAS.398.1651H}
{Huang}, S. \& {Gu}, Q. 2009, \mnras, 398, 1651

\bibitem[{{Hurley}(1992)}]{1992AIPC..265....3H}
{Hurley}, K. 1992, in AIP, ed. {W.S. Paciesas} \& {G.J. Fishman}, Vol. 265, 3

\bibitem[{{Hurley} {et~al.}(1995){Hurley}, {Dingus}, {Mukherjee}, {Sreekumar},
  {Kouveliotou}, {Meegan}, {Fishman}, {Band}, {Ford}, {Bertsch}, {Cline},
  {Fichtel}, {Hartman}, {Hunter}, {Thompson}, {Kanbach}, {Mayer-Hasselwander},
  {von Montigny}, {Sommer}, {Lin}, {Nolan}, {Michelson}, {Kniffen}, {Mattox},
  {Schneid}, {Boer}, \& {Niel}}]{1995Natur.374...94H}
{Hurley}, K., {Dingus}, B.~L., {Mukherjee}, R., {et~al.} 1995, \nat, 374, 94

\bibitem[{{Hurley} {et~al.}(2009)}]{2008IPN090323}
{Hurley}, K. {et~al.} 2009, GRB Coordinates Network, 9023

\bibitem[{{Kaneko} {et~al.}(2008){Kaneko}, {Gonz{\'a}lez}, {Preece}, {Dingus},
  \& {Briggs}}]{2008ApJ...677.1168K}
{Kaneko}, Y., {Gonz{\'a}lez}, M.~M., {Preece}, R.~D., {Dingus}, B.~L., \&
  {Briggs}, M.~S. 2008, \apj, 677, 1168

\bibitem[{{Kann} {et~al.}(2006){Kann}, {Klose}, \& {Zeh}}]{2006ApJ...641..993K}
{Kann}, D.~A., {Klose}, S., \& {Zeh}, A. 2006, \apj, 641, 993

\bibitem[{{Kann} {et~al.}(2009{\natexlab{a}}){Kann}, {Klose}, {Zhang},
  {Malesani}, {Nakar}, {Wilson}, {Butler}, {Antonelli}, {Chincarini}, {Cobb},
  {Covino}, {D'Avanzo}, {D'Elia}, {Della Valle}, {Ferrero}, {Fugazza},
  {Gorosabel}, {Israel}, {Mannucci}, {Piranomonte}, {Schulze}, {Stella},
  {Tagliaferri}, \& {Wiersema}}]{2007arXiv0712.2186K}
{Kann}, D.~A., {Klose}, S., {Zhang}, B., {et~al.} 2009{\natexlab{a}}, ArXiv
  e-prints, 0712.2186v2

\bibitem[{{Kann} {et~al.}(2008){Kann}, {Klose}, {Zhang}, {Wilson}, {Butler},
  {Malesani}, {Nakar}, {Antonelli}, {Chincarini}, {Cobb}, {Covino}, {D'Avanzo},
  {D'Elia}, {Della Valle}, {Ferrero}, {Fugazza}, {Gorosabel}, {Israel},
  {Mannucci}, {Piranomonte}, {Schulze}, {Stella}, {Tagliaferri}, \&
  {Wiersema}}]{2008arXiv0804.1959K}
{Kann}, D.~A., {Klose}, S., {Zhang}, B., {et~al.} 2008, ArXiv e-prints,
  0804.1959

\bibitem[{{Kann} {et~al.}(2009{\natexlab{b}}){Kann}, {Laux}, {Ludwig}, \&
  {Stecklum}}]{2009GCN..9063....1K}
{Kann}, D.~A., {Laux}, U., {Ludwig}, F., \& {Stecklum}, S. 2009{\natexlab{b}},
  GRB Coordinates Network, 9063

\bibitem[{{Kann} {et~al.}(2009{\natexlab{c}}){Kann}, {Laux}, \&
  {Stecklum}}]{2009GCN..9033....1K}
{Kann}, D.~A., {Laux}, U., \& {Stecklum}, S. 2009{\natexlab{c}}, GRB
  Coordinates Network, 9033

\bibitem[{{Kann} {et~al.}(2009{\natexlab{d}}){Kann}, {Laux}, \&
  {Stecklum}}]{2009GCN..9041....1K}
{Kann}, D.~A., {Laux}, U., \& {Stecklum}, S. 2009{\natexlab{d}}, GRB
  Coordinates Network, 9041

\bibitem[{{Kashlinsky} {et~al.}(2005){Kashlinsky}, {Arendt}, {Mather}, \&
  {Moseley}}]{2005Natur.438...45K}
{Kashlinsky}, A., {Arendt}, R.~G., {Mather}, J., \& {Moseley}, S.~H. 2005,
  \nat, 438, 45

\bibitem[{{Kennea}(2009)}]{2009GCN..9045....1K}
{Kennea}, J. 2009, GRB Coordinates Network, 9045

\bibitem[{{Kennea} {et~al.}(2009{\natexlab{a}}){Kennea}, {Evans}, \&
  {Goad}}]{2009GCN..9024....1K}
{Kennea}, J., {Evans}, P., \& {Goad}, M. 2009{\natexlab{a}}, GRB Coordinates
  Network, 9024

\bibitem[{{Kennea} {et~al.}(2009{\natexlab{b}}){Kennea}, {Evans}, \&
  {Goad}}]{2009GCN..9046....1K}
{Kennea}, J., {Evans}, P., \& {Goad}, M. 2009{\natexlab{b}}, GRB Coordinates
  Network, 9046

\bibitem[{{Kennea} \& {Stratta}(2009)}]{XRT_090902B}
{Kennea}, J. \& {Stratta}, G. 2009, GRB Coordinates Network, 9868

\bibitem[{{Kneiske} {et~al.}(2004){Kneiske}, {Bretz}, {Mannheim}, \&
  {Hartmann}}]{2004A&A...413..807K}
{Kneiske}, T.~M., {Bretz}, T., {Mannheim}, K., \& {Hartmann}, D.~H. 2004, \aap,
  413, 807

\bibitem[{{Kong} {et~al.}(2010){Kong}, {Wong}, {Huang}, \&
  {Cheng}}]{2010MNRAS.402..409K}
{Kong}, S.~W., {Wong}, A.~Y.~L., {Huang}, Y.~F., \& {Cheng}, K.~S. 2010,
  \mnras, 402, 409

\bibitem[{{Kouveliotou} {et~al.}(1993)}]{1993ApJ...413L.101K}
{Kouveliotou}, C. {et~al.} 1993, \apjl, 413, L101

\bibitem[{{Kr{\"u}hler} {et~al.}(2009{\natexlab{a}}){Kr{\"u}hler}, {Greiner},
  {Afonso}, {Burlon}, {Clemens}, {Filgas}, {Kann}, {Klose}, {K{\"u}pc{\"u}
  Yolda{\c s}}, {McBreen}, {Olivares}, {Rau}, {Rossi}, {Schulze}, {Szokoly},
  {Updike}, \& {Yolda{\c s}}}]{2009A&A...508..593K}
{Kr{\"u}hler}, T., {Greiner}, J., {Afonso}, P., {et~al.} 2009{\natexlab{a}},
  \aap, 508, 593

\bibitem[{{Kr{\"u}hler} {et~al.}(2009{\natexlab{b}}){Kr{\"u}hler}, {Greiner},
  {McBreen}, {Klose}, {Rossi}, {Afonso}, {Clemens}, {Filgas}, {Yolda{\c s}},
  {Szokoly}, \& {Yolda{\c s}}}]{2009ApJ...697..758K}
{Kr{\"u}hler}, T., {Greiner}, J., {McBreen}, S., {et~al.} 2009{\natexlab{b}},
  \apj, 697, 758

\bibitem[{{Kr{\"u}hler} {et~al.}(2008){Kr{\"u}hler}, {K{\"u}pc{\"u} Yolda{\c
  s}}, {Greiner}, {Clemens}, {McBreen}, {Primak}, {Savaglio}, {Yolda{\c s}},
  {Szokoly}, \& {Klose}}]{2008ApJ...685..376K}
{Kr{\"u}hler}, T., {K{\"u}pc{\"u} Yolda{\c s}}, A., {Greiner}, J., {et~al.}
  2008, \apj, 685, 376

\bibitem[{{Kuin} \& {Hoversten}(2009)}]{2009GCN9342......1K}
{Kuin}, N.~P.~M. \& {Hoversten}, E.~A. 2009, GRB Coordinates Network, 9342

\bibitem[{{Kuin} {et~al.}(2009){Kuin}, {Oates}, {de Pasquale}, {Hoversten}, \&
  {Marshall}}]{2009GCN..9351....1K}
{Kuin}, N.~P.~M., {Oates}, S., {de Pasquale}, M., {Hoversten}, E.~A., \&
  {Marshall}, F. 2009, GRB Coordinates Network, 9351

\bibitem[{{Kumar} \& {Barniol Duran}(2009{\natexlab{a}})}]{2009arXiv0910.5726K}
{Kumar}, P. \& {Barniol Duran}, R. 2009{\natexlab{a}}, ArXiv e-prints,
  0910.5726

\bibitem[{{Kumar} \& {Barniol Duran}(2009{\natexlab{b}})}]{2009arXiv0905.2417K}
{Kumar}, P. \& {Barniol Duran}, R. 2009{\natexlab{b}}, \mnras, 400, L75

\bibitem[{{Lazzati} {et~al.}(2002){Lazzati}, {Rossi}, {Covino}, {Ghisellini},
  \& {Malesani}}]{2002A&A...396L...5L}
{Lazzati}, D., {Rossi}, E., {Covino}, S., {Ghisellini}, G., \& {Malesani}, D.
  2002, \aap, 396, L5

\bibitem[{{Lee} \& {Ramirez-Ruiz}(2007)}]{2007NJPh....9...17L}
{Lee}, W.~H. \& {Ramirez-Ruiz}, E. 2007, New Journal of Physics, 9, 17

\bibitem[{{Levan} {et~al.}(2007){Levan}, {Jakobsson}, {Hurkett}, {Tanvir},
  {Gorosabel}, {Vreeswijk}, {Rol}, {Chapman}, {Gehrels}, {O'Brien}, {Osborne},
  {Priddey}, {Kouveliotou}, {Starling}, {vanden Berk}, \&
  {Wiersema}}]{2007MNRAS.378.1439L}
{Levan}, A.~J., {Jakobsson}, P., {Hurkett}, C., {et~al.} 2007, \mnras, 378,
  1439

\bibitem[{{Levan} {et~al.}(2006{\natexlab{a}}){Levan}, {Tanvir}, {Fruchter},
  {Rol}, {Fynbo}, {Hjorth}, {Williams}, {Bergeron}, {Bersier}, {Bremer},
  {Grav}, {Jakobsson}, {Nilsson}, {Olszewski}, {Priddey}, {Rafferty}, \&
  {Rhoads}}]{2006ApJ...648L...9L}
{Levan}, A.~J., {Tanvir}, N.~R., {Fruchter}, A.~S., {et~al.}
  2006{\natexlab{a}}, \apjl, 648, L9

\bibitem[{{Levan} {et~al.}(2006{\natexlab{b}}){Levan}, {Wynn}, {Chapman},
  {Davies}, {King}, {Priddey}, \& {Tanvir}}]{2006MNRAS.368L...1L}
{Levan}, A.~J., {Wynn}, G.~A., {Chapman}, R., {et~al.} 2006{\natexlab{b}},
  \mnras, 368, L1

\bibitem[{{Levesque} {et~al.}(2010){Levesque}, {Bloom}, {Butler}, {Perley},
  {Cenko}, {Prochaska}, {Kewley}, {Bunker}, {Chen}, {Chornock}, {Filippenko},
  {Glazebrook}, {Lopez}, {Masiero}, {Modjaz}, {Morgan}, \&
  {Poznanski}}]{2009arXiv0907.1661L}
{Levesque}, E.~M., {Bloom}, J.~S., {Butler}, N.~R., {et~al.} 2010, \mnras, 401,
  963

\bibitem[{{Lipkin} {et~al.}(2004){Lipkin}, {Ofek}, {Gal-Yam}, {Leibowitz},
  {Poznanski}, {Kaspi}, {Polishook}, {Kulkarni}, {Fox}, {Berger}, {Mirabal},
  {Halpern}, {Bureau}, {Fathi}, {Price}, {Peterson}, {Frebel}, {Schmidt},
  {Orosz}, {Fitzgerald}, {Bloom}, {van Dokkum}, {Bailyn}, {Buxton}, \&
  {Barsony}}]{2004ApJ...606..381L}
{Lipkin}, Y.~M., {Ofek}, E.~O., {Gal-Yam}, A., {et~al.} 2004, \apj, 606, 381

\bibitem[{{Longo} {et~al.}(2009){Longo}, {Moretti}, {Barbiellini}, {Vallazza},
  {Giuliani}, {Cutini}, {Pittori}, {Marisaldi}, {Bulgarelli}, {Gianotti},
  {Trifoglio}, {di Cocco}, {Labanti}, {Fuschino}, {Galli}, {Chen},
  {Mereghetti}, {Perotti}, {Caraveo}, {Evangelista}, {Del}, {Feroci},
  {Donnarumma}, {Pacciani}, {Soffitta}, {Costa}, {Lazzarotto}, {Lapshov},
  {Rapisarda}, {Pellizzoni}, {Pilia}, {Vercellone}, {Tavani}, {Pucella},
  {D'Ammando}, {Vittorini}, {Argan}, {Trois}, {Piano}, {Sabatini}, {Picozza},
  {Morselli}, {Prest}, {Lipari}, {Zanello}, {Rappoldi}, {Cattaneo}, {Giommi},
  {Santolamazza}, {Verrecchia}, \& {Salotti}}]{2009GCN..9343....1L}
{Longo}, F., {Moretti}, E., {Barbiellini}, G., {et~al.} 2009, GRB Coordinates
  Network, 9343

\bibitem[{{Malesani} {et~al.}(2009){Malesani}, {Goldoni}, {Fynbo},
  {et~al.}}]{Xshooter_090926}
{Malesani}, D., {Goldoni}, P., {Fynbo}, J., {et~al.} 2009, GRB Coordinates
  Network, 9942

\bibitem[{{Malesani} {et~al.}(2004)}]{2004ApJ...609L...5M}
{Malesani}, D. {et~al.} 2004, \apjl, 609, L5

\bibitem[{{Marshall} \& {Hoversten}(2009)}]{2009GCN..9332....1M}
{Marshall}, F.~E. \& {Hoversten}, E.~A. 2009, GRB Coordinates Network, 9332

\bibitem[{{Mattingly}(2005)}]{2005LRR.....8....5M}
{Mattingly}, D. 2005, Living Reviews in Relativity, 8, 5

\bibitem[{{Mazets} {et~al.}(1981)}]{1981Ap&SS..80..119M}
{Mazets}, E.~P. {et~al.} 1981, \apss, 80, 119

\bibitem[{{McEnery} {et~al.}(2009{\natexlab{a}}){McEnery}, {Chiang}, \&
  {Hanabata}}]{McEnery09_GCN9985}
{McEnery}, J., {Chiang}, J., \& {Hanabata}, Y. 2009{\natexlab{a}}, {GCN
  Circular} 9985

\bibitem[{{McEnery} {et~al.}(2010){McEnery}, {Chiang}, {Omodei}, \&
  {Nakamori}}]{McEnery10_GCN10333}
{McEnery}, J., {Chiang}, J., {Omodei}, N., \& {Nakamori}, T. 2010, {GCN
  Circular} 10333

\bibitem[{{McEnery} {et~al.}(2009{\natexlab{b}}){McEnery}, {Cutini}, {Ohno},
  {Koerding}, \& {Connaughton}}]{2009GCN..9044....1M}
{McEnery}, J., {Cutini}, S., {Ohno}, M., {Koerding}, E., \& {Connaughton}, V.
  2009{\natexlab{b}}, GRB Coordinates Network, 9044

\bibitem[{{Meegan} {et~al.}(2009){Meegan}, {Lichti}, {Bhat}, {Bissaldi},
  {Briggs}, {Connaughton}, {Diehl}, {Fishman}, {Greiner}, {Hoover}, {van der
  Horst}, {von Kienlin}, {Kippen}, {Kouveliotou}, {McBreen}, {Paciesas},
  {Preece}, {Steinle}, {Wallace}, {Wilson}, \&
  {Wilson-Hodge}}]{2009arXiv0908.0450M}
{Meegan}, C., {Lichti}, G., {Bhat}, P.~N., {et~al.} 2009, \apj, 702, 791

\bibitem[{{M{\'e}sz{\'a}ros}(2006)}]{2006RPPh...69.2259M}
{M{\'e}sz{\'a}ros}, P. 2006, Reports on Progress in Physics, 69, 2259

\bibitem[{{Metcalfe} {et~al.}(2003){Metcalfe}, {Kneib}, {McBreen}, {Altieri},
  {Biviano}, {Delaney}, {Elbaz}, {Kessler}, {Leech}, {Okumura}, {Ott},
  {Perez-Martinez}, {Sanchez-Fernandez}, \& {Schulz}}]{2003A&A...407..791M}
{Metcalfe}, L., {Kneib}, J., {McBreen}, B., {et~al.} 2003, \aap, 407, 791

\bibitem[{{Michelson}(1996)}]{1996SPIE.2806...31M}
{Michelson}, P.~F. 1996, in Presented at the Society of Photo-Optical
  Instrumentation Engineers (SPIE) Conference, Vol. 2806, Society of
  Photo-Optical Instrumentation Engineers (SPIE) Conference Series, ed. B.~D.
  {Ramsey} \& T.~A. {Parnell}, 31--40

\bibitem[{{Nakar}(2007)}]{2007PhR...442..166N}
{Nakar}, E. 2007, \physrep, 442, 166

\bibitem[{{Nakar} \& {Granot}(2007)}]{2007MNRAS.380.1744N}
{Nakar}, E. \& {Granot}, J. 2007, \mnras, 380, 1744

\bibitem[{{Norris} {et~al.}(1984){Norris}, {Cline, T.~L.}, {Desai, U.~D.},
  {et~al.}}]{1984Natur.308..434N}
{Norris}, J.~P., {Cline, T.~L.}, {Desai, U.~D.}, {et~al.} 1984, \nat, 308, 434

\bibitem[{{Nousek} {et~al.}(2006){Nousek}, {Kouveliotou}, {Grupe}, {Page},
  {Granot}, {Ramirez-Ruiz}, {Patel}, {Burrows}, {Mangano}, {Barthelmy},
  {Beardmore}, {Campana}, {Capalbi}, {Chincarini}, {Cusumano}, {Falcone},
  {Gehrels}, {Giommi}, {Goad}, {Godet}, {Hurkett}, {Kennea}, {Moretti},
  {O'Brien}, {Osborne}, {Romano}, {Tagliaferri}, \&
  {Wells}}]{2006ApJ...642..389N}
{Nousek}, J.~A., {Kouveliotou}, C., {Grupe}, D., {et~al.} 2006, \apj, 642, 389

\bibitem[{{Nysewander} {et~al.}(2009){Nysewander}, {Fruchter}, \&
  {Pe'er}}]{2009ApJ...701..824N}
{Nysewander}, M., {Fruchter}, A.~S., \& {Pe'er}, A. 2009, \apj, 701, 824

\bibitem[{{Oates}(2009)}]{2009GCN..9048....1O}
{Oates}, S.~R. 2009, GRB Coordinates Network, 9048

\bibitem[{{Ohmori} {et~al.}(2009){Ohmori}, {Noda}, {Sonoda}, {Yamauchi},
  {Kono}, {Hayashi}, {Daikyuji}, {Nishioka}, {Ohno}, {Suzuki}, {Kokubun},
  {Takahashi}, {Yamaoka}, {Sugita}, {Nakagawa}, {Tamagawa}, {Hong}, {Vasquez},
  {Uehara}, {Hanabata}, {Fukazawa}, {Iwakiri}, {Tashiro}, {Terada}, {Endo},
  {Onda}, {Sugasahara}, {Urata}, {Enoto}, {Nakazawa}, \&
  {Makishima}}]{2009GCN..9355....1O}
{Ohmori}, N., {Noda}, K., {Sonoda}, E., {et~al.} 2009, GRB Coordinates Network,
  9355

\bibitem[{{Ohno} {et~al.}(2009){Ohno}, {Cutini}, {McEnery}, {Chiang}, \&
  {Koerding}}]{2009GCN..9021....1O}
{Ohno}, M., {Cutini}, S., {McEnery}, J., {Chiang}, J., \& {Koerding}, E. 2009,
  GRB Coordinates Network, 9021

\bibitem[{{Ohno} {et~al.}(2008){Ohno}, {McEnery}, \& {Pelassa}}]{LAT_090217}
{Ohno}, M., {McEnery}, J., \& {Pelassa}, V. 2008, GRB Coordinates Network, 8903

\bibitem[{{Ohno} \& {Pelassa}(2009)}]{2009GCN..9334....1O}
{Ohno}, M. \& {Pelassa}, V. 2009, GRB Coordinates Network, 9334

\bibitem[{{Olivares} {et~al.}(2009{\natexlab{a}}){Olivares}, {Afonso},
  {Greiner}, {McBreen}, {Kr\"uhler}, {Rau}, {Yoldas}, \& G.}]{Olivares_090902B}
{Olivares}, F., {Afonso}, F., {Greiner}, J., {et~al.} 2009{\natexlab{a}}, GRB
  Coordinates Network, 9874

\bibitem[{{Olivares} {et~al.}(2009{\natexlab{b}}){Olivares}, {Klose},
  {Kr\"uhler}, \& {Greiner}}]{2009GCN..9352....1O}
{Olivares}, F., {Klose}, S., {Kr\"uhler}, T., \& {Greiner}, J.
  2009{\natexlab{b}}, GRB Coordinates Network, 9352

\bibitem[{{Olofsson} {et~al.}(2009){Olofsson}, {Ergon}, {Malesani}, {Fynbo},
  {Jakobsson}, {Tanvir}, {Wiersema}, \& {Levan}}]{2009GCN..9338....1O}
{Olofsson}, G., {Ergon}, M., {Malesani}, D., {et~al.} 2009, GRB Coordinates
  Network, 9338, 1

\bibitem[{{Omodei}(2008)}]{2008GCN..8407....1O}
{Omodei}, N. 2008, GRB Coordinates Network, 8407

\bibitem[{{Omodei} {et~al.}(2009){Omodei}, {Granot}, {Meszaros}, {McEnery},
  {Piron}, {Razzaque}, {Tajima}, {Vasileiou}, \&
  {Williams}}]{2009GCN..9350....1O}
{Omodei}, N., {Granot}, J., {Meszaros}, P., {et~al.} 2009, GRB Coordinates
  Network, 9350

\bibitem[{{Palma} {et~al.}(2009){Palma}, {Bari)}, {Omodei}, {McEnery}, \&
  {Vasileiou}}]{Palma09_GCN10163}
{Palma}, F.~d., {Bari)}, I., {Omodei}, N., {McEnery}, J., \& {Vasileiou}, V.
  2009, {GCN Circular} 10163

\bibitem[{{Panaitescu}(2005)}]{2005MNRAS.362..921P}
{Panaitescu}, A. 2005, \mnras, 362, 921

\bibitem[{{Panaitescu} {et~al.}(2006){Panaitescu}, {M{\'e}sz{\'a}ros},
  {Gehrels}, {Burrows}, \& {Nousek}}]{2006MNRAS.366.1357P}
{Panaitescu}, A., {M{\'e}sz{\'a}ros}, P., {Gehrels}, N., {Burrows}, D., \&
  {Nousek}, J. 2006, \mnras, 366, 1357

\bibitem[{{Pandey} {et~al.}(2009){Pandey}, {Zheng}, {Yuan}, \&
  {Akerlof}}]{Pandey_090902B}
{Pandey}, S.~B., {Zheng}, W., {Yuan}, F., \& {Akerlof}, C. 2009, GRB
  Coordinates Network, 9878

\bibitem[{{Perley} {et~al.}(2009{\natexlab{a}}){Perley}, {Kleiser}, \&
  {Rex}}]{Perley_090902B}
{Perley}, D., {Kleiser}, I.~K.~W., \& {Rex}, J.~M. 2009{\natexlab{a}}, GRB
  Coordinates Network, 9870

\bibitem[{{Perley} {et~al.}(2008){Perley}, {Bloom}, {Butler}, {Pollack},
  {Holtzman}, {Blake}, {Kocevski}, {Vestrand}, {Li}, {Foley}, {Bellm}, {Chen},
  {Prochaska}, {Starr}, {Filippenko}, {Falco}, {Szentgyorgyi}, {Wren},
  {Wozniak}, {White}, \& {Pergande}}]{2008ApJ...672..449P}
{Perley}, D.~A., {Bloom}, J.~S., {Butler}, N.~R., {et~al.} 2008, \apj, 672, 449

\bibitem[{{Perley} {et~al.}(2009{\natexlab{b}}){Perley}, {Cenko}, {Bloom},
  {Chen}, {Butler}, {Kocevski}, {Prochaska}, {Brodwin}, {Glazebrook},
  {Kasliwal}, {Kulkarni}, {Lopez}, {Ofek}, {Pettini}, {Soderberg}, \&
  {Starr}}]{2009AJ....138.1690P}
{Perley}, D.~A., {Cenko}, S.~B., {Bloom}, J.~S., {et~al.} 2009{\natexlab{b}},
  \aj, 138, 1690

\bibitem[{{Perri} \& {Stratta}(2009)}]{2009GCN..9031....1P}
{Perri}, M. \& {Stratta}, G. 2009, GRB Coordinates Network, 9031

\bibitem[{{Piran}(2004)}]{2004RvMP...76.1143P}
{Piran}, T. 2004, Reviews of Modern Physics, 76, 1143

\bibitem[{{Piron} {et~al.}(2009){Piron}, {Longo}, {Iafrate}, {Cheung}, {Tajima
  }, \& {Connaughton}}]{090626_LAT}
{Piron}, F., {Longo}, F., {Iafrate}, G., {et~al.} 2009, GRB Coordinates
  Network, 9584

\bibitem[{{Polletta} {et~al.}(2007){Polletta}, {Tajer}, {Maraschi},
  {Trinchieri}, {Lonsdale}, {Chiappetti}, {Andreon}, {Pierre}, {Le F{\`e}vre},
  {Zamorani}, {Maccagni}, {Garcet}, {Surdej}, {Franceschini}, {Alloin},
  {Shupe}, {Surace}, {Fang}, {Rowan-Robinson}, {Smith}, \&
  {Tresse}}]{2007ApJ...663...81P}
{Polletta}, M., {Tajer}, M., {Maraschi}, L., {et~al.} 2007, \apj, 663, 81

\bibitem[{{Prochaska} {et~al.}(2006){Prochaska}, {Bloom}, {Chen}, {Foley},
  {Perley}, {Ramirez-Ruiz}, {Granot}, {Lee}, {Pooley}, {Alatalo}, {Hurley},
  {Cooper}, {Dupree}, {Gerke}, {Hansen}, {Kalirai}, {Newman}, {Rich}, {Richer},
  {Stanford}, {Stern}, \& {van Breugel}}]{2006ApJ...642..989P}
{Prochaska}, J.~X., {Bloom}, J.~S., {Chen}, H.-W., {et~al.} 2006, \apj, 642,
  989

\bibitem[{{Racusin} {et~al.}(2008){Racusin}, {Karpov}, {Sokolowski}, {Granot},
  {Wu}, {Pal'Shin}, {Covino}, {van der Horst}, {Oates}, {Schady}, {Smith},
  {Cummings}, {Starling}, {Piotrowski}, {Zhang}, {Evans}, {Holland}, {Malek},
  {Page}, {Vetere}, {Margutti}, {Guidorzi}, {Kamble}, {Curran}, {Beardmore},
  {Kouveliotou}, {Mankiewicz}, {Melandri}, {O'Brien}, {Page}, {Piran},
  {Tanvir}, {Wrochna}, {Aptekar}, {Barthelmy}, {Bartolini}, {Beskin}, {Bondar},
  {Bremer}, {Campana}, {Castro-Tirado}, {Cucchiara}, {Cwiok}, {D'Avanzo},
  {D'Elia}, {Della Valle}, {de Ugarte Postigo}, {Dominik}, {Falcone}, {Fiore},
  {Fox}, {Frederiks}, {Fruchter}, {Fugazza}, {Garrett}, {Gehrels},
  {Golenetskii}, {Gomboc}, {Gorosabel}, {Greco}, {Guarnieri}, {Immler},
  {Jelinek}, {Kasprowicz}, {La Parola}, {Levan}, {Mangano}, {Mazets},
  {Molinari}, {Moretti}, {Nawrocki}, {Oleynik}, {Osborne}, {Pagani}, {Pandey},
  {Paragi}, {Perri}, {Piccioni}, {Ramirez-Ruiz}, {Roming}, {Steele}, {Strom},
  {Testa}, {Tosti}, {Ulanov}, {Wiersema}, {Wijers}, {Winters}, {Zarnecki},
  {Zerbi}, {M{\'e}sz{\'a}ros}, {Chincarini}, \&
  {Burrows}}]{2008Natur.455..183R}
{Racusin}, J.~L., {Karpov}, S.~V., {Sokolowski}, M., {et~al.} 2008, \nat, 455,
  183

\bibitem[{{Racusin} {et~al.}(2009){Racusin}, {Liang}, {Burrows}, {Falcone},
  {Sakamoto}, {Zhang}, {Zhang}, {Evans}, \& {Osborne}}]{2009ApJ...698...43R}
{Racusin}, J.~L., {Liang}, E.~W., {Burrows}, D.~N., {et~al.} 2009, \apj, 698,
  43

\bibitem[{{Rau} {et~al.}(2009{\natexlab{a}}){Rau}, {Connaughton}, \&
  {Briggs}}]{2009GCN..9057....1R}
{Rau}, A., {Connaughton}, V., \& {Briggs}, M. 2009{\natexlab{a}}, GRB
  Coordinates Network, 9057

\bibitem[{{Rau} {et~al.}(2009{\natexlab{b}}){Rau}, {McBreen}, {Kr\"uhler}, \&
  {Greiner}}]{2009_arne}
{Rau}, A., {McBreen}, S., {Kr\"uhler}, T., \& {Greiner}, J. 2009{\natexlab{b}},
  GRB Coordinates Network, 9353

\bibitem[{{Razzaque} {et~al.}(2009){Razzaque}, {Dermer}, \&
  {Finke}}]{2009arXiv0908.0513R}
{Razzaque}, S., {Dermer}, C.~D., \& {Finke}, J.~D. 2009, ArXiv e-prints,
  0908.0513

\bibitem[{{Rees} \& {Meszaros}(1998)}]{1998ApJ...496L...1R}
{Rees}, M.~J. \& {Meszaros}, P. 1998, \apjl, 496, L1

\bibitem[{{Roming} {et~al.}(2005){Roming}, {Kennedy}, {Mason}, {Nousek}, {Ahr},
  {Bingham}, {Broos}, {Carter}, {Hancock}, {Huckle}, {Hunsberger}, {Kawakami},
  {Killough}, {Koch}, {McLelland}, {Smith}, {Smith}, {Soto}, {Boyd},
  {Breeveld}, {Holland}, {Ivanushkina}, {Pryzby}, {Still}, \&
  {Stock}}]{2005SSRv..120...95R}
{Roming}, P.~W.~A., {Kennedy}, T.~E., {Mason}, K.~O., {et~al.} 2005, Space
  Science Reviews, 120, 95

\bibitem[{{Rossi} {et~al.}(2008){Rossi}, {de Ugarte Postigo}, {Ferrero},
  {Kann}, {Klose}, {Schulze}, {Greiner}, {Schady}, {Filgas}, {Gonsalves},
  {K{\"u}pc{\"u} Yolda{\c s}}, {Kr{\"u}hler}, {Szokoly}, {Yolda{\c s}},
  {Afonso}, {Clemens}, {Bloom}, {Perley}, {Fynbo}, {Castro-Tirado},
  {Gorosabel}, {Kub{\'a}nek}, {Updike}, {Hartmann}, {Giuliani}, {Holland},
  {Hanlon}, {Bremer}, {French}, {Melady}, \&
  {Garc{\'{\i}}a-Hern{\'a}ndez}}]{2008A&A...491L..29R}
{Rossi}, A., {de Ugarte Postigo}, A., {Ferrero}, P., {et~al.} 2008, \aap, 491,
  L29

\bibitem[{{Salvaterra} {et~al.}(2009){Salvaterra}, {Della Valle}, {Campana},
  {Chincarini}, {Covino}, {D'Avanzo}, {Fernandez-Soto}, {Guidorzi}, {Mannucci},
  {Margutti}, {Thoene}, {Antonelli}, {Barthelmy}, {De Pasquale}, {D'Elia},
  {Fiore}, {Fugazza}, {Hunt}, {Maiorano}, {Marinoni}, {Marshall}, {Molinari},
  {Nousek}, {Pian}, {Racusin}, {Stella}, {Amati}, {Andreuzzi}, {Cusumano},
  {Fenimore}, {Ferrero}, {Giommi}, {Guetta}, {Holland}, {Hurley}, {Israel},
  {Mao}, {Markwardt}, {Masetti}, {Pagani}, {Palazzi}, {Palmer}, {Piranomonte},
  {Tagliaferri}, \& {Testa}}]{2009arXiv0906.1578S}
{Salvaterra}, R., {Della Valle}, M., {Campana}, S., {et~al.} 2009, Nature, 461,
  1258

\bibitem[{{Sari} {et~al.}(1999){Sari}, {Piran}, \&
  {Halpern}}]{1999ApJ...519L..17S}
{Sari}, R., {Piran}, T., \& {Halpern}, J.~P. 1999, \apjl, 519, L17

\bibitem[{{Sari} {et~al.}(1998){Sari}, {Piran}, \&
  {Narayan}}]{1998ApJ...497L..17S}
{Sari}, R., {Piran}, T., \& {Narayan}, R. 1998, \apjl, 497, L17

\bibitem[{{Savage} \& {Sembach}(1991)}]{1991ApJ...379..245S}
{Savage}, B.~D. \& {Sembach}, K.~R. 1991, \apj, 379, 245

\bibitem[{{Savaglio} {et~al.}(2009){Savaglio}, {Glazebrook}, \&
  {LeBorgne}}]{2009ApJ...691..182S}
{Savaglio}, S., {Glazebrook}, K., \& {LeBorgne}, D. 2009, \apj, 691, 182

\bibitem[{{Scargle} {et~al.}(2008){Scargle}, {Norris}, \&
  {Bonnell}}]{2008ApJ...673..972S}
{Scargle}, J.~D., {Norris}, J.~P., \& {Bonnell}, J.~T. 2008, \apj, 673, 972

\bibitem[{{Schlegel} {et~al.}(1998){Schlegel}, {Finkbeiner}, \&
  {Davis}}]{sch98}
{Schlegel}, D.~J., {Finkbeiner}, D.~P., \& {Davis}, M. 1998, \apj, 500, 525

\bibitem[{{Seaton}(1979)}]{sea79}
{Seaton}, M.~J. 1979, \mnras, 187, 73P

\bibitem[{{Skrutskie} {et~al.}(2006)}]{skr06}
{Skrutskie}, M.~F. {et~al.} 2006, \aj, 131, 1163

\bibitem[{{Spitzer}(1978)}]{1978ppim.book.....S}
{Spitzer}, L. 1978, {Physical processes in the interstellar medium} (New York
  Wiley)

\bibitem[{{Stecker} {et~al.}(2006){Stecker}, {Malkan}, \&
  {Scully}}]{2006ApJ...648..774S}
{Stecker}, F.~W., {Malkan}, M.~A., \& {Scully}, S.~T. 2006, \apj, 648, 774

\bibitem[{{Stratta} {et~al.}(2009){Stratta}, {D'Elia}, \&
  {Perri}}]{XRT2_090902B}
{Stratta}, G., {D'Elia}, V., \& {Perri}, M. 2009, GRB Coordinates Network, 9876

\bibitem[{{Swenson} \& {Siegel}(2009)}]{UVOT_090902B}
{Swenson}, C.~A. \& {Siegel}, M.~H. 2009, GRB Coordinates Network, 9869

\bibitem[{{Swenson} \& {Stratta}(2009)}]{UVOT2_090902B}
{Swenson}, C.~A. \& {Stratta}, G. 2009, GRB Coordinates Network, 9877

\bibitem[{{Tanvir} {et~al.}(2009){Tanvir}, {Fox}, {Levan}, {Berger},
  {Wiersema}, {Fynbo}, {Cucchiara}, {Kr\"uhler}, {Gehrels}, {Bloom}, {Greiner},
  {Evans}, {Rol}, {Olivares}, {Hjorth}, {Jakobsson}, {Farihi}, {Willingale},
  {Starling}, {Cenko}, {Perley}, {Maund}, {Duke}, {Wijers}, {Adamson}, {Allan},
  {Bremer}, {Burrows}, {Castro Tirado}, {Cavanagh}, {de Ugarte Postigo},
  {Dopita}, {Fatkhullin}, {Fruchter}, {Foley}, {Gorosabel}, {Holland},
  {Kennea}, {Kerr}, {Klose}, {Krimm}, {Komarova}, {Kulkarni}, {Moskvitin},
  {Naylor}, {Penprase}, {Perri}, {Podsiadlowski}, {Roth}, {Rutledge},
  {Sakamoto}, {Schady}, {Schmidt}, {Soderberg}, {Sollerman}, {Stephens},
  {Stratta}, {Ukwatta}, {Watson}, {Westra}, {Wold}, \&
  {Wolf}}]{2009arXiv0906.1577T}
{Tanvir}, N.~R., {Fox}, D.~B., {Levan}, A.~J., {et~al.} 2009, Nature, 461, 1254

\bibitem[{{Tinney} {et~al.}(1998){Tinney}, {Stathakis}, {Cannon}, \&
  {Galama}}]{1998IAUC.6896....3W}
{Tinney}, C., {Stathakis}, R., {Cannon}, R., \& {Galama}, T. 1998, \iaucirc,
  6896, 3

\bibitem[{{Tody}(1993)}]{1993ASPC...52..173T}
{Tody}, D. 1993, in Astronomical Society of the Pacific Conference Series,
  Vol.~52, Astronomical Data Analysis Software and Systems II, ed. R.~J.
  {Hanisch}, R.~J.~V. {Brissenden}, \& J.~{Barnes}

\bibitem[{{Troja} {et~al.}(2008){Troja}, {King}, {O'Brien}, {Lyons}, \&
  {Cusumano}}]{2008MNRAS.385L..10T}
{Troja}, E., {King}, A.~R., {O'Brien}, P.~T., {Lyons}, N., \& {Cusumano}, G.
  2008, \mnras, 385, L10

\bibitem[{{Uehara} {et~al.}(2009){Uehara}, {Takahashi}, \&
  {McEnery}}]{LAT_090926}
{Uehara}, T., {Takahashi}, H., \& {McEnery}, J. 2009, GRB Coordinates Network,
  9934

\bibitem[{{Updike} {et~al.}(2009{\natexlab{a}}){Updike}, {Klose}, {Clemens}, \&
  {Greiner}}]{2009GCN..9054....1U}
{Updike}, A., {Klose}, S., {Clemens}, C., \& {Greiner}, J. 2009{\natexlab{a}},
  GRB Coordinates Network, 9054

\bibitem[{{Updike} {et~al.}(2009{\natexlab{b}}){Updike}, {Filgas}, {Kr\"uhler},
  {Greiner}, \& {McBreen}}]{2009GCN..9026....1U}
{Updike}, A.~C., {Filgas}, R., {Kr\"uhler}, T., {Greiner}, J., \& {McBreen}, S.
  2009{\natexlab{b}}, GRB Coordinates Network, 9026

\bibitem[{{Updike} {et~al.}(2008){Updike}, {Haislip}, {Nysewander}, {Fruchter},
  {Kann}, {Klose}, {Milne}, {Williams}, {Zheng}, {Hergenrother}, {Prochaska},
  {Halpern}, {Mirabal}, {Thorstensen}, {van der Horst}, {Starling}, {Racusin},
  {Burrows}, {Kuin}, {Roming}, {Bellm}, {Hurley}, {Li}, {Filippenko}, {Blake},
  {Starr}, {Falco}, {Brown}, {Dai}, {Deng}, {Xin}, {Qiu}, {Wei}, {Urata},
  {Nanni}, {Maiorano}, {Palazzi}, {Greco}, {Bartolini}, {Guarnieri},
  {Piccioni}, {Pizzichini}, {Terra}, {Misra}, {Bhatt}, {Anupama}, {Fan},
  {Jiang}, {Wijers}, {Reichart}, {Eid}, {Bryngelson}, {Puls}, {Goldthwaite}, \&
  {Hartmann}}]{2008ApJ...685..361U}
{Updike}, A.~C., {Haislip}, J.~B., {Nysewander}, M.~C., {et~al.} 2008, \apj,
  685, 361

\bibitem[{{van der Horst } {et~al.}(2009){van der Horst }, {Kamble}, {Wijers},
  \& {Kouveliotou}}]{Alexander_090902B}
{van der Horst }, A.~J., {Kamble}, A.~P., {Wijers}, R.~A.~M.~J., \&
  {Kouveliotou}, C. 2009, GRB Coordinates Network, 9883

\bibitem[{{van der Horst}(2009)}]{2009GCN..9047....1V}
{van der Horst}, A.~J. 2009, GRB Coordinates Network, 9047

\bibitem[{{Vedrenne} \& {Atteia}(2009)}]{2009grbb.book.....V}
{Vedrenne}, G. \& {Atteia}, J.~L. 2009, {Gamma-Ray Bursts: The brightest
  explosions in the Universe} (Springer Praxis Books, Astronomy and Planetary
  Sciences Jointly published with Praxis Publishing)

\bibitem[{{Willmer} {et~al.}(2006){Willmer}, {Faber}, {Koo}, {Weiner},
  {Newman}, {Coil}, {Connolly}, {Conroy}, {Cooper}, {Davis}, {Finkbeiner},
  {Gerke}, {Guhathakurta}, {Harker}, {Kaiser}, {Kassin}, {Konidaris}, {Lin},
  {Luppino}, {Madgwick}, {Noeske}, {Phillips}, \& {Yan}}]{2006ApJ...647..853W}
{Willmer}, C.~N.~A., {Faber}, S.~M., {Koo}, D.~C., {et~al.} 2006, \apj, 647,
  853

\bibitem[{{Wilson-Hodge} {et~al.}(2008){Wilson-Hodge}, {Connaughton}, {Longo},
  \& {Omodei}}]{2008GCN..8723....1W}
{Wilson-Hodge}, C., {Connaughton}, V., {Longo}, F., \& {Omodei}, N. 2008, GRB
  Coordinates Network, 8723

\bibitem[{{Zeh} {et~al.}(2004){Zeh}, {Klose}, \&
  {Hartmann}}]{2004ApJ...609..952Z}
{Zeh}, A., {Klose}, S., \& {Hartmann}, D.~H. 2004, \apj, 609, 952

\bibitem[{{Zeh} {et~al.}(2006){Zeh}, {Klose}, \& {Kann}}]{2006ApJ...637..889Z}
{Zeh}, A., {Klose}, S., \& {Kann}, D.~A. 2006, \apj, 637, 889

\bibitem[{{Zhang} {et~al.}(2006){Zhang}, {Fan}, {Dyks}, {Kobayashi},
  {M{\'e}sz{\'a}ros}, {Burrows}, {Nousek}, \& {Gehrels}}]{2006ApJ...642..354Z}
{Zhang}, B., {Fan}, Y.~Z., {Dyks}, J., {et~al.} 2006, \apj, 642, 354

\bibitem[{{Zhang} \& {M{\'e}sz{\'a}ros}(2004)}]{2004IJMPA..19.2385Z}
{Zhang}, B. \& {M{\'e}sz{\'a}ros}, P. 2004, International Journal of Modern
  Physics A, 19, 2385

\bibitem[{{Zhang} {et~al.}(2009){Zhang}, {Zhang}, {Virgili}, {Liang}, {Kann},
  {Wu}, {Proga}, {Lv}, {Toma}, {M{\'e}sz{\'a}ros}, {Burrows}, {Roming}, \&
  {Gehrels}}]{Zhang080913}
{Zhang}, B., {Zhang}, B., {Virgili}, F.~J., {et~al.} 2009, \apj, 703, 1696

\end{thebibliography}

\end{document}